\documentclass[12pt,fleqn,reqno]{amsart} 

\usepackage{lipsum}
\makeatletter
\g@addto@macro{\endabstract}{\@setabstract}
\newcommand{\authorfootnotes}{\renewcommand\thefootnote{\@fnsymbol\c@footnote}}%
\makeatother

\usepackage{epsfig}
\usepackage{graphicx}
\usepackage{enumerate}
\usepackage{color} 
\usepackage[toc,page]{appendix}

\usepackage{setspace}

\usepackage[dvipsnames]{xcolor}
\usepackage{bm,bbm}
\usepackage{cancel}
\usepackage{enumitem}

\usepackage{physics}

\usepackage{amsmath,euscript}
\usepackage{amsthm}
\usepackage{amssymb}
\usepackage{graphicx}
\usepackage{mathrsfs}
\usepackage{epsf}
\usepackage{xcolor}
\usepackage{psfrag}
\usepackage{enumerate}
\usepackage{amsfonts,amssymb,amsthm,amsmath} 
\usepackage{hyperref}

\usepackage[numbers]{natbib}
\usepackage{enumitem} 
\usepackage{xspace}
\usepackage[margin=1.37in]{geometry}

\usepackage{changes}
\definechangesauthor[name={SB}, color=magenta]{sb}
\definechangesauthor[name={JF}, color=blue]{jf}
\definechangesauthor[name={ML}, color=red]{ml}
\definechangesauthor[name={MS}, color=green]{ms}

\usepackage{soul}

\newtheorem{thm}{Theorem}

\newtheorem{lemma}[thm]{Lemma}
\newtheorem{prop}[thm]{Proposition}
\newtheorem{cor}[thm]{Corollary}
\newtheorem{Def}[thm]{Definition}

\theoremstyle{remark}
\newtheorem{remark}[thm]{Remark}

\theoremstyle{definition}

\numberwithin{thm}{section}
\numberwithin{equation}{section}

\usepackage{soul}
\setstcolor{red}

\definecolor{green}{rgb}{0.0, 0.5, 0.5}
\definecolor{yellow}{rgb}{0.5, 0.5, 0}
\definecolor{lgray}{gray}{0.9}
\definecolor{llgray}{gray}{0.95}
\definecolor{lllgray}{gray}{0.975}

\newcommand{\nc}{\newcommand}

\nc{\la}{\label}
\nc{\ba}{\begin{array}}
\nc{\ea}{\end{array}}
\nc{\bs}{\begin{split}}
\nc{\es}{\end{split}}

\newcommand{\R}{\mathbb{R}}
\newcommand{\C}{\mathbb{C}}
\newcommand{\Z}{\mathbb{Z}}

\newcommand{\cB}{\mathcal{B}}

\newcommand{\cF}{\mathcal{F}}

\newcommand{\cH}{\mathcal{H}}
\newcommand{\cI}{\mathcal{I}}

\newcommand{\cS}{\mathcal{S}}

\newcommand{\sD}{\mathscr{D}}

\newcommand{\sF}{\mathscr{F}}

\nc{\al}{\alpha}
\nc{\del}{\delta}
\nc{\h}{\delta}
\nc{\G}{\Gamma}
\nc{\et}{\eta} 
\nc{\g}{\gamma}
\nc{\gam}{\gamma}
\nc{\ka}{\kappa}
\nc{\lam}{\lambda}
\nc{\Lam}{\Lambda}
\nc{\Om}{\Omega}
\nc{\om}{\omega}

\nc{\ta}{\tau}
\nc{\w}{\omega}
\nc{\io}{\iota}
\nc{\z}{\zeta}
\nc{\s}{\sigma}
\nc{\Si}{\Sigma}
\nc{\vphi}{\varphi}
\nc{\ran}{\rangle}
\nc{\lan}{\langle}

\newcommand{\one}{\mathbbm1} 

\newcommand{\bbone}{\mathbbm1}

\nc{\bfone}{{\bf 1}}
\nc{\1}{{\bf 1}}

\newcommand{\Ran}{\operatorname{Ran}}
\newcommand{\Null}{\operatorname{Null}}

\renewcommand{\Re}{\mathrm{Re}} 
\renewcommand{\Im}{\mathrm{Im}} 

\newcommand{\dist}{\mathrm{dist}}

\usepackage{centernot}

\usepackage[normalem]{ulem}

\newcommand{\DETAILS}[1]{}

\nc{\den}{\text{den}}
\nc{\ex}{\text{xc}}
\nc{\Ex}{\text{Xc}}

\usepackage{soul}

\begin{document}
\title[Approach to Equilibrium in Markovian open quantum systems]{Approach to Equilibrium in Markovian open quantum systems}
\maketitle

\begin{center}
	\authorfootnotes
	Dong Hao Ou Yang\footnote{\textit{A corresponding author Email}: ouyang@math.lmu.de}\textsuperscript{1} and Israel Michael Sigal\footnote{\textit{A corresponding author Email}: im.sigal@utoronto.ca}\textsuperscript{2} \par \bigskip
\end{center}
\textsuperscript{1}LMU Munich, Dept of Mathematics, Theresienstr. 39, 80333 Munich, Germany \\
\textsuperscript{2}Dept of Mathematics, University of Toronto, Toronto, M5S 2E4, Canada \par\bigskip

	\begin{abstract}
		In this paper, we study the evolution of Markovian open quantum systems, whose dynamics are governed by the von Neumann-Lindblad equations.  Our goal is to prove the return-to-equilibrium property for systems of infinite degrees of freedom under quantum detailed balance condition.
	\end{abstract}

	\tableofcontents
	
	\section{Introduction}\label{sec:intro}
	\subsection{Von Neumann-Lindblad equation}
	Consider a quantum system described by a quantum Hamiltonian $H$ on a Hilbert space $\cH$.  Assume this system interacts with a (quantum) environment and we trace/integrate out the environment's degrees of freedom.  The resulting physical system is called the \textit{open quantum system}.  Its states are given by density operators $\rho$ (i.e. positive, trace-class operators, $\rho=\rho^{*}\geq 0$, $\Tr(\rho)<\infty$)  on $\cH$, and its dynamic, $\beta_{t}:\rho_{0}\mapsto\rho_{t}$, results from a unitary dynamics of the total system by tracing out the environment.  Under the assumption that $\beta_{t}$ is Markovian and uniformly continuous, $\beta_{t}(\rho_{0})$ satisfies the {\it von Neumann-Lindblad (vNL) equation} (or the \textit{quantum master equation})
	\begin{align}\label{vNLeq}
		&\frac{\partial\rho}{\partial t}=-i[H,\rho]+\sum_{j}\left(W_{j}\rho W_{j}^{*}-\frac{1}{2}\{W_{j}^{*}W_{j},\rho\}\right), 
	\end{align}
	with the initial condition $\rho|_{t=0}=\rho_{0}$, where  $H$ is, as above, the quantum Hamiltonian of a proper quantum system, which is bounded in this case, $W_{j}$ are bounded operators on $\cH$ such that $\sum\nolimits_{j}W_{j}^{*}W_{j}$ converges weakly and $\{\cdot,\cdot\}$ is the anti-commutator.   
	(Here and in what follows we use the units in which the Planck constant is set to $2\pi$ and speed of light, to one: $\hbar=1$ and $c=1$.)   The converse statement is also true.  Moreover, it is shown in \cite{Davies1} that for $H$ self-adjoint and $W_{j}$ $H$-bounded for each $j$, the operator on the r.h.s. of \eqref{vNLeq} generates a one-parameter, positive contraction semigroup $\beta_{t}$.  Such semigroup defines a weak solution to \eqref{vNLeq} (in the sense explained below) on the Schatten space $\cS_{1}$ of trace-class operators on $\cH$.  The meaning of the expression on the r.h.s. of \eqref{vNLeq} is discussed in Section \ref{sec:generators}.


	
	The set of operators $W_j$, called {\it jump} or \textit{Kraus-Lindblad operators}, is what is left over from the  interaction with the environment. If we set them to zero, then the second term on the r.h.s. of \eqref{vNLeq} drops out and \eqref{vNLeq} reduces to 
	the von Neumann equation
	\begin{equation}\label{vN eqn}
		\frac{\partial\rho}{\partial t}=-i[H,\rho],\hspace{0.5cm}\rho_{t=0}=\rho_{0},
	\end{equation}
	for the resulting closed system. The latter extends the Schr\"odinger equation to quantum statistics (see \cite{GS} for some definitions).  Put differently, Markovian open quantum systems present an extension of quantum mechanics incorporating terms resulting from interaction with the environment.
	
	
	Due to the presence of jump operators, the evolution described by \eqref{vNLeq} is generally dissipative, i.e., the energy of the system will decrease (``dissipating into environment").  
	
	
	We will call the operators due to the two terms on the r.h.s. of \eqref{vNLeq}  as the \textit{von Neumann operator} and \textit{Lindblad operator}, respectively.
	
	
	
	It is assumed that the Markovian open quantum dynamics approximate open quantum dynamics coming from quantum systems weakly coupled to an environment.  So far this is proven, under some technical conditions, for the simplest environment given by the free massless fermion quantum field and simplest interaction (linear in field) in the van Hove limit, see \cite{Davies-QME,JP,KB} for finite-dimensional systems and  \cite{Davies-QME2,Davies-QME3}, for infinite-dimensional ones.
	

	In quantum computations, the vNL equation is also used for preparation of the Gibbs and ground states (a quantum Monte-Carlo), see \cite{CB, CKB, Cubitt, DCL, KB, SM, TOVPV, YungGuzik}.
	
	
	The vNL equation is closely related to the non-commutative analogue of the dissipative stochastic equations in a separable Hilbert space $\cH$ (see \cite{Holevo})
	\begin{align}\label{stochastic}
		\partial_{t}\psi(t)&=\sum_{j}W_{j}\psi(t)dw_{j}(t)-K\psi(t).
	\end{align}
	Here $w_{j}(t)$, $j=1,2,...$, are independent standard Wiener processes, $K=-iH-\frac{1}{2}\sum_{j\geq 1}W_{j}^{*}W_{j}$, with $H$ and $W_{j}$'s as given above.  Let $\psi(t)$ be a solution to \eqref{stochastic} with an initial condition $\psi_{0}\in\cH$ and $\textbf{E}$ stands for expectation w.r.t. the Wiener measure.  Under some technical conditions, the equation
	\begin{align}\label{stochastic2}
		\langle\psi_{0},\beta_{t}'(A)\psi_{0}\rangle_{\cH}&=\textbf{E}\langle\psi(t),A\psi(t)\rangle_{\cH},\quad t\geq 0,
	\end{align}
	defines the quantum dynamical semigroup $\beta_{t}'$ on the space $\cB(\cH)$ of bounded operators on $\cH$, which is dual to $\beta_{t}$ (see \cite{Holevo}, Section 3). 
	

	
	
	In this paper, we are interested in the long time behaviour of solutions of the vNL equation on an infinite-dimensional Hilbert space $\cH$, under the assumptions 
	\begin{enumerate}
		\item[$(H)$] $H$ is a self-adjoint operator on a Hilbert space $\cH$ such that $\Tr(e^{-H/\tau})<\infty$ for some $\tau>0$;
		\item[$(W)$] $W_{j}$ are bounded operators on $\cH$ such that the sum $\sum_{j}W_{j}^{*}W_{j}$ converges in the weak$^{*}$ topology,
	\end{enumerate}
	and, under the \textit{quantum detailed balance condition $(QDB)$} (see Subsection \ref{sec:QDB} for the definition and Appendix \ref{sec:jump-oprs} for a class of the operators $H$ and $W_{j}$'s satisfying this condition and for its interpretation).
	

\DETAILS{	\begin{enumerate}
		\item[$(H)$] $H$ is a self-adjoint operator on a Hilbert space $\cH$;
		\item[$(W)$] $W_{j}$ are bounded operators on $\cH$ such that the sum $\sum_{j\geq 1}W_{j}^{*}W_{j}$ converges weakly on $\cH$.
	\end{enumerate}
	For our main results, we assume also
	\begin{enumerate}
		\item[$(Tr)$] $\Tr(e^{-H/\tau})<\infty$ for some $\tau>0$,
	\end{enumerate}
	and, the \textit{quantum detailed balance condition $(QDB)$} (see Subsection \ref{sec:$(QDB)$} for the definition). 
	
	\vspace{3cm}

	\begin{align}\label{S*}
		\cS_{\tau}&:=\{\lambda\in\cS_{1}\mid \Tr(\rho_{\tau}^{-1}|\lambda|^{2})<\infty\},
	\end{align}
	with the inner product
	\begin{align}\label{st-ip}
		\langle \lambda,\mu\rangle_{\rm st,\tau}&:=\Tr(\lambda^{*}\mu\rho_{\tau}^{-1}).
	\end{align}
	In this paper, we take $\rho_{\tau}$ to be the Gibbs state  $\rho_{\tau}=e^{-H/\tau}/Z(\tau)$, with $Z(\tau):=\Tr(e^{-H/\tau})$, at a temperature $\tau>0$.  This is where Condition $(Tr)$ is used.}
	


	
	To state our main results,  let $\rho_{\tau}:=e^{-H/\tau}/Z(\tau)$, $Z(\tau)=\Tr(e^{-H/\tau})$, be the Gibbs state at a temperature $\tau$ and $\rho_{\tau}$, we define the Hilbert space 
	\begin{align}\label{Stau}
		\cS_{\tau}&:=\{\lambda\in\cS_{1}\mid \Tr(\lambda^{*}\lambda\rho_{\tau}^{-1})<\infty\},
	\end{align}
	with the inner product 
\begin{align}\label{ip-st}
		\langle \lambda,\mu\rangle_{\rm st,\tau}&:=\Tr(\lambda^{*}\mu\rho_{\tau}^{-1}).
	\end{align}
		
	
	With the definition of weak solution given in \eqref{weak-sol} below, we begin with the following existence result:
	\begin{thm}\label{thm:vNL-exist} 
		Assume Conditions $(H)$, $(W)$ and $(QDB)$ hold.  Then the vNL equation \eqref{vNLeq} has a unique weak solution for any initial condition in $\cS_{\tau}$ and unique strong solution for any initial condition in the set 
		\begin{align}
			\sD&=\{\rho\in\cS_{\tau}\mid \rho\sD(H)\subseteq\sD(H)\text{ and }[H,\rho]\in\cS_{\tau}\}.
		\end{align}
	\end{thm}
	This theorem is proven in Subsection \ref{sec:pf-vNL-HL-exist}.  
 The next theorem describes the long-time behavior of solutions of vNL equation \eqref{vNLeq} under general Conditions $(H)$, $(W)$ 
 and $(QDB)$.
\DETAILS{		\begin{thm}\label{thm:mains}
			Assume Conditions $(H)$, $(W)$, $(Tr)$ and $(QDB)$ hold.  Then, the quantum evolution $\beta_{t}(\lambda)$ converges, in the ergodic sense, to the space $\cS_{\rm stat}$ of stationary states of $\beta_{t}$:
			\begin{align}
				s\text{-}\lim_{T\rightarrow\infty}\frac{1}{T}\int_{0}^{T}\beta_{t}dt&=P,
			\end{align}
			strongly in $\cS_{\tau}$, where $P$ is the orthogonal projection onto $\cS_{\rm stat}$.
		\end{thm}}

	\begin{thm}\label{thm:main}
		Assume Conditions $(H)$, $(W)$ and $(QDB)$. 
		Then, for all initial condition $\lambda\in\cS_{\tau}$, the vNL quantum evolution $\beta_{t}(\lambda)$ converges, in the ergodic sense, to the space $\cS_{\rm stat}:=\{\rho\in\cS_{\tau}\mid \beta_{t}(\rho)=\rho\}$ of stationary states of $\beta_{t}$ (or \eqref{vNLeq}):
		\begin{align}\label{main}
			s\text{-}\lim_{T\rightarrow\infty}\frac{1}{T}\int_{0}^{T}\beta_{t}dt&=P,
		\end{align}
		strongly in $\cS_{\tau}$, where $P$ is the orthogonal projection onto $\cS_{\rm stat}$.
	\end{thm}
		This theorem is proven in Subsection \ref{sec:mains-pf}.
		This result shows the approach to stationary states for systems of infinite number of degrees of freedom.  We expect it could be extended to unbounded operators $W_{j}$'s.
	

		If $\dim\cS_{\rm stat}>1$, then Theorem \ref{thm:main} is relevant to the theory of measurement.  There
		\begin{itemize}
			\item the environment is the measuring apparatus, and
			\item the space $\cS_{\rm stat}$ of stationary states of \eqref{vNLeq} can be identified with the space of possible outcomes of the measurement process.  
		\end{itemize}
		\eqref{main} gives the probabilities of various measurement outcomes (depending on the initial conditions).




	It follows from $(QDB)$ (see Eq. \eqref{vNLeq2} and \eqref{L-zeroEV} below) that the Gibbs state $\rho_{\tau}$ is a stationary state of \eqref{vNLeq}, or $\beta_{t}$, so that $\beta_{t}(\rho_{\tau})=\rho_{\tau}$.  Condition $(U)$, formulated below,  then guarantees that $\rho_{\tau}$ is the unique stationary solution to \eqref{vNLeq} (see Proposition \ref{prop:simp-zero}).  This sharpens Theorem \ref{thm:main} giving raise to the term ``return to equilibrium". 

	
	\begin{thm}\label{thm:main1}
		Assume Conditions $(H)$, $(W)$ and $(QDB)$ w.r.t. $\rho_{\tau}$ hold.  Suppose further that
		\begin{enumerate}
			\item[$(U)$] $[W_{j},A]=[W_{j}^{*},A]=0$ $\forall j$ $\Rightarrow$ $A=\one$.
		\end{enumerate}
		Then, for all $\lambda\in\cS_{\tau}$ with $\Tr\lambda=1$, the vNL quantum evolution $\beta_{t}(\lambda)$ converges in $\cS_{\tau}$ to the Gibbs state $\rho_{\tau}$ in the ergodic sense:
		\begin{align}
			\lim_{T\rightarrow\infty}\frac{1}{T}\int_{0}^{T}\beta_{t}(\lambda)dt&=\rho_{\tau}.
		\end{align}
	\end{thm}


	This result shows the approach to stationary states for systems of infinite number of degrees of freedom.  We expect it could be extended to unbounded operators $W_{j}$'s.
	
	
	\begin{remark}
		By considering the space $\cB$ of bounded operators on $\cH$ as a $C^{*}$-algebra, Condition $(U)$ is satisfied if the collection $\{W_{j},W_{j}^{*}\}_{j\geq 1}$ of operators generates the whole $C^{*}$-algebra $\cB$.
	\end{remark}
	
	
	\begin{remark}
		For $\cH=L^{2}(\R^{d})$, the collection $\{W_{j}\}_{j=1}^{2d}$ of operators with $W_{2k-1}=x_{k}/\langle x\rangle$ and $W_{2k}=p_{k}/\langle p\rangle$, $k=1,...,d$, on $L^{2}(\R^{d})$ satisfies Condition $(U)$, where $p_{j}:=i\hbar\partial_{x_{j}}$ are the quantum momentum operators, $\langle x\rangle\equiv \sqrt{1+|x|^{2}}$ and similarly for $\langle p\rangle$.  
		%
	\end{remark}
	

	
	
As was mentioned above, 	 states for a quantum open system are given by density operators $\rho$, i.e. positive, trace-class operators, $\rho=\rho^{*}\geq 0$, $\Tr(\rho)<\infty$ (usually normalized as $\Tr(\rho)=1$), on $\cH$. The set of states is denoted by  $\cS_{1}^{+}:=\{\lambda\in\cS_{1}\mid \lambda\geq 0\}$, where $\cS_{1}$	is the Schatten space of trace-class operators.

	
	
More generally, states can be defined as  positive, normalized, continuous linear functionals on $\cB$. 
 Any density operator $\rho$ defines a state on $\cB$ by $A\mapsto\Tr(A\rho)$.  The duality between observables and states is given by the coupling
	\begin{align}\label{BS-coupling}
		(A,\rho)&:=\Tr(A\rho).
	\end{align}
	
	

		In addition to the von Neumann-Lindblad evolution, we also consider its dual evolution (w.r.t. the coupling $(A,\rho):=\Tr(A\rho)$)
	\begin{align}\label{HLeq}
		\partial_tA_t=i[H,A_{t}]+\sum_{j\geq 1}\left(W_{j}^{*}AW_{j}-\frac{1}{2}\{W_{j}^{*}W_{j},A\}\right),
	\end{align}
	where, recall, $\{A,B\}:=AB+BA$, on the space $\cB$ of observables.  The meaning of the expression on the r.h.s. of \eqref{HLeq} and the existence theory for \eqref{HLeq} are discussed in the next section.  Here, we mention briefly that the first term on the r.h.s. is defined on the set $\{A\in\cB\mid A:\mathscr{D}(H)\rightarrow\mathscr{D}(H)\}$, which is dense in the strongly operator topology on $\cB$, while the sum in the second term converges weakly for any bounded $A$, by Condition $(W)$, as shown in Corollary \ref{cor:G'-bdd} of Subsection \ref{sec:vNL-HL-op}.
	
	
	We call \eqref{HLeq} the \textit{Heisenberg-Lindblad The (HL) equation}.  HL equation \eqref{HLeq} is dual to vNL equation \eqref{vNLeq} w.r.t. the coupling $(A,\rho)=\Tr(A\rho)$ between the observables $A\in\cB$ and states $\rho\in\cS_{1}^{+}$.  

	
	Denote by $\beta_{t}'$ the flow generated by HL equation \eqref{HLeq}.  It is dual to the flow $\beta_{t}$ for vNL equation:
	\begin{align}\label{trace-preserving}
		\Tr(\beta_{t}'(A)\rho)&=\Tr(A\beta_{t}(\rho))\quad\quad\forall A\in\cB,\quad\rho\in\cS_{1}.
	\end{align}
	
		
Depending on the viewpoint, the HL equation can be considered as a more fundamental one. 

		We study 
		HL equation \eqref{HLeq} on 
		 the Hilbert space $\cB_{\tau}$ defined as the completion of $\cB$ w.r.t. the norm $\|A\|_{\tau}:=\sqrt{\Tr(A^{*}A\rho_{\tau})}$ 
		 with the inner product on $\cB$
		\begin{align}\label{ip-obs-2}
			\langle A,B\rangle_{\tau}&:=\Tr(A^{*}B\rho_{\tau}).
		\end{align}

We have the following results for the HLE.

			\begin{thm}\label{thm:HL-exist}
			Suppose 
			Conditions $(H)$, $(W)$ 
			and $(QDB)$.  Then, HL equation \eqref{HLeq} has a unique solution on $\cB_{\tau}$ for any initial condition in 
 the strongly dense set $\sD':=\{A\in\cB\mid A:\sD(H)\rightarrow\sD(H)\}$ (containing the strongly dense set $(H+i)^{-1}\cB(H-i)^{-1}$).		\end{thm}
		This theorem is proven in Subsection \ref{sec:pf-vNL-HL-exist}. 
		

For the next result, we define the Hilbert space $\cB_{\tau, 1}$  as the completion of $\cB$ w.r.t. the norm $\|A\|_{\tau, 1}:=\sqrt{\Tr(A^{*}\rho_{\tau}A)}$ 
		 with the inner product
\begin{align}\label{tau1-ip}
	\langle A,B\rangle_{\tau, 1}&=\Tr(A^{*}\rho_{\tau} B).
\end{align}
Note that  $\langle A,B\rangle_{\tau, 1}=\langle B^{*},A^{*}\rangle_{\tau}$.  As shown in Lemma \ref{lem:2.7a} below, the space $\cB_{\tau, 1}$ is dual to the space $\cS_{\tau}$ w.r. to the coupling \eqref{BS-coupling}. Theorem \ref{thm:main} implies via this duality the following result.  

		Let $P'$ is the orthogonal projector in $\cB_{\tau, 1}$ onto  the space $\cS_{\rm stat}':=\{A\in \cB_{\tau, 1}\mid \beta_{t}'(A)=A\}$$=\Null(L')$ of stationary states of $\beta'_{t}$ (or \eqref{HLeq}) and $P'^{\perp}=\one-P'$.  
		\begin{thm}\label{thm:RTE-dual1}
			Assume Conditions $(H)$, $(W)$ 
			  and $(QDB)$ hold.  Then, the dual quantum evolution $\beta_{t}'(A)$ converges to $\Null(G')$ in the ergodic sense:
			\begin{align}\label{RTE-dual2}
				s\text{-}\lim_{T\rightarrow\infty}\frac{1}{T}\int_{0}^{T}\beta_{t}'dt&=P',
			\end{align}
			strongly in $\cB_{\tau, 1}$.
		\end{thm}
		One can also prove a similar result on the space $\cB_{\tau}$, see Appendix \ref{sec:pf-RTE-dual1'} and Remark \ref{thm:RTE-dual1-Btau}.

	Note finally that, as follows from \eqref{HLeq}, $\one$ is a stationary solution for \eqref{HLeq}.


		\begin{thm}\label{thm:RTE-dual3}
			Assume Conditions $(H)$, $(W)$,
			 $(QDB)$ and $(U)$ (see Theorem \ref{thm:main1}).
			Then, for all $A\in\cB_{\tau, 1}$, we have
			\begin{align}\label{RTE-dual3}
				\lim_{T\rightarrow\infty}\frac{1}{T}\int_{0}^{T}\beta_{t}'(A)dt&=\Tr(A\rho_{\tau})\cdot\one.
			\end{align}
		\end{thm}
		
		Proofs of Theorems 
		\ref{thm:RTE-dual1} 
		 and \ref{thm:RTE-dual3} are given in Subsections 
		 \ref{sec:pf-RTE-dual3} and \ref{sec:uniq-stationary}, respectively.  

%
		
%
%
		

	\subsection{Remarks on related works}\label{sec:liter-rem} 
	To our knowledge, the vNL equation \eqref{vNLeq} was first studied by Davies in \cite{Davies-CMP,Davies-QME,Davies-QME2,Davies-QME3,Davies} for bounded $H$ and $W_{j}$'s and in \cite{Davies1}, for unbounded operators $H$ and $W_{j}$'s bounded relatively to $H$.
	

	The existence theory for the related non-commutative stochastic equation \eqref{stochastic} was developed in \cite{Holevo}.
	
	
	The fact that the generators of the norm continuous quantum dynamic semigroups are given by the r.h.s. of \eqref{vNLeq} was first derived in \cite{Lind3} for infinite dimensional $\cH$, and in \cite{GKS} for the finite dimensional $\cH$.
	
	
	
	Quantum detailed balance condition $(QDB)$ was considered in \cite{Ali,CM1,CM2,FrGo,MM}.  A specific form of the generators of quantum dynamic semigroups under $(QDB)$ was derived in \cite{Ali,CM1,CM2,Davies-QME,MM} for a finite-dimensional $\cH$.  For the Hilbert spaces $\cH$ with $\dim\cH=\infty$, a formula for bounded Lindblad operator satisfying $(QDB)$ was obtained in \cite{FGKV}.
	
	

	A finite dimensional version of Theorem \ref{thm:main} was proven by Spohn in \cite{Spohn1}, who also provided an algebraic condition on the $W_{j}$'s such that, under $(QDB)$, the vNL equation \eqref{vNLeq} has the unique stationary solution $\rho_{\rm st}$.  Later, the same condition was obtained in \cite{Spohn2} using the concept of entropy production.  Under the same condition, it was shown by Frigerio in \cite{Fri} that the solutions to \eqref{vNLeq} converge to $\rho_{\rm st}$ strongly as $t\rightarrow\infty$ for infinite dimensional $\cH$ and bounded $H$ and $W_{j}$'s, provided that the corresponding dynamic semigroup $\beta_{t}$ has a faithful normal stationary state and satisfies a ``weak coupling" condition\footnote{This condition says essentially that $\beta_{t}$ is the reduced dynamics of a system coupled weakly to a reservoir initially in a KMS state.}.  The latter says that $\beta_{t}$ describes the reduced dynamics of a system coupled weakly with a reservoir in a KMS state.
	

For finite-dimensional quantum systems coupled to a free massless fermionic or bosonic quantum field, the return to equilibrium was shown in \cite{JP1, JP2} and \cite{BFS} respectively, see \cite{JP3} for a review.

	\subsection{Organization of paper}\label{sec:outline}
	This paper is organized as follows.  In Section \ref{sec:generators}, we outline the existence theory for the vNL and HL equations on the spaces $\cS_{1}$ and $\cB$, respectively, and define the quantum detailed balance condition.  Section \ref{sec:G'} is preparatory.  It contains the most involved proofs of the paper establishing properties of the dual Lindblad generator $G'$.  In Section \ref{sec:pfs}, we prove our main results, Theorems \ref{thm:vNL-exist}, \ref{thm:main}, \ref{thm:main1}, \ref{thm:HL-exist} and \ref{thm:RTE-dual3} (recall that Theorem \ref{thm:RTE-dual1} is proven by a duality argument after Eq. \eqref{tau1-ip}).  Remarks on spaces, etc, are given in Section \ref{sec:remark-ext}.
	
	
	
	A class of jump operators $W_{j}$'s that satisfy $(QDB)$ are provided in Appendix \ref{sec:jump-oprs}.  In Appendix \ref{sec:pf-exist}, we provide a streamlined proof of a known result on the existence of solutions of the vNL equation.  We collect some extensions of Theorem \ref{thm:main} and \ref{thm:RTE-dual1} in Appendix \ref{sec:main-1}.  In Appendix \ref{sec:GNS}, we collect some known results on the GNS representation used in the main text.
	
%
%

	
	
%

	\bigskip
	
	\textit{Notation.} Throught out this paper, we fix the Hilbert space $\cH$ and denote $\cB(\cH)$ by $\cB$ whose norm $\|\cdot\|$ is given by the operator norm.  The identity element in $\cB$ is denoted by $\one$.  The norm and inner product on $\cH$ are denoted by $\|\cdot\|$ and $\langle\cdot,\cdot\rangle$.  (One should not confuse the norm on $\cH$ with the operator norm on $\cB$.)
	
	
	The $p$-th Schatten space of operators on the Hilbert space $\mathcal{H}$ (see \cite{Schatten}) is denoted by $\cS_{p}$. It  is a complex Banach space w.r.t. the norm $\|\cdot\|_{\cS_{p}}$ given by
	\begin{align}\label{p-norm}
		\|\kappa\|_{\cS_{p}}:=(\Tr|\kappa|^{p})^{1/p},\quad\quad\text{where }|\kappa|:=(\kappa^{*}\kappa)^{1/2}.
	\end{align}
	In particular, $\cS_{1},\cS_{2}$ are spaces of trace-class operators and Hilbert-Schmidt (HS) operators on $\cH$, respectively.  The latter is a Hilbert space, equipped with the HS-inner product:
	\begin{align}
		\langle\kappa,\sigma\rangle_{\cS_{2}}:=\Tr(\kappa^{*}\sigma).
	\end{align}

	\smallskip


\noindent\textbf{Acknowledgements}.  The first author is grateful to George Elliott and the second aurthor, to Lin Lin, for many stimulating discussions.  The authors are grateful to the referees for many constructive remarks.  The research for this paper was supported in part by NSERC through Discovery Grant  No. NA7901.

\smallskip
 
\noindent\textbf{Conflict of Interest}.  The authors have no conflicts to disclose.

\smallskip

\noindent\textbf{Data Availability}.
Data sharing is not applicable to this article as no new data were created or analyzed in this study.
 \smallskip
	
	\section{vNL and HL generators and QDBC}\label{sec:generators} 
	

	
	\subsection{Von Neumann-Lindblad and Heisenberg-Lindblad generators}\label{sec:vNL-HL-op}
	We write vNL equation \eqref{vNLeq} as
	\begin{align}\label{vNLeq2}
		\partial_{t}\rho_{t}=L \rho_{t},
	\end{align}
	where the operator $L$, called the \textit{vNL operator}, is defined as
	\begin{align}\label{L}
		&L=L_0+G, \qquad L_0\rho:= -i[H, \rho],\\
		\label{G}  &G(\rho)=\sum_{j\geq 1}\left(W_{j}\rho W_{j}^{*}-\frac{1}{2}\{W_{j}^{*}W_{j},\rho\}\right), 
	\end{align} 
	with the operator $L_{0}$ defined on the dense set $(H+i)^{-1}\cS_{1}(H-i)^{-1}\subset\cS_{1}$ or, more generally, on the dense domain (see \cite{Davies}, p.82)
	\begin{align}
		\sD(L_{0})&:=\{\rho\in\cS_{1}\mid \rho(\sD(H))\subseteq\sD(H)\text{ and }H\rho-\rho H\text{ defined on }\sD(H)\nonumber\\
		&\quad\quad\quad\quad\quad\quad\text{extends to } \cH \text{ as an element in } \cS_{1}\},
	\end{align}
	and, by Proposition \ref{prop:G-well-def} of Appendix \ref{sec:pf-exist}, the series on the r.h.s. converging absolutely in $\cS_{1}$ and yields the bounded operator $G:\cS_{1}\rightarrow\cS_{1}$.  (The operator $\sum_{j\geq 1}W_{j}^{*}W_{j}$ is defined by Condition $(W)$ and $\sum_{j\geq 1}W_{j}\rho W_{j}^{*}$, with $\rho\geq 0$, is the $\cS_{1}$-limit of monotonically increasing sequence of the positive partial sum operators.)  Hence the domain of $L$ is given by 
	\begin{align}\label{domain-L}
		\sD(L)=\sD(L_{0}).
	\end{align}
	We will call the operator $L_{0}$ the \textit{von Neumann (vN) operator}  and $G$ the \textit{Lindblad operator}.

	
	Similarly to \eqref{vNLeq2}, we write the HL equation \eqref{HLeq} as
	\begin{align}\label{HLeq3}
		\partial_{t}A_{t}&=L'A_{t},
	\end{align}
	where $L'$ is the dual operator of $L$ w.r.t. the coupling $(A, \rho):= \Tr(A \rho)$, i.e. 
		\begin{align}\label{rhoA-coupl}
			\Tr(A L(\rho))= \Tr(L'(A) \rho)
		\end{align}
		for all $\rho\in\sD(L)$ and $A\in\sD(L')$.  We have $L'=L_{0}'+G'$, where $L_{0}'$ and $G'$ are the dual operators to $L_{0}$ and $G$, w.r.t. the coupling $(A,\rho)=\Tr(A\rho)$, i.e.  $\Tr(L_{0}'(A)\rho)=\Tr(AL_{0}(\rho))$ and  $\Tr(G'(A)\rho)=\Tr(AG(\rho))$. Explicitly, we have
		\begin{align}\label{L0'}
			L_{0}'(A)=i[H,A],
		\end{align}
		defined on the strongly dense set $\sD(L_{0}'):=\{A\in\cB\mid A:\sD(H)\rightarrow\sD(H)\}$ (containing the strongly dense set $(H+i)^{-1}\cB(H-i)^{-1}$), and $G'$ is given by 
		\begin{align}\label{G'}
			G'(A)&=\sum_{j\geq 1}\left(W_{j}^{*}AW_{j}-\frac{1}{2}\{W_{j}^{*}W_{j},A\}\right).
		\end{align}

		
		The operator $G'$ is well-defined on $\cB$ as a weak limit of partial sums as follows from the next lemma:
		\begin{lemma}\label{lem:Phi'-bdd}
			Under Condition $(W)$, for each $A\in\cB$, the operator
			 \begin{equation}\label{Phi'}
				\Phi'(A)=\sum\limits_{j=1}^{\infty}W_j^*A W_j, 
			\end{equation}
			is defined as the weak limit of the bounded operators $\Phi_{N}'(A)=\sum_{j=1}^{N}W_{j}^{*}AW_{j}$ and yields a bounded map $\Phi'$ on $\cB$.
		\end{lemma}
		\begin{proof}
			We observe that, in \eqref{G'}, $\sum_{j\geq 1}W_{j}^{*}W_{j}$ is defined by Condition $(W)$ as a bounded operator on $\cH$.  Hence the second term in the sum on the r.h.s. of \eqref{G'} yields a bounded operator on $\cB$.  We have the estimate
		\begin{align}
			|\sum_{j=M}^{N}&\langle \varphi,W_{j}^{*}AW_{j}\psi\rangle|\leq \sum_{j=M}^{N}|\langle\varphi,W_{j}^{*}AW_{j}\psi\rangle|\nonumber\\
			&\leq \|A\|\big(\sum_{j=M}^{N}\|W_{j}\varphi\|^{2}\big)^{1/2}\big(\sum_{j=M}^{N}\|W_{j}\psi\|^{2}\big)^{1/2}\rightarrow 0,
		\end{align}
		for all $A\in\cB$ and for every $\varphi,\psi\in\cH$, as $M,N\rightarrow\infty$, since, by Condition $(W)$, 
		\begin{align}
			\sum_{j=M}^{N}\|W_{j}\psi\|^{2}&=\sum_{j=M}^{N}\langle\psi,W_{j}^{*}W_{j}\psi\rangle\rightarrow 0,
		\end{align}
		as $M,N\rightarrow\infty$, and similarly for $\sum_{j=M}^{N}\|W_{j}\varphi\|^{2}$.  Hence, $\Phi_{N}'(A)=\sum_{j=1}^{N}W_{j}^{*}AW_{j}$ converges weakly on $\cH$ and the limiting operator, $\Phi'(A)=\sum_{j\geq 1}W_{j}^{*}AW_{j}$, is bounded and satisfies
		\begin{align}\label{WAW-bdd}
			\|\Phi'(A)\|\leq \|A\|\|\Phi'(\one)\|.
		\end{align}
		Note that bound \eqref{WAW-bdd} also follows from the fact that the map $A\mapsto\sum_{j\geq 1}W_{j}^{*}AW_{j}$ is positive.
	\end{proof}
	
	
	\begin{cor}\label{cor:G'-bdd}
		Under Condition $(W)$, expression \eqref{Phi'} defines a bounded map on $\cB$, and the map $G'$ can be written as
		\begin{align}
			G'(A)&=\Phi'(A)-\frac{1}{2}\{\Phi'(\one),A\}.
		\end{align}
	\end{cor}
		
\begin{remark}\label{rem: Phi'}
By the Krauss' theorem, $\Phi'$ is a completely positive map on $\mathcal B$.  On the other hand, the complete positivity implies that $\Phi'$ is of the form \eqref{Phi'}   for some bounded operators $W_j,j=1,2,\cdots,$ satisfying (W). We think of \eqref{Phi'} as a `coordinate' representation of a completely positive map $\Phi'$.

Now, the positivity of  $\Phi'$  implies \eqref{WAW-bdd} (by applying $\Phi'$ to the operator $B=\|A\|\mathbbm1-A\geq 0$, for any self-adjoint operator $A$, and then extending this bound to non-self-adjoint operators).
 
The HL generator $L'$ on the r.h.s. of the HL equation ~\eqref{HLeq3} (cf. \eqref{HLeq}) can be written  in a `coordinate-free' form as 
    \begin{equation}\label{L'-Phi'}
    L'A=i[H,A]+\Phi'(A)-\frac{1}{2}\left\{\Phi'(\mathbbm1), A\right\}.
    \end{equation}
\end{remark}

\begin{remark} If one thinks of the algebra of observables $\mathcal B\equiv\mathcal B(\mathcal H)$ and the Heisenberg (resp. Heisenberg-Lindblad) dynamics on it as primary objects, then one might define the state space as the dual $\mathcal B'$ of $\mathcal B$ with the dynamics given by the von Neumann (resp. von Neumann-Lindblad) dynamics. Then $\mathcal S_1$ is a proper, closed subspace of $\mathcal B'$  invariant under the von Neumann and von Neumann-Lindblad dynamics. By restricting the von Neumann dynamics further to the invariant subspace of $\mathcal S_1$ of rank $1$ orthogonal projections one arrives at a formulation equivalent to the standard quantum mechanics. For closed systems, the latter extends uniquely to von Neumann dynamics on $\mathcal S_1$ and then on $\mathcal B'$. For open systems, this is not ture any more: the minimal state space for the vNL dynamics is $\mathcal S_1$.   
\end{remark}


		We say that $\rho_{t}$ is a \textit{weak solution}\footnote{Another possible definition of the weak solution is the one satisfying
			\begin{align}
				\int\Tr((\partial_{t}A_{t})\rho_{t})dt&=\int\Tr((L'A_{t})\rho_{t})dt
			\end{align}
			for any differentiable family $A_{t}\in\sD(L')$.  Using the facts proven below that $L_{0}$ is anti-self-adjoint and $G$ is bounded, one can also defined the mild solution of \eqref{HLeq} with the initial condition $\rho|_{t=0}=\rho_{0}\in\cS_{1}$ as a solution to the integral equation
			\begin{align}\label{weak-sol'}
				\rho_{t}=e^{L_{0}t}\rho+\int_{0}^{t}e^{L_{0}(t-s)}G\rho_{s}ds.
			\end{align}}  of \eqref{vNLeq}  in $\cS_{1}$ if, for any observable $A$ in $\sD(L')$, $\rho_{t}$ satisfies
		\begin{align}\label{weak-sol}\partial_{t}\Tr(A\rho_{t})= \Tr((L'A) \rho_{t}).\end{align}
		

		If $L$ generates a weakly continuous semigroup $\beta_{t}=e^{Lt}$ on the space $\cS_{1}$, then, for every initial condition $\rho_{0}\in\cS_{1}$, Eq. \eqref{vNLeq2} has a weak solution $\rho_{t}=\beta_{t}(\rho_{0})$ and, for any $\rho_{0}\in\sD(L)$, a strong one.  Similarly, for the operator $L'$.
		

		
		\begin{thm}\label{thm:exist-uniq-S} \cite{Davies-CMP,Davies,Davies1}			Assume Conditions $(H)$ and $(W)$.  Then 
			the vNL operator $L$ generates a weakly continuous, completely positive, bounded semigroup, $\beta_t$, on $\cS_{1}$ and therefore vNL equation \eqref{vNLeq} has a unique weak solution for any initial condition in $\cS_{1}$ and a unique strong solution for any initial condition in $\sD(L)$. 
		\end{thm}
		 For readers' convenience, we present a proof of this theorem in Appendix \ref{sec:pf-exist}.
		 
\begin{prop}\label{thm:exist-uniq-B}The HL operator $L'$ generates a weakly continuous, completely positive bounded semigroup and therefore HL equation \eqref{HLeq} has a unique weak solution for any initial condition in $\cB$ and a unique strong solution for  any initial condition in $\sD(L')$. \end{prop}
		
		\begin{proof}[Proof of Proposition \ref{thm:exist-uniq-B}]
			We begin with the following lemma:
			\begin{lemma}\label{lem:S-w*con}
				Let $H$ be self-adjoint.  Then the group $e^{L_{0}'t}A:=e^{iHt}Ae^{-iHt}$ on $\cB$ is weakly$^{*}$ (ultraweakly) continuous in $t$, i.e., for all $A\in\cB$ and $\rho\in\cS_{1}$, 
				\begin{enumerate}
					\item[(a)] the function $t\mapsto\Tr((e^{L_{0}'t}A)\rho)$ is continuous;
					\item[(b)] the dual group $e^{L_{0}t}$, defined by $\Tr(A(e^{L_{0}t}\rho))=\Tr((e^{L_{0}'t}A)\rho)$, maps $\cS_{1}$ into $\cS_{1}$,
				\end{enumerate}
				where we used that $\cS_{1}$ is the predual of $\cB$ (see Definition 3.1.2 of \cite{BrRo1}).
			\end{lemma}
			\begin{proof}
				For property (b), by duality and the cyclicity of trace, we have the dual group $e^{L_{0}t}$ of $e^{L_{0}'t}$ is given by $e^{L_{0}t}\rho=e^{-iHt}\rho e^{iHt}$.  Since $e^{\pm iHt}$ are bounded, $e^{L_{0}t}$ maps $\cS_{1}$ into itself.
				
				For property (a), by the duality $\Tr((e^{L_{0}'t}A)\rho)=\Tr(Ae^{L_{0}t}\rho)$ for every $A\in\cB$ and $\rho\in\cS_{1}$, it is equivalent to the weak continuity of the dual group $e^{L_{0}t}$ on $\cS_{1}$.  The latter follows from (in fact, by Corollary 3.1.8 of \cite{BrRo1}, is equivalent to) the strong continuity of $e^{L_{0}t}$ on $\cS_{1}$, which is shown in Lemma \ref{lem:S-str-con}.
			\end{proof}


			By Corollary \ref{cor:G'-bdd}, $G'$ is bounded on $\cB$, while $L_{0}'$ generates a one-parameter weakly$^{*}$ continuous group $e^{L_{0}'t}$ of bounded operators on $\cB$.  Hence, Theorem 3.1.33 of \cite{BrRo1} (see Appendix \ref{sec:pf-exist} for the formulation of this theorem) implies that the HL operator $L'=L_{0}'+G'$ generates a bounded semigroup $\beta_{t}'$ on $\cB$.  Therefore, the HL equation has a unique soluton in $\cB$ for any given initial condition in $\sD(L')$.  
		\end{proof}


		As apparent from expressions \eqref{L0'} and \eqref{G'}, $L_{0}'$ and $G'$, and therefore $L'$, has the eigenvalue $0$ (with the eigenvector $\one$), 
		\begin{align}\label{L'-zeroEV}
			L'\one=0,\quad L_{0}'\one=0,\quad\text{and}\quad G'\one =0.
		\end{align}

		
		
		\subsection{Quantum detailed balance condition}\label{sec:QDB}
		\DETAILS{We say that the vNL operator $L=L_0+G$, or what is the same, the HL operator $L'=L_{0}'+G'$, satisfies the {\it quantum detailed balance condition $(QDB)$} w.r.t. a strictly positive, density operator  $\rho_{\tau}$ if 
		\begin{enumerate}
			\item[$(QDB)$] (a) $L_0\rho_{\tau}=0$, and (b)  the Lindblad operators $G$ and $G'$ satisfy
			\begin{align}\label{(QDB)-G} 
				G(A \rho_{\tau})=(G' A)\rho_{\tau}\quad\quad\text{for all }A\in\cB.
			\end{align}
		\end{enumerate}
		

		In this paper, we assume that $\rho_{\tau}$ is the {\it Gibbs state} at a temperature $\tau>0$: 
		\begin{align}\label{Gibbs-*} 
			\rho_{\tau}= e^{-H/\tau} / Z(\tau), \quad\text{where }Z(\tau) := \Tr e^{-H/\tau}<\infty.
		\end{align}
		Hence Condition (a) is automatically satisfied.  This is the only place where Condition $(Tr)$ is used.}
		

			Consider the operator $G'$ given in \eqref{G'}. It is the dual of $G$ w.r.to the coupling $\Tr(A\rho)$: For all $A\in\cB$ and $\rho\in\cS_{1}$, 
			\begin{align}
				\Tr(G'(A)\rho)=\Tr(AG(\rho)).
			\end{align}
			We say that the Lindblad operator $G$, or what is the same, its dual $G'$, satisfies the {\it quantum detailed balance condition $(QDB)$} if
		\begin{enumerate}
			\item[$(QDB)$] the Lindblad operators $G$ and $G'$ satisfy
			\begin{align}\label{(QDB)} 
				G(A \rho_{\tau})=(G' A)\rho_{\tau}\quad\quad\text{for all }A\in\cB,
			\end{align}
		\end{enumerate}
		where, recall, $\rho_{\tau}$ is the {\it Gibbs state} at a temperature $\tau$, see above \eqref{Stau}, which we reproduce here: 
		\begin{align}\label{Gibbs} 
			\rho_{\tau}= e^{-H/\tau} / Z(\tau), \quad\text{where }Z(\tau):=\Tr (e^{-H/\tau}),\quad \tau>0.
		\end{align}  
		
		
		Note that, by the definitions, we have
		\begin{align}\label{L0-zero}
			L_{0}(\rho_{\tau})=0.
		\end{align}

		
		By the explicit formula \eqref{G'}, we have $G'(\one)=0$, which, together with \eqref{(QDB)}, implies that 
		\begin{align}\label{G0eigeneq}
			G(\rho_{\tau})=0.
		\end{align} 
		Since $L=L_0+G$, relations \eqref{L0-zero} and \eqref{G0eigeneq} yield  
		\begin{align}\label{L-zeroEV}
			&L(\rho_{\tau})=0,
		\end{align}
		i.e. $\rho_{\tau}$ is a stationary state and $0$ is an eigenvalue of $L$.  
		
		\begin{remark}
			Instead of $\rho_{\tau}$, we could have taken a density operator of the form $\rho_{\tau}=f(H)$ with any positive, continuous function satisfying $\Tr f(H)<\infty$.  


\DETAILS{By the explicit formula \eqref{G'}, $G' \one=0$ (see also \eqref{L'-zeroEV}).  Then, \eqref{(QDB)} implies that 
		\begin{align}\label{G0eigeneq-*}
			G\rho_{\tau}=0.
		\end{align} 
		Since $L=L_0+G$, the relations $L_0\rho_{\tau}=0$ and Eq. \eqref{G0eigeneq} yield  
		\begin{align}\label{rho*-zeroEV}
			&L\rho_*=0,
		\end{align}
		i.e. $\rho_*$ is a stationary state and $0$ is an eigenvalue of $L$.  
		\begin{remark}
			$\rho_{\tau}=f(H)$ for any (reasonable) functions satisfies Condition (a).  Under some conditions on $H$, the converse is also true.
%
%
	}
		
		On the other hand, any $\rho_{*}>0$ can be written in the form of \eqref{Gibbs}.  Indeed, since $\rho_{*}>0$, it can be written as $\rho_{*}=e^{-H_{*}/\tau}$ for the self-adjoint operator $H_{*}:=-\tau\ln \rho_{*}$.  Hence, $\rho_{*}$ is of the form \eqref{Gibbs} with $H=H_{*}+\mu$ and $Z=\Tr(e^{-(H_{*}-\mu)}/\tau)=e^{\mu/\tau}$ for any $\mu\in\R$.  
\end{remark}
		
		
		According to quantum mechanics, in the state $\rho_{\tau}=\sum_{s}\rho_{s}P_{s}$, the ``levels" (or pure states) $P_{s}:=\ket{\psi_{s}}\bra{\psi_{s}}$ (or $\psi_{s}$) occur with probabilities $\rho_{s}=e^{-\varepsilon_{s}/\tau}$, where $\varepsilon_{s}$ are the eigenvalues of $H_{\tau}$ and $P_{s}$, the corresponding eigenprojections.  In the system described by $L$,  the probability of the transition from the level $P_{s}$ to the level $P_{s'}$ is given by
		\begin{align}
			\text{Prob}(s\rightarrow s')&=\Tr(P_{s'}L(P_{s}))=\Tr(P_{s'}G(P_{s}))=\Tr(G'(P_{s'})P_{s})\nonumber\\
			&=\Tr(\Phi(P_{s'})P_{s})-\Tr(\Phi(\one)P_{s})\delta_{ss'},
		\end{align}
		where $\Phi(\rho):=\sum_{j}W_{j}\rho W_{j}^{*}$ for $\rho\in\cS_{1}$.  Let $K_{rr',ss'}:=\Tr(P_{sr}\Phi(P_{r's'}))$, where $P_{rs}:=\ket{\psi_{r}}\bra{\psi_{s}}$.  Then we have
		\begin{align}
			\text{Prob}(s\rightarrow s')=K_{ss'}-\delta_{ss'}\sum_{\ell}K_{s\ell}=:C_{ss',ss'}\equiv C_{ss'}.
		\end{align}
		Hence, the QDB relation $\rho_{s}C_{ss'}=C_{s's}\rho_{s'}$ says that the levels $P_{s}$ are at the equilibrium, i.e., the average number of particles leaving each $P_{s}$ is equal to the average number of particles coming to it.  Condition $(QDB)$ is closely related to the Einstein's $AB$ coefficients.

		

		\subsection{Remarks about spaces and inner products}\label{sec:spaces}
		\begin{lemma}\label{lem:2.7a}
			The space $\cB_{\tau,1}$ is dual to $\cS_{\tau}$ ($\cB_{\tau,1}=\cS_{\tau}'$) w.r.t. the coupling $\Tr(A\rho)$
		\end{lemma}
		\begin{proof}
			The statement follows from the inequality
			\begin{align}\label{coupling-ineq}
				|\Tr(A\lambda)|&\leq \|A\|_{\tau,1}\|\lambda\|_{\tau,1}.
			\end{align}
			Indeed, the inequality \eqref{coupling-ineq} shows that, for any $\lambda\in\cS_{\tau}$, the map $A\mapsto\Tr(A\lambda)$ can be extended to a bounded, linear functional $f_{A}(\lambda)$ on $\cB_{\tau,1}$.  
			
			Now, for all $A\in\cB_{\tau,1}$, the map $\lambda\mapsto f_{A}(\lambda)$ is a bounded, linear functional on $\cS_{\tau}$.  This shows that $\cB_{\tau,1}$ can be identified with $\cS_{\tau}'$ by the correspondence $A\mapsto f_{A}$.  
			
			To prove \eqref{coupling-ineq}, we use the Cauchy-Schwarz type inequality to obtain
			\begin{align}
				|\Tr(A\lambda)|&=|\Tr(\rho_{\tau}^{1/2}A\lambda\rho_{\tau}^{-1/2})|\leq \|A^{*}\rho_{\tau}^{1/2}\|_{\cS_{2}}\|\lambda\rho_{\tau}^{-1/2}\|_{\cS_{2}}.
			\end{align}
			Since $\|A^{*}\rho_{\tau}^{1/2}\|_{\cS_{2}}=\|A\|_{\tau,1}$ and $\|\lambda\rho_{\tau}^{-1/2}\|_{\cS_{2}}=\|\lambda\|_{\rm st,\tau}$, \eqref{coupling-ineq} follows.  This completes the proof.
		\end{proof}

			\begin{remark}\label{rem:B*-r}
			(i) We show below (see Theorem \ref{thm:G'-property}) that $L=L_{0}+G$ and $L'=L_{0}'+G'$ are the decompositions of $L$ and $L'$ into anti-self-adjoint and self-adjoint parts w.r.t. the inner products \eqref{ip-st}and \eqref{ip-obs-2}, respectively.
			
			
		(ii)	 $G'$ is symmetric in the inner product \eqref{tau1-ip}. 
		Indeed, using the fact that $G'$ is a $*$-map  and is symmetric w.r.t. \eqref{ip-obs-2} and using the relation  $\langle A,B\rangle_{\tau, 1}=\langle B^{*},A^{*}\rangle_{\tau}$, we obtain
	\begin{align}
		\langle G'(A),B\rangle_{\tau,1}&=\langle B^{*},G'(A^{*})\rangle_{\tau}=\langle G'(B^{*}),A^{*}\rangle_{\tau}=\langle A,G'(B)\rangle_{\tau,1}.
	\end{align}

			
			(iii) The norm associated with the inner product \eqref{ip-st} satisfies 
			\begin{align}\label{tr-st*}
				\|\lambda\|_{\rm st,\tau}&\geq \|\lambda\|_{\cS_{1}}(\Tr\rho_{\tau})^{-1/2}.
			\end{align}
			Indeed, using the non-Abelian Cauchy-Schwarz inequality, we obtain
			\begin{align}
				\|\lambda\|_{\cS_{1}}&=\Tr|\lambda|=\Tr(|\lambda|\rho_{\tau}^{-1/2}\rho_{\tau}^{1/2})\leq\||\lambda|\rho_{\tau}^{-1/2}\|_{\cS_{2}}\|\rho_{\tau}^{1/2}\|_{\cS_{2}}\nonumber\\
				&=(\Tr(\rho_{\tau}^{-1/2}\lambda^{*}\lambda\rho_{\tau}^{-1/2}))^{1/2}(\Tr\rho_{\tau})^{1/2}\nonumber\\
				&=\|\lambda\|_{\rm st,\tau}(\Tr\rho_{\tau})^{1/2}.
			\end{align}
			\end{remark}
			The space $\cS_{\tau}$ is a point $r=0$ in the family of Hilbert spaces 
			\begin{align}\label{Sr}
				\cS_{\tau}^{(r)}&=\{\lambda\in\cS_{1}\mid \rho_{\tau}^{-r/2}\lambda\rho_{\tau}^{-(1-r)/2}\in\cS_{2}\},
			\end{align}
			$r\in [0,1]$, with the inner products
			\begin{align}\label{ip-st-r}
				\langle \lambda, \sigma\rangle_{{\rm st},\tau,r}:=\Tr(\lambda^{*} \rho_{\tau}^{-r}\sigma \rho_{\tau}^{-1+r}).
			\end{align}
	
	\begin{remark}
		The inner products \eqref{ip-st-r}, $r\in [0,1]$, are special examples of the so-called quantum $\chi^{2}$-divergence (c.f. \cite{TKRWV}).
	\end{remark}


Inner product \eqref{ip-obs-2} is a member of the family of inner products on $\cB$: For each $r\in [0,1]$, we define the inner product
\begin{align}
	\langle A,B\rangle_{\tau,r}&=\Tr(A^{*}\rho_{\tau}^{r}B\rho_{\tau}^{1-r}).
\end{align}
Note that $\langle A,B\rangle_{\tau,0}=\langle A,B\rangle_{\tau}$ and $\langle A,B\rangle_{\tau,1}=\langle B^{*},A^{*}\rangle_{\tau}$. 




	\section{Heisenberg and dual Lindblad generators $L_0'$ and $G'$}\label{sec:G'}

	 Recall from Section \ref{sec:vNL-HL-op} (and, in particular, \eqref{L'-Phi'}) that the HL operator  $L' $, acting on observables, 
is given by
\begin{align}
	\label{L'}&L'=L_0'+G',\hspace{0.5cm}L_{0}'A=i[H,A],\\ 
	\label{G'-Phi'}&G' (A):=\Phi'(A)-\frac{1}{2}\left\{\Phi'(\mathbbm1), A\right\}, 
\end{align}
where $\Phi'(A)=\sum_{j\geq 1}W_{j}^{*}AW_{j}$, a bounded, completely positive map, with $\Phi'(\bbone)=\sum_{j\geq 1}W_{j}^{*}W_{j}$ a bounded operator by Condition (W) (see Remark \ref{rem: Phi'}).


\smallskip
We begin with the following proposition:
\begin{prop}\label{prop:G'-property-B*}
	Suppose Conditions $(W)$ and $(QDB)$ hold.  Then $G'$ is bounded on $\cB_{\tau}$.
\end{prop}
Before  proceeding to the proof of Proposition \ref{prop:G'-property-B*}, we prove the following technical result:
\begin{lemma}\label{lem:comm-G'-qdbc}
	Suppose Conditions $(W)$ and $(QDB)$ w.r.t. $\rho_{\tau}>0$ hold.  Then there exists a family of operators $\{\widetilde{W}_{j}\}_{j\geq 1}$, satisfying Condition $(W)$, such that 
	\begin{align}
		G'(A)&=\sum_{j\geq 1}\big(\widetilde{W}_{j}^{*}A\widetilde{W}_{j}-\frac{1}{2}\{\widetilde{W}_{j}^{*}\widetilde{W}_{j},A\}\big)
	\end{align}
	and the series $\sum_{j\geq 1}\widetilde{W}_{j}^{*}\widetilde{W}_{j}$ commutes with $\rho_{\tau}$.
\end{lemma}
\begin{proof}[Proof of Lemma \ref{lem:comm-G'-qdbc}]
	Recall that $\alpha_{t}(A):=\rho_{\tau}^{it}A\rho_{\tau}^{-it}$ for $t\in\R$.  By the same argument in the proof of Theorem \ref{thm:Gibbs}, if $G'$ satisfies $(QDB)$ w.r.t. $\rho_{\tau}$, then $G'$ commutes with $\alpha_{t}$ for each $t\in\R$.  Then, by the argument in \cite{FGKV}, for each $A\in\cB$ and $T>0$, we have
	\begin{align}\label{G'-alphat2}
		G'(A)&=\frac{1}{2T}\int_{-T}^{T}(\alpha_{t}\circ G'\circ\alpha_{-t})(A)dt\nonumber\\
		&=\frac{1}{2T}\int_{-T}^{T}(\alpha_{t}\circ\Phi'\circ\alpha_{-t})(A)dt-\frac{1}{4T}\int_{-T}^{T}\alpha_{t}\{Y,\alpha_{-t}(A)\}dt\nonumber\\
		&=\frac{1}{2T}\sum_{j\geq 1}\int_{-T}^{T}W_{j,t}^{*}AW_{j,t}dt-\frac{1}{4T}\int_{-T}^{T}\{\alpha_{t}(Y),A\}dt,
	\end{align}
	where $W_{j,t}:=\alpha_{t}(W_{j})$, 
	and, recall, $Y=\Phi'(\one)=\sum_{j\geq 1}W_{j}^{*}W_{j}$.  Since the l.h.s. of \eqref{G'-alphat2} is independent of $T$, we can take $T\rightarrow\infty$ on the r.h.s. of \eqref{G'-alphat2}.  
	

	Next, for each $T>0$, we define
	\begin{align}
		\Phi_{T}'(A)&:=\frac{1}{2T}\sum_{j\geq 1}\int_{-T}^{T}W_{j,t}^{*}AW_{j,t}dt.
	\end{align}
	We claim that the limit $\lim_{T\rightarrow\infty}\Phi_{T}'$ exists in the strong sense: there exists a bounded map $\Phi_{\infty}'$ such that, for each $A\in\cB$,
		\begin{align}\label{stronglimit2}
		\Phi_{\infty}'(A)&=\lim_{T\rightarrow\infty}\Phi_{T}'(A).
	\end{align}
	To justify this claim, we follow an argument in \cite{Davies4}.  Let $\{\psi_{s}\}_{s\geq 1}$ be an orthonormal basis in $\cH$ consisting of eigenvectors of $\rho_{\tau}$, with corresponding eigenvalues $\rho_{s}$, and $P_{ss'}:=\ket{\psi_{s}}\bra{\psi_{s'}}$.  Then, for each $x\in\R$, we define
	\begin{align}
		Q_{x}^{T}&:=\frac{1}{2T}\int_{-T}^{T}\alpha_{t}e^{ixt}dt
	\end{align} 
	so that $\|Q_{x}^{T}\|\leq 1$ for all $x$ and $T$ and, for each $s,s'$,
	\begin{align}
		\lim_{T\rightarrow\infty}Q_{x}^{T}P_{ss'}&=\lim_{T\rightarrow\infty}\frac{1}{2T}\int_{-T}^{T}e^{ixt}\alpha_{t}(P_{ss'})dt\nonumber\\
		&=\lim_{T\rightarrow\infty}\frac{1}{2T}\int_{-T}^{T}e^{i(x-\omega_{ss'})t}P_{ss'}dt=\delta_{x,\omega_{ss'}}P_{ss'},
	\end{align}
	where $\omega_{ss'}:=\log(\rho_{s}\rho_{s'}^{-1})$.  Since the span of $\{P_{ss'}\}_{s,s'}$ is dense in $\cB$ (in the weak operator topology), the limit of $Q_{x}^{T}$ as $T\rightarrow\infty$ exists, which we denote the limit by $Q_{x}$.  
	
	
	Since, for each $T>0$ (cf. the proof of \eqref{WAW-bdd}),
	\begin{align}
		\|\Phi_{T}'(A)\|&\leq \frac{1}{2T}\int_{-T}^{T}\big\|\sum_{j\geq 1}W_{j,t}^{*}AW_{j,t}\big\|dt\leq \frac{1}{2T}\int_{-T}^{T}\|A\|\big\|\sum_{j\geq 1}\alpha_{t}(W_{j}^{*}W_{j})\|dt\nonumber\\
		&=\frac{1}{2T}\int_{-T}^{T}\|A\|\|Y\|dt=\|A\|\|Y\|,
	\end{align}
	to show the limit of $\Phi_{T}'$ exists, it suffices to show the limit exists on a dense subset.  Since the span of $\{P_{ss'}\}$ is dense in $\cB$ (in the weak operator topology), we have, for each $s,s'$, 
	\begin{align}
		\lim_{T\rightarrow\infty}\Phi_{T}'(P_{ss'})&=\lim_{T\rightarrow\infty}\frac{1}{2T}\int_{-T}^{T}e^{-i\omega_{ss'}t}\alpha_{t}(\Phi'(P_{ss'}))dt\nonumber\\
		&=Q_{-\omega_{ss'}}(\Phi'(P_{ss'})).
	\end{align}
	Therefore, the limit $\Phi_{\infty}'$ of $\Phi_{T}'$ as $T\rightarrow\infty$ exists in the sense of \eqref{stronglimit2}.
	
	Now, since $\Phi_{T}'$ is a completely positive map for each $T$, $\Phi_{\infty}'$ is also completely positive so that, by \cite{Kr2}, Theorem 3.3, there exists a collection of bounded operators $\{\widetilde{W}_{j}\}_{j\geq 1}$ on $\cH$ such that, for all $A\in\cB$,
	\begin{align}
		\Phi_{\infty}'(A)&=\sum_{j\geq 1}\widetilde{W}_{j}^{*}A\widetilde{W}_{j},
	\end{align}
	and the series $\sum_{j\geq 1}\widetilde{W}_{j}^{*}\widetilde{W}_{j}$ converges weakly in $\cB$.  Similarly, we have
	\begin{align}
		\lim_{T\rightarrow\infty}\frac{1}{2T}\int_{-T}^{T}\{\alpha_{t}(Y),A\}dt&=\{\widetilde{Y},A\},
	\end{align}
	where
	\begin{align}
		\widetilde{Y}:=\lim_{T\rightarrow\infty}\frac{1}{2T}\int_{-T}^{T}\alpha_{t}(Y)dt.
	\end{align}
	Since
	\begin{align}
		0&=G'(\one)=\Phi_{\infty}'(\one)-\frac{1}{2}\{\widetilde{Y},\one\}=\Phi_{\infty}'(\one)-\widetilde{Y},
	\end{align}
	we must have
	\begin{align}
		\widetilde{Y}&=\Phi_{\infty}'(\one)=\sum_{j\geq 1}\widetilde{W}_{j}^{*}\widetilde{W}_{j}.
	\end{align}
	We note that $\Phi_{\infty}'$ commutes with $\alpha_{t}$, which implies that, for all $t\in\R$,
	\begin{align}
		\widetilde{Y}=\Phi_{\infty}'(\one)=\alpha_{t}\Phi_{\infty}'(\alpha_{-t}(\one))=\alpha_{t}(\widetilde{Y})=\rho_{\tau}^{it}\widetilde{Y}\rho_{\tau}^{-it}.
	\end{align}
	Therefore, $\widetilde{Y}$ commutes with $\rho_{\tau}$.
\end{proof}


Now, we are ready to proceed directly to the proof of Proposition \ref{prop:G'-property-B*}.
\begin{proof}[Proof of Proposition \ref{prop:G'-property-B*}]
	By Lemma \ref{lem:comm-G'-qdbc}, we can write 
	\begin{align}\label{G'(A)-cc}		G'(A)&=\widetilde{\Phi}'(A)-\frac{1}{2}\{\widetilde{\Phi}'(\one),A\}\big),
	\end{align}
	where $\widetilde{\Phi}'(A):=\sum_{j\geq 1} \widetilde{W}_{j}^{*}A\widetilde{W}_{j}$ and the series $\widetilde{\Phi}'(\one)=\sum_{j\geq 1}\widetilde{W}_{j}^{*}\widetilde{W}_{j}$ commutes with $\rho_{\tau}$.  For notational simplicity, we drop the tilde over $\widetilde{W}_{j}$'s and $\widetilde{Y}\equiv \widetilde{\Phi}'(\one)$.
	
	
	By \eqref{G'(A)-cc} and triangle inequality, we have
	\begin{align}\label{G'(A)-norm}
		\|G'(A)\|_{\tau}&\leq \|\Phi'(A)\|_{\tau}+\frac{1}{2}\|\{Y,A\}\|_{\tau}.
	\end{align}
	First, we estimate $\Phi'(A)$.  We claim that, for all $A\in\cB$, 
	\begin{align}\label{gen-Schwarz-2}
		\Phi'(A)^{*}\Phi'(A)&\leq \|Y\|\Phi'(A^{*}A).
	\end{align}
	Indeed, for any $u,v\in\cH$ and for any $A\in\cB$, we have
	\begin{align}\label{proof-gen-Schwarz}
		|\langle u,\Phi'(A)v\rangle|&\leq |\sum_{j\geq 1}\langle W_{j}u,AW_{j}v\rangle|\leq \left(\sum_{j\geq 1}\|W_{j}u\|^{2}\right)^{1/2}\left(\sum_{j\geq 1}\|AW_{j}v\|^{2}\right)^{1/2}\nonumber\\
		&=\left(\sum_{j\geq 1}\langle u,W_{j}^{*}W_{j}u\rangle\right)^{1/2}\left(\sum_{j\geq 1}\langle v,W_{j}^{*}A^{*}AW_{j}v\rangle\right)^{1/2}\nonumber\\
		&=\langle u,\Phi'(\one)u\rangle^{1/2}\langle v,\Phi(A^{*}A)v\rangle^{1/2}\nonumber\\
		&\leq \|Y\|^{1/2}\|u\|\langle v,\Phi'(A^{*}A)v\rangle^{1/2}.
	\end{align}
	By taking $u=\Phi'(A)v$, \eqref{proof-gen-Schwarz} implies that
	\begin{align}
		\|\Phi'(A)v\|^{2}&\leq \|Y\|\langle v,\Phi'(A^{*}A)v\rangle,
	\end{align}
	which, since $\|\Phi'(A)v\|^{2}=\langle v,\Phi'(A)^{*}\Phi'(A)v\rangle$, gives
	\begin{align}
		\langle v,\Phi'(A)^{*}\Phi'(A)v\rangle&\leq \|Y\|\langle v,\Phi'(A^{*}A)v\rangle
	\end{align}
	for all $v\in\cH$.  This is equivalent to \eqref{gen-Schwarz-2}.
	

	Since $\rho_{\tau}>0$, inequality \eqref{gen-Schwarz-2}, together with cyclicity of trace, gives
	\begin{align}\label{phi(A)-norm}
		\|\Phi'(A)\|_{\tau}^{2}&=\Tr(\Phi'(A)^{*}\Phi'(A)\rho_{\tau})\leq \|Y\|\Tr(\Phi'(A^{*}A)\rho_{\tau})\nonumber\\
		&=\|Y\|\Tr(A^{*}A\Phi(\rho_{\tau})).
	\end{align}
	where $\Phi(\rho):=\sum_{j\geq 1}W_{j}\rho W_{j}^{*}$, the dual of $\Phi$.  Since $G(\rho_{\tau})=0$ (see \eqref{(QDB)}) and, by Lemma \ref{lem:comm-G'-qdbc}, $\rho_{\tau}$ and $Y$ commute, \eqref{G} implies that
	\begin{align}\label{qdbc-stationary}
		\Phi(\rho_{\tau})&=\frac{1}{2}\{Y,\rho_{\tau}\}=\rho_{\tau}Y=\rho_{\tau}^{1/2}Y\rho_{\tau}^{1/2},
	\end{align}
	where, for the last equality in \eqref{qdbc-stationary}, we use that $[\rho_{\tau},Y]=0$ implies $[\rho_{\tau}^{1/2},Y]=0$.  Next, using cyclicity of trace and Lemma \ref{lem:comm-G'-qdbc}, we obtain
	\begin{align}
		\Tr(A^{*}A\Phi(\rho_{\tau}))&=\Tr( A^{*}A\rho_{\tau}Y)=\Tr(A^{*}A\rho_{\tau}^{1/2}Y\rho_{\tau}^{1/2})\nonumber\\
		&=\Tr(A\rho_{\tau}^{1/2}Y\rho_{\tau}^{1/2}A^{*})\nonumber\\
		&\leq \|Y\|\Tr(A\rho_{\tau}A^{*})=\|Y\|\|A\|_{\tau}^{2},
	\end{align}
	so that, by substituting \eqref{qdbc-stationary} into the last term in \eqref{phi(A)-norm},
	\begin{align}
		\|\Phi'(A)\|_{\tau}&\leq \|Y\|\|A\|_{\tau}.
	\end{align}
	
	Finally, for the second term on the r.h.s. of \eqref{G'(A)-norm}, we use the triangle inequality, cyclicty of trace and Lemma \ref{lem:comm-G'-qdbc} to obtain
	\begin{align}\label{cY-A-norm}
		\|\{Y,A\}\|_{\tau}^{2}&\leq \|Y A\|_{\tau}^{2}+\|AY\|_{\tau}^{2}\nonumber\\
		&=\Tr(A^{*}Y^{2}A\rho_{\tau})+\Tr(Y A^{*}AY\rho_{\tau})\nonumber\\
		&=\Tr(\rho_{\tau}^{1/2}A^{*}Y^{2}A\rho_{\tau}^{1/2})+\Tr(\rho_{\tau}^{1/2}Y A^{*}AY \rho_{\tau}^{1/2})\nonumber\\
		&=\Tr(\rho_{\tau}^{1/2}A^{*}Y^{2}A\rho_{\tau}^{1/2})+\Tr(A\rho_{\tau}^{1/2}Y^{2}\rho_{\tau}^{1/2}A^{*}).
	\end{align}
	Using the operator inequality
	\begin{align}
		A^{*}Y^{2}A&\leq \|Y\|^{2}A^{*}A,\quad\quad A\rho_{\tau}^{1/2}Y^{2}\rho_{\tau}^{1/2}A^{*}\leq \|Y\|^{2}A\rho_{\tau}A^{*},
	\end{align}
	we find
	\begin{align}\label{cY-A-norm-2}
		\|\{Y,A\}\|_{\tau}&\leq \|Y\|\|A\|_{\tau}.
	\end{align} 
	
	
	By substituting \eqref{phi(A)-norm} and \eqref{cY-A-norm-2} into \eqref{G'(A)-norm}, we obtain
	\begin{align}
		\|G'(A)\|_{\tau}&\leq 2\|Y\|\|A\|_{\tau}
	\end{align}
	so that $G'$ is bounded on $\cB$, hence is bounded on $\cB_{\tau}$, since $\cB$ is dense in $\cB_{\tau}$.
\end{proof}



		
		\begin{thm}\label{thm:G'-property}
			Suppose Conditions $(H)$, $(W)$ and $(QDB)$ hold.  Then, on $\cB_{\tau}$, (a) the Lindblad operator $G'$ is self-adjoint, (b) $G'\leq 0$, (c) $0$ is an eigenvalue of $G'$, and (d) $L_{0}'$ is anti-self-adjoint.
		\end{thm}

		\begin{proof}[Proof of Theorem \ref{thm:G'-property}]
			(a) We begin with
				\begin{lemma}\label{lem:qdbc-sym}
				Given any $\rho_{\tau}\in\cS_{1}^{+}$, $\rho_{\tau}>0$, the dual Lindblad operator $G'$ satisfies \eqref{(QDB)} w.r.t. $\rho_{\tau}$ if and only if $G'$ is symmetric w.r.t. the inner product \eqref{ip-obs-2}:
				\begin{align}\label{G'-symm}
					\langle A,G'(B)\rangle_{\tau}&=\langle G'(A),B\rangle_{\tau}
				\end{align}
				for all $A,B\in\cB$.  
			\end{lemma}
			\begin{proof}
				Suppose $G'$ satisfies \eqref{(QDB)}.  Since $G'$ is a $*$-map and is dual operator of $G$, relation \eqref{(QDB)} yields
				\begin{align}\label{G'-symm-cal}
					\langle G'(A),&B\rangle_{\tau}=\Tr((G'(A))^{*}B\rho_{\tau})=\Tr(G'(A^{*})B\rho_{\tau})\nonumber\\
					&=\Tr(A^{*}G(B\rho_{\tau}))=\Tr(A^{*}G'(B)\rho_{\tau})=\langle A,G'(B)\rangle_{\tau}
				\end{align}
				for all $A,B\in\cB$, which proves \eqref{G'-symm}.  Conversely, going in the opposite direction in \eqref{G'-symm-cal} shows \eqref{G'-symm} implies \eqref{(QDB)}.
			\end{proof}	
			
			
			By Proposition \ref{prop:G'-property-B*} and Lemma \ref{lem:qdbc-sym}, $G'$ is bounded and symmetric w.r.t. \eqref{ip-obs-2} on $\cB$ so that $G'$ is symmetric on $\cB_{\tau}$ using a density argument of $\cB$ in $\cB_{\tau}$.  Therefore, $G'$ is self-adjoint on $\cB_{\tau}$.
			

			
			(b) First, we introduce the \textit{dissipation function} on $\cB$ (see \cite{Lind3})
			\begin{align}\label{dissip-G'}
				D_{G'}(A,B)&:=G'(A^{*}B)-G'(A)^{*}B-A^{*}G'(B).
			\end{align}
			\begin{lemma}\label{lem:diss-G'2}
				For any $A,B\in\cB$, we have
				\begin{align}\label{diss-G'2}
					D_{G'}(A,B)&=\sum_{j\geq 1}[W_{j},A]^{*}[W_{j},B],
				\end{align}
				where the sum on the r.h.s. of \eqref{diss-G'2} converges weakly on $\cH$.
			\end{lemma}
			\begin{proof}
				By definitions \eqref{G'} and \eqref{dissip-G'}, Corollary \ref{cor:G'-bdd} and a straightforward calculation, we have
				\begin{align}
					D_{G'}(A,B)&=\sum_{j\geq 1}(W_{j}^{*}A^{*}BW_{j}-\frac{1}{2}\{W_{j}^{*}W_{j},A^{*}B\})\nonumber\\
					&\quad\quad-\sum_{j\geq 1}(W_{j}^{*}A^{*}W_{j}B-\frac{1}{2}\{W_{j}^{*}W_{j},A^{*}\}B)\nonumber\\
					&\quad\quad-\sum_{j\geq 1}(A^{*}W_{j}^{*}BW_{j}-\frac{1}{2}A^{*}\{W_{j}^{*}W_{j},B\})\nonumber\\
					&=\sum_{j\geq 1}(W_{j}^{*}A^{*}BW_{j}-W_{j}^{*}A^{*}W_{j}B-A^{*}W_{j}^{*}BW_{j}+A^{*}W_{j}^{*}W_{j}B)\nonumber\\
					&=\sum_{j\geq 1}[A^{*},W_{j}^{*}][W_{j},B],
				\end{align}
				where each sum above converges weakly on $\cH$.  Since $[A^{*},W_{j}^{*}]=[W_{j},A]^{*}$ for each $j$, this gives \eqref{diss-G'2}.
			\end{proof}
			\begin{lemma}\label{lem:G'A-ip}
				For all $A,B\in\cB$,
				\begin{align}\label{G'A-ip}
					\langle A,G'(B)\rangle_{\tau}&=-\frac{1}{2}\Tr(D_{G'}(A,B)\rho_{\tau})\nonumber\\
					&=-\frac{1}{2}\sum_{j\geq 1}\Tr([W_{j},A]^{*}[W_{j},B]\rho_{\tau}),
				\end{align}
				where the sum on the r.h.s. of \eqref{G'A-ip} converges weakly on $\cH$.
			\end{lemma}
			\begin{proof}
				Since $G'$ is self-adjoint on $\cB_{\tau}$,  $G(\rho_{\tau})=0$ and $\Tr(AG(\rho))=\Tr(G'(A)\rho)$, we have, by \eqref{dissip-G'}, for all $A,B\in\cB$,
				\begin{align}
					\langle A,G'(B)\rangle_{\tau}&=\frac{1}{2}(\langle A,G'(B)\rangle_{\tau}+\langle G'(A),B\rangle_{\tau})\nonumber\\
					&=\frac{1}{2}(\Tr(A^{*}G'(B)\rho_{\tau})+\Tr(G'(A)^{*}B\rho_{\tau})-\Tr(A^{*}BG(\rho_{\tau})))\nonumber\\
					&=\frac{1}{2}\Tr((A^{*}G'(B)+G'(A)^{*}B-G'(A^{*}B))\rho_{\tau})\nonumber\\
					&=-\frac{1}{2}\Tr(D_{G'}(A,B)\rho_{\tau}).
				\end{align}
				By Lemma \ref{lem:diss-G'2}, we obtain \eqref{G'A-ip}.
			\end{proof}
			
			
			By Lemma \ref{lem:G'A-ip}, we have, for any $A\in\cB$,
			\begin{align}\label{nc-Laplacian}
				\langle A,G'(A)\rangle_{\tau}=-\frac{1}{2}\sum_{j\geq 1}\Tr([W_{j},A]^{*}[W_{j},A]\rho_{\tau}).
			\end{align}
			Since $[W_{j},A]^{*}[W_{j},A]\geq 0$ for all $A\in\cB$ and $j\geq 1$, we have that $\langle A,G'(A)\rangle_{\tau}\leq 0$ for all $A\in \cB$.  By a density argument, since $G'$ is bounded on $\cB_{\tau}$, we conclude that $G'\leq 0$ on $\cB_{\tau}$.  This proves Theorem \ref{thm:G'-property} (b).
			
			
			
			(c) We show that $L_{0}'$ is anti-symmetric on $\cB_{\tau}$ under $(QDB)$.  Indeed, since $L_{0}\rho_{\tau}=0$ under $(QDB)$, then, for all $A,B\in\sD(L_{0}')$, we have
			\begin{align}\label{qdbc-L0'-symm}
				\langle L_{0}'A,B\rangle_{\tau}&=\Tr((i[H,A])^{*}B\rho_{\tau})=\Tr(i[H,A^{*}]B\rho_{\tau})\nonumber\\
				&=\Tr(i[H,A^{*}B]\rho_{\tau})-\Tr(A^{*}(i[H,B])\rho_{\tau})\nonumber\\
				&=\Tr(A^{*}B(L_{0}\rho_{\tau}))-\langle A,L_{0}'B\rangle_{\tau}\nonumber\\
				&=-\langle A,L_{0}'B\rangle_{\tau}.
			\end{align}
		
			
			(d) The fact that $0$ is an eigenvalue of $G'$ on $\cB_{\tau}$ follows from the computation:
			\begin{align}
				G'(\one)&=\sum_{j\geq 1}(W_{j}^{*}W_{j}-\frac{1}{2}\{W_{j}^{*}W_{j},\one\})=0.
			\end{align}
			This completes the proof of Theorem \ref{thm:G'-property}.
			\end{proof}
			
%
%


	\subsection{Commutativity of $L_{0}'$ and $G'$}\label{sec:com-L0'-G'}
		
		\begin{thm}\label{thm:Gibbs}
			Assume Conditions $(H)$, $(W)$ and $(QDB)$. 
			Then the semigroups $e^{L_{0}'t}$ and $e^{G't}$ on $\cB_{\tau}$ commute, and, consequently, $\beta_{t}'=e^{L_{0}'t}e^{G't}$.
		\end{thm}
		

		
		Theorem \ref{thm:Gibbs} was proven in \cite{CM1} for the finite-dimensional case, and was sketched for the infinite-dimensional case in \cite{FGKV}.  In Subsections \ref{sec:*-map}--\ref{sec:qdbc2}, we give a detailed proof of this theorem.  
		
		
		
		We begin with some generalities.  We use the following definition (see \cite{BrRob}).
		\begin{Def}\label{def:strong-com}
			\normalfont
			We say two unbounded, self-adjoint operators $H_{1}$ and $H_{2}$ \textit{commute strongly} if the spectral projectors of $H_{1}$ and $H_{2}$ commute, or, equivalently, if $e^{iH_{1}s}$ and $e^{iH_{2}t}$ commute for all $s,t\in\R$.
		\end{Def}
		
		
		\begin{lemma}[\cite{BrRob}, Lemma 1]\label{lem:com-unbdd}
			For a bounded operator $B$ and unbounded self-adjoint operator $A$, the following conditions are equivalent:
			\begin{enumerate}
				\item[(i)] $B$ commutes strongly with $A$.
				\item[(ii)] There is some core $\sD$ for $A$ such that $B\sD\subseteq\sD$ and $BA\psi=AB\psi$ for all $\psi\in\sD$.
				\item[(iii)] The domain $\sD(A)$ is invariant under $B$ and, for all $\psi\in\sD(A)$, $BA\psi=AB\psi$.
			\end{enumerate}
		\end{lemma}

		
		\begin{remark}\label{rem:strong-comm}
			The statements (ii)--(iii) for strong commutativity can be extended for general closed (not necessarily self-adjoint) operators $A$.
		\end{remark}
		
		
		\begin{remark}\label{rem:comm-adj}
			Note that $T\subset S$ implies that $S^{*}\subset T^{*}$ for general operators $T$ and $S$ on a Hilbert space.  If a bounded operator $B$ commutes strongly with some operator $A$, we have $B^{*}A^{*}\subset A^{*}B^{*}$, i.e., $B^{*}$ commutes strongly with $A^{*}$.
		\end{remark}
		
		

		\subsubsection{Consequence of $G'$ being a $*$-map}\label{sec:*-map}
		We have shown in Appendix \ref{sec:GNS} that the triple $(\cS_{2},\pi_{\tau},\Omega_{\tau})$, where $\cS_{2}$ is the space of Hilbert-Schmidt operators on $\cH$, $\Omega_{\tau}:=\rho_{\tau}^{1/2}$ and $\pi_{\tau}$ is the linear representation of left action by $\cB$ on $\cS_{2}$, given explicitly by
		\begin{align}
			\pi_{\tau}(A)\kappa&=A\kappa\quad\quad\text{for all }A\in\cB,
		\end{align}
		yields the GNS representation associated with the pair $(\cB,\omega_{\tau})$.  Note that $\Omega_{\tau}$ is cyclic for $\pi_{\tau}(\cB)$ in $\cS_{2}$ (see Appendix \ref{sec:GNS} for details).
		
		
		
		The Lindblad operator $G'$ induces a linear operator $\widehat{G'}$, defined by
		\begin{align}
			\widehat{G'}(\pi_{\tau}(A)\Omega_{\tau})&:=\pi_{\tau}(G'(A))\Omega_{\tau},\quad\quad\text{for all } A\in\cB,
		\end{align}
		on the dense set in $\cS_{2}$, given by
		\begin{align}
			\sF:=\pi_{\tau}(A)\Omega_{\tau}.
		\end{align}
		Since $G'$ is bounded on $\cB$, $\widehat{G'}$ extends to a bounded operator on $\cS_{2}$.  We retain the same notation $\widehat{G'}$ also for this extension.
		
		
		
		Define the anti-unitary operator $J$ on $\cS_{2}$ by
		\begin{align}
			\label{mod-conj2}&J\kappa=\kappa^{*},\quad\text{ for all }\kappa\in\cS_{2}.
		\end{align}
		Note that $J=J^{*}=J^{-1}$ (see Proposition \ref{prop:prop-GNS} (d) for the proof).  
		
		
		
		A key to the proof of Theorem \ref{thm:Gibbs} is the following theorem.
		\begin{thm}\label{thm:GNS-TT2}(\cite{Haag}, Section V.1.4, Theorem 1.4.2)
			Let $\alpha_{t}(A):=e^{iH_{\tau}t}Ae^{-iH_{\tau}t}$, where $H_{\tau}=-\ln\rho_{\tau}$.  Then, there is a (unbounded) self-adjoint operator $L_{\tau}$ on $\cS_{2}$ such that
			\begin{align}
				\label{mod-auto2}&\pi_{\tau}(\alpha_{t}(A))=e^{iL_{\tau}t}\pi_{\tau}(A)e^{-iL_{\tau}t},\quad L_{\tau}\Omega_{\tau}=0.
			\end{align}
			Moreover, we have the following relation for operators $J$ and $L_{\tau}$:  For any $A\in\cB$,  $\pi_{\tau}(A)\Omega_{\tau}\in\sD(e^{-L_{\tau}/2})$ and
			\begin{align}
				\label{mod2}&Je^{-L_{\tau}/2}(\pi_{\tau}(A)\Omega_{\tau})=\pi_{\tau}(A^{*})\Omega_{\tau},\quad J\Omega_{\tau}=\Omega_{\tau}.
			\end{align}
		\end{thm}
		
		
		
		We define the anti-linear operator $S$ on $\cS_{2}$ by
		\begin{align}
			S&=Je^{-L_{\tau}/2}.
		\end{align}
		By relation \eqref{mod2}, we have
		\begin{align}\label{S-op}
			S(\pi_{\tau}(A)\Omega_{\tau})&=\pi_{\tau}(A^{*})\Omega_{\tau}.
		\end{align}
		\begin{lemma}\label{lem:F-core}
			The set $\sF$ is a core for $S$.
		\end{lemma}
		\begin{proof}
			Since $J$ is bounded and invertible, we have $\sD(S)=\sD(e^{-L_{\tau}/2})$.  Thus, to show that $\sF$ is a core for $S$, it suffices to show that it is a core for $e^{-L_{\tau}/2}$.  Since $L_{\tau}$ is self-adjoint on $\cS_{2}$, it generates a one-parameter group $e^{iL_{\tau}t}$, $t\in\R$, of unitary operators on $\cS_{2}$.  By \eqref{mod-auto2}, for each $t\in\R$ and $A\in\cB$, we have
			\begin{align}
				e^{iL_{\tau}t}(\pi_{\tau}(A)\Omega_{\tau})&=(e^{iL_{\tau}t}\pi_{\tau}(A)e^{-iL_{\tau}t})\Omega_{\tau}=\pi_{\tau}(\alpha_{t}(A))\Omega_{\tau}\in\sF.
			\end{align} 
			Thus, $e^{iL_{\tau}t}\sF\subseteq\sF$ for each $t\in\R$ so that, by the density of $\sF$ in $\cS_{2}$, $\sF$ is a core for $L_{\tau}$ and therefore for $S$.
		\end{proof}
		
		
		
		Recall that an operator $K$ on $\cB$ is called a $*$-map if it commutes with taking the adjoint: $K(A^{*})=(KA)^{*}$ for every $A\in\cB$.
		
		
		In the next proposition, we relate $G'$ and $S$:
		\begin{prop}\label{prop:*-map-commS}
			The Lindblad operator $G'$ is a $*$-map on $\cB$ if and only if its induced operator $\widehat{G'}$ on $\cS_{2}$ commutes strongly with $S$.
		\end{prop}
		\begin{proof}
			Suppose $G'$ is a $*$-map.  Then, by Lemma \ref{lem:F-core}, since $\sF$ is a core for $S$ and $\widehat{G'}\sF\subseteq \sF$, for each $A\in\cB$, we have, by \eqref{S-op},
			\begin{align}
				S\widehat{G'}(\pi_{\tau}(A)\Omega_{\tau})&=S(\pi_{\tau}(G'A)\Omega_{\tau})=\pi_{\tau}((G'A)^{*})\Omega_{\tau}\nonumber\\
				&=\pi_{\tau}(G'(A^{*}))\Omega_{\tau}=\widehat{G'}(\pi_{\tau}(A^{*})\Omega_{\tau})=\widehat{G'}S(\pi_{\tau}(A)\Omega_{\tau}).
			\end{align}
			Thus, $\widehat{G'}$ commutes strongly with $S$.
			
			
			Conversely, by the same computations above, we obtain, for each $A\in\cB$,
			\begin{align}
				(G'(A^{*}))\rho_{\tau}^{1/2}&=\pi_{\tau}(G'(A^{*}))\Omega_{\tau}=\pi_{\tau}((G'A)^{*})\Omega_{\tau}=(G'A)^{*}\rho_{\tau}^{1/2}.
			\end{align}
			Since $\rho_{\tau}>0$, we have $G'(A^{*})=(G'A)^{*}$ for all $A\in\cB$.  Hence, $G'$ is a $*$-map.
		\end{proof}
		
		\smallskip
		
		\subsubsection{Consequence of Condition $(QDB)$}\label{sec:qdbc2}
		Now, we impose Condition $(QDB)$ on $G'$.  An immediate consequence of Condition $(QDB)$ is that the induced operator $\widehat{G'}$ is self-adjoint on $\cS_{2}$:
		\begin{lemma}\label{lem:G'-sa}
			The Lindblad operator $G'$ satisfies Condition $(QDB)$ w.r.t. $\rho_{\tau}$ if and only if the induced operator $\widehat{G'}$ is self-adjoint on $\cS_{2}$.
		\end{lemma}
		\begin{proof}
			Suppose $G'$ satisfies Condition $(QDB)$.  For any $A,B\in\cB$, we have
			\begin{align}
				\langle\widehat{G'}&(\pi_{\tau}(A)\Omega_{\tau}),\pi_{\tau}(B)\Omega_{\tau}\rangle_{\cS_{2}}=\Tr((G'A)^{*}B\rho_{\tau})=\Tr(A^{*}G(B\rho_{\tau}))\nonumber\\
				&=\Tr(A^{*}G'(B)\rho_{\tau})=\langle\pi_{\tau}(A)\Omega_{\tau},\widehat{G'}(\pi_{\tau}(B)\Omega_{\tau})\rangle_{\cS_{2}}.
			\end{align}
			Since $\sF$ is dense in $\cS_{2}$ and $\widehat{G'}$ is bounded, we can extend this equality to the entire space $\cS_{2}$.  Thus, $\widehat{G'}$ is self-adjont on $\cS_{2}$.
			
			
			The converse follows from the same computation so that $\widehat{G'}$ is self-adjoint on $\cS_{2}$ implies that $G'$ satisfies Condition $(QDB)$.
		\end{proof}
		
		
		Using Proposition \ref{prop:*-map-commS} and Lemma \ref{lem:G'-sa}, we establish the strong commutativity between $\widehat{G'}$ and $e^{-L_{\tau}}$:
		\begin{thm}(\cite{FGKV})\label{thm:G'-L*-comm}
			Suppose the Lindblad operator $G'$ satisfies Condition $(QDB)$ w.r.t. $\rho_{\tau}$.  Then, the induced operator $\widehat{G'}$ commutes strongly with $e^{-L_{\tau}}$.
		\end{thm}
%
%
		
		
		\begin{proof}[Proof of Theorem \ref{thm:Gibbs}]
			Suppose $\rho_{\tau}=e^{-H/\tau}/Z(\tau)$.  In this case, we have $H_{\tau}=\beta H+\ln Z(\tau)$ so that $\alpha_{t}=e^{\beta L_{0}'t}$.  Then, by Theorem \ref{thm:G'-L*-comm}, $\widehat{G'}$ commutes with $e^{iL_{\tau}t}$ so that, for all $A\in\cB$,
			\begin{align}
				\pi_{\tau}(e^{L_{0}'t}G'A)\Omega_{\tau}&=\pi_{\tau}(\alpha_{t/\beta}(G'A))\Omega_{\tau}=e^{iL_{\tau}t/\beta}(\pi_{\tau}(G'A))\Omega_{\tau}\nonumber\\
				&=e^{iL_{\tau}t/\beta}\widehat{G'}(\pi_{\tau}(A)\Omega_{\tau})=\widehat{G'}e^{iL_{\tau}t/\beta}(\pi_{\tau}(A)\Omega_{\tau})\nonumber\\
				&=\widehat{G'}(\pi_{\tau}(e^{L_{0}'t}A)\Omega_{\tau})=\pi_{\tau}(G'e^{L_{0}'t}A)\Omega_{\tau}.
			\end{align}
			Since $\sF$ is dense in $\cS_{2}$, we have $e^{L_{0}'t}$ and $G'$ commute on $\cB$.  Since $e^{L_{0}'t}$ and $G'$ are bounded and $\cB$ is dense in $\cB_{\tau}$, the commutativity between $e^{L_{0}'t}$ and $G'$ can be extended to the entire space $\cB_{\tau}$.  Also, by the duality relation, we obtain the strong commutativity between $L_{0}$ and $G$.  This proves Theorem \ref{thm:Gibbs}.
		\end{proof}
		




\section{Proofs of main results}\label{sec:pfs}

\subsection{Proof of  Theorems \ref{thm:vNL-exist} and  \ref{thm:HL-exist}}\label{sec:pf-vNL-HL-exist}
Recall the definitions of the spaces $\cB_{\tau}$ and $\cS_{\tau}$ in \eqref{Stau} and above \eqref{ip-obs-2}. 
\begin{proof}[Proof of  Theorem \ref{thm:HL-exist}] By Proposition \ref{prop:G'-property-B*}, $G'$ is bounded on $\cB_{\tau}$.  Also, since $L_{0}'$ is anti-self-adjoint on $\cB_{\tau}$, it generates a one-parameter strongly continuous group $e^{L_{0}'t}$ of unitary operators on $\cB_{\tau}$.  Then by Theorem \ref{thm:exist-evol}, 
the HL operator $L'$ generates a bounded semigroup $\beta_{t}'$, and therefore the HL equation has a unique solution in $\cB_{\tau}$ for any given initial condition in $\sD(L')$. 
\end{proof}



\begin{remark}\label{uni-eL0'}
	The unitarity of $e^{L_{0}'t}$ can be proven directly as follows.  For all $A,B\in\cB_{\tau}$, 
	\begin{align}
		\langle e^{L_{0}'t}A,e^{L_{0}'t}B\rangle_{\tau}&=\Tr(e^{iHt}A^{*}Be^{-iHt}\rho_{\tau})=\Tr(A^{*}Be^{-iHt}\rho_{\tau}e^{iHt})\nonumber\\
		&=\Tr(A^{*}B\rho_{\tau})=\langle A,B\rangle_{\tau}
	\end{align}
	by cyclicity of trace and the fact that $L_{0}'\rho_{\tau}=0$ by $(QDB)$.  Hence, it is unitary.
\end{remark}

%

The semigroup, $\beta_{t}'$, generated by $L'$, is dual to $\beta_{t}$.  Indeed, the semigroup $\beta_{t}'$ has the generator $L'$: 
\begin{align}\label{HL-gen}
	\partial_{t}|_{t=0}\Tr(\beta_{t}'(A)\rho)&=\partial_{t}|_{t=0}\Tr(A\beta_{t}(\rho))=\Tr(AL(\rho))=\Tr(L'(A)\rho).
\end{align}
Given that $G'$ is bounded on $\cB$, the domain of $L'$ is the same as the domain $\sD(L_{0}')$ of $L_{0}'$, where 
\begin{align}\label{domain-L'}
	\sD(L_{0}')&:=\{A\in\cB\mid A(\sD(H))\subset\sD(H)\text{ and } HA-AH\nonumber\\
	&\quad\quad\quad\quad\quad\quad\quad\text{defined on }\sD(H)\text{ extends to }\cH\nonumber\\
	&\quad\quad\quad\quad\quad\quad\quad\text{as an element in }\cB\}.
\end{align}

Theorem \ref{thm:vNL-exist} follows from the following result: 
		\begin{thm}\label{thm:semigroup-L}
			Assume Conditions $(H)$, $(W)$ 
			 and $(QDB)$ hold.  Then, the vNL operator generates a bounded semigroup, $\beta_{t}$, on $\cS_{\tau}$.  Consequently, the vNL equation \eqref{vNLeq} has a unique weak solution for any initial condition in $\cS_{\tau}$ and unique strong solution for any initial condition in $\sD(L)$ (the domain of $L$ in $\cS_{\tau}$).
		\end{thm}

	

The proof Theorem \ref{thm:semigroup-L} follow the same lines as that of Theorem \ref{thm:HL-exist}. 
First, we define a subspace of the Schatten space $\cS_{1}$:
		\begin{align}\label{hat-St}
			\widehat{\cS}_{\tau}:=\{A\rho_{\tau}\mid A\in\cB\}.
		\end{align}
		Then, by definition, for all $A,B\in\cB$, we have
		\begin{align}\label{rel-ips}
			\langle A\rho_{\tau},B\rho_{\tau}\rangle_{\rm st,\tau}&=\langle A,B\rangle_{\tau}.
		\end{align}
		\begin{lemma}\label{lem:hatS*-dense}
			$\widehat{\cS}_{\tau}$ is a dense subspace of the space $\cS_{\tau}$ defined in \eqref{Stau}.  
		\end{lemma}
		\begin{proof}
			For each $A\in\cB$, we have
			\begin{align}
				\|A\rho_{\tau}\|_{\rm st,\tau}^{2}&=\Tr(\rho_{\tau}^{-1}(\rho_{\tau}A^{*}A\rho_{\tau}))=\Tr(A^{*}A\rho_{\tau})<\infty.
			\end{align}
			Hence $\widehat{\cS}_{\tau}\subseteq\cS_{\tau}$.  Furthermore, if $\lambda\in\cS_{\tau}$ such that $\lambda\perp\widehat{\cS}_{\tau}$, then, for all $A\in\cB$,
			\begin{align}\label{den-Stau}
				0&=\langle A\rho_{\tau}, \lambda\rangle_{\rm st,\tau}=\Tr(A^{*}\lambda).
			\end{align}
			Since $\lambda\in\cS_{1}$ and $\cB$ is the dual space of $\cS_{1}$, \eqref{den-Stau} implies that $\lambda=0$.  Hence, $\widehat{\cS}_{\tau}$ is dense in $\cS_{\tau}$.
		\end{proof}

			 Now, we have
		\begin{lemma}\label{lem:S*-dense}
			The map $\varphi:A\in\cB\mapsto A\rho_{\tau}\in\widehat{\cS}_{\tau}$ can be extended to a unitary map from $\cB_{\tau}$ to $\cS_{\tau}$ (for which we keep the symbol $\varphi$),
			\begin{align}\label{phi-unit}
				\langle\varphi(A),\varphi(B)\rangle_{\rm st,\tau}&=\langle A,B\rangle_{\tau}\quad\text{ for all }A,B\in\cB_{\tau}.
			\end{align}
		\end{lemma}
		\begin{proof}
%
%
%
			By \eqref{rel-ips} and the definition of $\varphi$, the relation \eqref{phi-unit} holds for any $A,B\in\cB$.  Hence, $\varphi$ is an isometry.  Since $\varphi$ is invertible with the inverse $\varphi^{-1}(\lambda)=\lambda\rho_{\tau}^{-1}$ and since $\cB$ and $\widehat{\cS}_{\tau}$ are dense in $\cB_{\tau}$ and $\cS_{\tau}$, respectively, we can extend this map to a unitary map $\varphi:\cB_{\tau}\rightarrow\cS_{\tau}$ (using the same symbol).
		\end{proof}

		
		The significance of the map $\varphi$ 
		lies in the fact that $(QDB)$ \eqref{(QDB)} can be written as
		\begin{align}\label{(QDB)-2}
			L_{0}\circ \varphi&=-\varphi\circ L_{0}',\quad\quad G\circ\varphi=\varphi\circ G'.
		\end{align}
		This will be used to translate our results for the HL equation to ones on the vNL equation. 
		
		
		Equations \eqref{phi-unit} and \eqref{(QDB)-2} imply that, for all $A,B\in\cB_{\tau}$,
		\begin{align}\label{G-phi-rel}
			\langle G(\varphi(A)),\varphi(B)\rangle_{\rm st,\tau}&=\langle G'(A),B\rangle_{\tau}.
		\end{align}
		

\begin{prop}\label{prop:bdd-sa-G}
	Assume Conditions $(W)$ and $(QDB)$ hold.  Then, $G$ is bounded and self-adjoint on $\cS_{\tau}$.
\end{prop}
\begin{proof}
	For boundedness, by \eqref{(QDB)}, we have, for any $\lambda\in\widehat{\cS}_{\tau}$,
	\begin{align}
		\|G(\lambda)\|_{\rm st,\tau}&=\|G'(A)\rho_{\tau}\|_{\rm st,\tau}=\|G'(A)\|_{\tau}\nonumber\\
		&\leq \|G'\|\|A\|_{\tau}=\|G'\|\|A\rho_{\tau}\|_{\rm st,\tau},
	\end{align}
	where $A=\lambda\rho_{\tau}^{-1}\in\cB$.  Since, by Lemma \ref{lem:S*-dense}, $\widehat{\cS}_{\tau}$ is dense in $\cS_{\tau}$, then $G$ is bounded on $\cS_{\tau}$.
	
	
	
	Now, we show that $G$ is symmetric on $\cS_{\tau}$.  By Lemma \ref{lem:qdbc-sym} and cyclicity of trace, we have, for any $\lambda,\mu\in\widehat{\cS}_{\tau}$, we take $\lambda=A\rho_{\tau}$ and $\mu=B\rho_{\tau}$ so that
	\begin{align}
		\langle\lambda,G(\mu)\rangle_{\rm st,\tau}&=\Tr(\lambda^{*}G(\mu)\rho_{\tau}^{-1})=\Tr(\rho_{\tau}A^{*}G(B\rho_{\tau})\rho_{\tau}^{-1})\nonumber\\
		&=\Tr(A^{*}G'(B)\rho_{\tau})=\langle A,G'(B)\rangle_{\tau}=\langle G'(A),B\rangle_{\tau}\nonumber\\
		&=\Tr((G'(A))^{*}B\rho_{\tau})=\Tr(\rho_{\tau}^{-1}(G'(A)\rho_{\tau})^{*}B\rho_{\tau})\nonumber\\
		&=\Tr(\rho_{\tau}^{-1}(G(A\rho_{\tau}))^{*}B\rho_{\tau})=\Tr(\rho_{\tau}^{-1}G(\lambda)^{*}\mu)\nonumber\\
		&=\langle G(\lambda),\mu\rangle_{\rm st,\tau}.
	\end{align}
	Therefore, $G$ is symmetric on $\widehat{\cS}_{\tau}$ and hence, by a density argument, $G$ is symmetric on $\cS_{\tau}$.  Since $G$ is bounded on $\cS_{\tau}$, then $G$ is self-adjoint on $\cS_{\tau}$.
\end{proof}


\begin{prop}\label{prop:asa-L0}
	Assume Conditions $(H)$ and $(QDB)$ hold.  Then, $L_{0}$ is anti-self-adjoint on $\cS_{\tau}$.
\end{prop}
\begin{proof}
	By Stone's theorem, it suffices to show that $L_{0}$ generates a strongly continuous group of unitary operators on $\cS_{\tau}$.  By definition of $L_{0}$, for any $\lambda\in\cS_{\tau}$, we have $e^{L_{0}t}\lambda=e^{-iHt}\lambda e^{iHt}$.  The group property of $e^{L_{0}t}$ follows from the group property of $e^{\pm iHt}$.  The group $e^{L_{0}t}$ is unitary on $\cS_{\tau}$ since, for any $\lambda,\mu\in\cS_{\tau}$ and $t\in\R$, by cyclicity of trace and Condition $(QDB)$, we have
	\begin{align}
		\langle e^{L_{0}t}\lambda,e^{L_{0}t}\mu\rangle_{\rm st,\tau}&=\Tr((e^{-iHt}\lambda e^{iHt})^{*}(e^{-iHt}\mu e^{-iHt})\rho_{\tau}^{-1})\nonumber\\
		&=\Tr(\lambda^{*}\mu\rho_{\tau}^{-1})=\langle\lambda,\mu\rangle_{\rm st,\tau}.
	\end{align}
	
	
	Now, we show $e^{L_{0}t}$ is strongly continuous on $\cS_{\tau}$.  Recall that $\cS_{\tau}$ is the completion of $\tilde{\cS}_{\tau}=\{\lambda\in\cS_{1}\mid \lambda\rho_{\tau}^{-1/2}\in\cS_{2}\}$ w.r.t. to the norm $\|\cdot\|_{\rm st,\tau}$.  Let $\lambda\in\tilde{\cS}_{\tau}$.  Then, there exists some $\kappa\in\cS_{2}$ such that $\kappa=\lambda\rho_{\tau}^{-1/2}$ so that, by Condition $(QDB)$, cyclicity of trace and the fact that $e^{-iHt}$ is strongly continuous on $\cH$, we have
	\begin{align}
		\|e^{L_{0}t}\lambda-\lambda\|_{\rm st,\tau}^{2}&=\Tr((e^{-iHt}\lambda e^{iHt}-\lambda)^{*}(e^{-iHt}\lambda e^{iHt}-\lambda)\rho_{\tau}^{-1})\nonumber\\
		&=\Tr((e^{-iHt}\kappa e^{iHt}-\kappa)^{*}(e^{-iHt}\kappa e^{iHt}-\kappa))\nonumber\\
		&=\|e^{-iHt}\kappa e^{iHt}-\kappa\|_{\cS_{2}}^{2}\rightarrow 0
	\end{align}
	as $t\rightarrow 0$.  By a density argument and unitary property of $e^{L_{0}t}$, we can extend this limit to the whole space $\cS_{\tau}$ so that $e^{L_{0}t}$ is strongly continuous.  Therefore, the generator $L_{0}$ is anti-self-adjoint on $\cS_{\tau}$.
\end{proof}



\begin{proof}[Proof of Theorem \ref{thm:semigroup-L}]
	Since the operator $L_{0}$ generates a one-parameter strongly continuous group $e^{L_{0}t}$ of unitary operators on $\cS_{\tau}$ (given explicitly by $e^{L_0 t}\rho =e^{-iHt}\rho e^{iHt}$), and, by Proposition \ref{prop:bdd-sa-G}, $G$ is bounded on $\cS_{\tau}$.  It follows from Theorem \ref{thm:exist-evol}, with $X=\cS_{\tau}$, that $L=L_{0}+G$ generates a one-parameter, strongly continuous semigroup on $\cS_{\tau}$, which implies the statement of Theorem \ref{thm:semigroup-L}.
\end{proof}

\begin{remark}
	One can also prove a version of Proposition \ref{prop:bdd-sa-G} with the space $\cS_{\tau}^{(1)}$ (see Remark \ref{rem:spaces}(a)) replacing $\cS_{\tau}$.
\end{remark}

\begin{remark}
	It is \textit{not} clear whether the semigroup $\beta_{t}$ is positive on $\cS_{\tau}$ (under $(QDB)$), i.e., $\langle\lambda,\beta_{t}(\lambda)\rangle_{\rm st,\tau}$ is not necessarily non-negative.  Indeed, if we follow the procedure in the proof of Lemma \ref{lem:pos-vNL-evol}, since $Y$ commutes with $\rho_{\tau}$ by Lemma \ref{lem:comm-G'-qdbc}, then we have $B(t)=e^{-iHt-Y t}$ commutes with $\rho_{\tau}$ so that, by cyclicity of trace,
	\begin{align}
		\langle \lambda, S_{t}(\lambda)\rangle_{\rm st,\tau}&=\Tr(\lambda^{*}B(t)\lambda B(t)^{*}\rho_{\tau}^{-1})=\Tr(\lambda^{*}B(t)\lambda\rho_{\tau}^{-1}B(t)^{*})\nonumber\\
		&=\Tr(B(t)^{*}\lambda^{*} B(t)\lambda\rho_{\tau}^{-1})=\langle \lambda B(t),B(t)\lambda\rangle_{\rm st,\tau}\nonumber\\
		&\neq \langle B(t)\lambda,B(t)\lambda\rangle_{\rm st,\tau}\geq 0.
	\end{align}
	Thus, $S_{t}$ is not necessarily positive on $\cS_{\tau}$ so that $\beta_{t}$ is not either.
\end{remark}


		
		\subsection{Proof of Theorem \ref{thm:main}}\label{sec:mains-pf}
				
		The proof of Theorem \ref{thm:main} follows from the following proposition:
		

		\begin{prop}\label{prop:key-prop-main}
			Assume Conditions $(H)$, $(W)$ and $(QDB)$ hold.  Then we have
			\begin{enumerate}
				\item[(i)] $G\leq 0$;
				\item[(ii)] $[L_{0},G]=0$.
			\end{enumerate}
			(Statement (ii) says that $[L_{0},G]=0$ is well-defined and vanishes.)
		\end{prop}
		
		
		\begin{remark}
			In order not to deal with the domain questions, we can interpret the relation in Proposition \ref{prop:key-prop-main} (ii) as $[e^{L_{0}t},G]=0$ for all $t\in\R$. 
		\end{remark}
		
		
		\begin{proof}[Proof of Proposition \ref{prop:key-prop-main}]
			First, we define the operator $L_{0}'$ dual to $L_{0}$:
			\begin{align}
				\Tr(L_{0}'(A)\rho)&=\Tr(AL_{0}(\rho)),\quad\forall A\in\sD(L_{0}'),\rho\in\sD(L_{0}).
			\end{align}
			

			To prove Proposition \ref{prop:key-prop-main} (i), we use that, by Theorem \ref{thm:G'-property}, $G'\leq 0$ on $\cB_{\tau}$, and use \eqref{rel-ips} to obtain, for all $\lambda=A\rho_{\tau}\in\widehat{\cS}_{\tau}$, 
			\begin{align}
				\langle\lambda,G(\lambda)\rangle_{\rm st,\tau}&=\langle A\rho_{\tau},G(A\rho_{\tau})\rangle_{\rm st,\tau}=\langle A\rho_{\tau},G'(A)\rho_{\tau}\rangle_{\rm st,\tau}\nonumber\\
				&=\langle A,G'(A)\rangle_{\tau}\leq 0.
			\end{align}
			Hence, $G\leq 0$ on $\widehat{\cS}_{\tau}$.  Again, since $\widehat{\cS}_{\tau}$ is dense in $\cS_{\tau}$, $G\leq 0$ on $\cS_{\tau}$, which gives (i).
			
			
			To prove Proposition \ref{prop:key-prop-main} (ii), we use that, by Theorem \ref{thm:Gibbs}, $[e^{L_{0}'t},G']=0$ and use \eqref{rel-ips} to obtain, for all $\lambda=A\rho_{\tau}\in\widehat{\cS}_{\tau}$, 
			\begin{align}\label{comm-L0-G}
				[e^{L_{0}t},G]\lambda&=[e^{L_{0}t},G]A\rho_{\tau}=e^{L_{0}t}G(A\rho_{\tau})-G(e^{L_{0}t}(A\rho_{\tau}))\nonumber\\
				&=(e^{-L_{0}'t}G'(A)-G'(e^{-L_{0}'t}A))\rho_{\tau}=([e^{-L_{0}'t},G]A)\rho_{\tau}=0.
			\end{align}
			By the density of $\widehat{\cS}_{\tau}$ in $\cS_{\tau}$, we can extend equality in \eqref{comm-L0-G} to all $\lambda\in\cS_{\tau}$.  Thus, $[G,L_{0}]=0$.
		\end{proof}
		\begin{prop}\label{prop:i-iv-ergodic-vNLE}
			Propositions \ref{prop:bdd-sa-G}, \ref{prop:key-prop-main} and Theorem \ref{thm:G'-property} imply Theorem \ref{thm:main}.
		\end{prop}
		\begin{proof}[Proof of Proposition \ref{prop:i-iv-ergodic-vNLE}]
			In what follows, for any operator $K$ on $\cS_{\tau}$, we denote by $P_{K}$ the orthogonal projection onto the subspace $\Null(K)$ and let $P_{K}^{\perp}:=\one-P_{K}$.
			
			
			We begin with an elementary lemma.
			\begin{lemma}\label{lem:ergodic-arg}
				Let $K$ be an operator on a Hilbert space $\cH$, which is either anti-self-adjoint or self-adjoint and negative.  Then we have
				\begin{align}\label{st-con-K}
					s\text{-}\lim_{T\rightarrow\infty}\frac{1}{T}\int_{0}^{T}e^{Kt}dt&=P_{K},
				\end{align}
				where $P_{K}$ is the orthogonal projection in $\cH$ onto $\Null(K)$.
			\end{lemma}
			\begin{proof}[Proof of Lemma \ref{lem:ergodic-arg}]
				We begin with the well-known relation 
				\begin{align}
					\cH&=\Null(K)\oplus\overline{\Ran(K^{*})},
				\end{align}
				valid for any operator $K$ on $\cH$.  In our case, $\Ran(K^{*})=\Ran(K)$.  Now, let $K^{\perp}:=K\big|_{\Null(K)^{\perp}}$.  The operator $K^{\perp}$ has an inverse with the dense domain $\sD((K^{\perp})^{-1})=\Ran(K)$.
				
				
				Next, let $\|\cdot\|$ denote the norm in $\cH$.  For each $\lambda\in\sD((K^{\perp})^{-1})$, we have
				\begin{align}
					\int_{0}^{T}e^{Kt}\lambda dt&=(e^{KT}-\one)(K^{\perp})^{-1}\lambda.
				\end{align}
				Since $\|e^{KT}\|\leq 1$, for all $T>0$, we have
				\begin{align}
					\|(e^{KT}-\one)(K^{\perp})^{-1}\lambda\|&\leq 2\|(K^{\perp})^{-1}\lambda\|.
				\end{align}
				Since $\|(K^{\perp})^{-1}\lambda\|<\infty$ for $\lambda\in\sD((K^{\perp})^{-1})$, we have, for such $\lambda$, as $T\rightarrow\infty$,
				\begin{align}\label{ergodic-K}
					\big\|\frac{1}{T}\int_{0}^{T}e^{Kt}\lambda dt\big\|&=\frac{1}{T}\|(e^{KT}-\one)(K^{\perp})^{-1}\lambda\|\leq \frac{2}{T}\|(K^{\perp})^{-1}\lambda\|\rightarrow 0.
				\end{align}
				Since the integral $\frac{1}{T}\int_{0}^{T}(e^{Kt}\lambda)dt$ is uniformly bounded in $T$ for all $\lambda\in\cH$, we can extend the convergence \eqref{ergodic-K} to all $\lambda\in\overline{\sD((K^{\perp})^{-1})}=\overline{\Ran(K)}=\Null(K)^{\perp}$.  
				
				
				Let $P_{K}^{\perp}:=\one-P_{K}$ and note that $P_{K}^{\perp}\lambda\in\Null(K)^{\perp}$.  Then, for all $\lambda\in\cH$, we have, as $T\rightarrow\infty$,
				\begin{align}
					\big\|\frac{1}{T}\int_{0}^{T}e^{Kt}\lambda  dt-P_{K}\lambda\big\|&=\big\|\frac{1}{T}\int_{0}^{T}e^{Kt}(P_{K}^{\perp}\lambda)dt\big\|\rightarrow 0,
				\end{align}
				which implies \eqref{st-con-K}.
			\end{proof}
			

			In the rest of the proof, we deal with the space $\cS_{\tau}$ and drop the subindex ``$\tau$" from notation for this space and its norm and inner product.  
			
			
			Let $\lambda\in\cS$.  Then, since $[L_{0},G]=0$ under $(QDB)$, we have
			\begin{align}\label{int-beta-lam}
				\frac{1}{T}\int_{0}^{T}\beta_{t}(\lambda)dt&=\frac{1}{T}\int_{0}^{T}(e^{L_{0}t}e^{Gt}\lambda)dt\nonumber\\
				&=\frac{1}{T}\int_{0}^{T}e^{L_{0}t}\big(e^{Gt}\lambda-P_{G}\lambda\big)dt+\frac{1}{T}\int_{0}^{T}e^{L_{0}t}P_{G}\lambda dt.
			\end{align}
			We denote the first and second term on the r.h.s. of \eqref{int-beta-lam} by $A_{T}$ and $B_{T}$, respectively.  
			
			
			For the term $A_{T}$, by using Cauchy-Schwarz inequality and the fact that $e^{L_{0}t}$ is unitary on $\cS$, we obtain
			\begin{align}\label{A_T}
				\|A_{T}\|&\leq \frac{1}{T}\int_{0}^{T}\|e^{Gt}\lambda-P_{G}\lambda\|dt\leq\frac{1}{\sqrt{T}}\big(\int_{0}^{T}\|e^{Gt}\lambda-P_{G}\lambda\|^{2} dt\big)^{1/2}\nonumber\\
				&=\frac{1}{\sqrt{T}}\left(\int_{0}^{T}\big(\langle e^{2Gt}\lambda,\lambda\rangle-\langle P_{G}\lambda,\lambda\rangle\big)dt\right)^{1/2}\nonumber\\
				&=\left(\langle \frac{1}{T}\int_{0}^{T}(e^{2Gt}\lambda-P_{G}\lambda)dt,\lambda\rangle\right)^{1/2}\nonumber\\
				&\leq \big\|\frac{1}{T}\int_{0}^{T}(e^{2Gt}\lambda-P_{G}\lambda) dt\big\|^{1/2}\|\lambda\|^{1/2}.
			\end{align}
			By Lemma \ref{lem:ergodic-arg} and the fact that $G\leq 0$, we have, as $T\rightarrow\infty$,
			\begin{align}
				\big\|\frac{1}{T}\int_{0}^{T}(e^{2Gt}\lambda-P_{G}\lambda) dt\big\|&\rightarrow 0.
			\end{align}
			Hence, $\|A_{T}\|\rightarrow 0$ as $T\rightarrow\infty$.
			

			Now, for the term $B_{T}$, since $L_{0}$ is anti-self-adjoint, by Lemma \ref{lem:ergodic-arg}, we obtain
			\begin{align}\label{B_T}
				\lim_{T\rightarrow\infty}B_{T}&=\lim_{T\rightarrow\infty}\frac{1}{T}\int_{0}^{T}(e^{L_{0}t}P_{G}\lambda)dt= P_{L_{0}}P_{G}\lambda.
			\end{align}
			Relations $\lim_{T\rightarrow\infty}A_{T}=0$ and \eqref{B_T}, together with \eqref{int-beta-lam}, imply 
			\begin{align}\label{ergodic-PL0G}
				\lim_{T\rightarrow\infty}\frac{1}{T}\int_{0}^{T}\beta_{t}dt&=P_{L_{0}}P_{G}.
			\end{align}
			

			To finish this proof of Theorem \ref{thm:main}, we use the following lemma: 
			\begin{lemma}\label{lem:null-space-L}
				Assume the Conditions $(H)$, $(W)$ 
				and $(QDB)$.  Then $\Null(L)=\Null(G)\cap\Null(L_{0})$.
			\end{lemma}
			\begin{proof}
				Since $L=L_{0}+G$, we have $\Null(G)\cap\Null(L_{0})\subseteq\Null(L)$.  Thus, it suffices to show the reverse inclusion.
				
				
				Let $\lambda\in\Null(L)$.  Since $L_{0}$ is anti-self-adjoint and $G\leq 0$ on $\cS_{\tau}$, we have
				\begin{align}
					0&=\langle \lambda,L\lambda\rangle+\langle L\lambda,\lambda\rangle=2\langle\lambda,G\lambda\rangle=-2\||G|^{1/2}\lambda\|^{2}.
				\end{align} 
				Hence, $\lambda\in\Null(|G|^{1/2})$, which implies that $\lambda\in\Null(G)$.  Hence, $\Null(L)\subseteq\Null(G)$.  
				
				
				Next, it follows that
				\begin{align}
					L_{0}\lambda&=L\lambda-G\lambda=0,
				\end{align}
				so that $\lambda\in\Null(L_{0})$.  Hence $\Null(L)\subseteq\Null(L_{0})$, which, together with the previous conclusion, implies that $\Null(L)\subseteq\Null(G)\cap\Null(L_{0})$.  Hence, the statement of Lemma \ref{lem:null-space-L} follows.
			\end{proof}
			
			
			Since, by Lemma \ref{lem:null-space-L}, $P=P_{L_{0}}P_{G}$, Eq. \eqref{ergodic-PL0G} yields (on the space $\cS$)
			\begin{align}\label{str-ergodic}
				s\text{-}\lim_{T\rightarrow\infty}\frac{1}{T}\int_{0}^{T}\beta_{t}dt&=P.
			\end{align}
		\end{proof}


		
		\subsection{Uniqueness of stationary state.  Proof of Theorems \ref{thm:main1} and \ref{thm:RTE-dual3}}\label{sec:uniq-stationary}
		The next result gives a sufficient condition for an eigenvalue of $G'$ at $0$ to be simple (see \cite{Evans,Fri} for different proofs):
		\begin{prop}\label{prop:simp-zero}
			Suppose Condition $(W)$, $(QDB)$ and $(U)$.  Then $0$ is a simple eigenvalue of $G'$ on $\cB_{\tau}$.
		\end{prop}
		\begin{proof}
		First, let $A\in\Null(G')\cap\cB$.  By \eqref{G'A-ip}, we have
		\begin{align}\label{G'A-ip-0}
			0&=\langle A,G'(A)\rangle_{\tau}=-\frac{1}{2}\sum_{j\geq 1}\|[W_{j},A]\|_{\tau}^{2}.
		\end{align}
		Since $\|[W_{j},A]\|\geq 0$ for all $j\geq 1$, \eqref{G'A-ip-0} implies that $[W_{j},A]=0$ for all $j\geq 1$.  
		
		
		Next, since $G'$ is a $*$-map, we have $G'(A^{*})=0$ if $A\in\Null(G')\cap\cB$ and, by \eqref{G'A-ip} again,
		\begin{align}\label{G'A*-ip-0}
			0&=\langle A^{*},G'(A^{*})\rangle_{\tau}=-\frac{1}{2}\sum_{j\geq 1}\|[W_{j},A^{*}]\|_{\tau}^{2}\nonumber\\
			&=-\frac{1}{2}\sum_{j\geq 1}\|[W_{j}^{*},A]\|_{\tau}^{2}
		\end{align}
		so that $[W_{j}^{*},A]=0$ for all $j\geq 1$.  By Condition $(U)$, we must have $A\in\C\cdot\one$.
		
		
		Since $G'(\one)=0$, we have $\Null(G')\cap\cB=\C\cdot\one$.  Since $\cB$ is dense in $\cB_{\tau}$, we conclude that $\Null(G')=\C\cdot \one$.
		\end{proof}

		\begin{proof}[Proof of Theorem \ref{thm:RTE-dual3}]
			Recall $P'$ denotes the orthogonal projection onto the subspace $\Null(L')$ in $\cB_{\tau}$ w.r.t. the inner product \eqref{ip-obs-2}.  Since $\Null(L')\subseteq\Null(G')$ and $\one\in\Null(L')$, Proposition \ref{prop:simp-zero} implies that, under Condition $(U)$, the orthogonal projection $P'$ onto $\Null(L')$ in $\cB_{\tau}$ is given by
			\begin{align}\label{proj-A}
				P'A&=\Tr(A\rho_{\tau})\cdot\one,\quad\quad\forall A\in\cB_{\tau}.
			\end{align}
			Then \eqref{proj-A} and Theorem \ref{thm:RTE-dual1} imply Theorem \ref{thm:RTE-dual3}.
		\end{proof}
		
		\begin{proof}[Proof of Theorem \ref{thm:main1}]
			By Lemma \ref{lem:S*-dense} and Eq. \eqref{(QDB)-2}, $\dim\Null(G)=\dim\Null(G')$.  Hence, under Condition $(U)$, by Proposition \ref{prop:simp-zero}, $\dim\Null(G)=1$.  Since $\rho_{\tau}\in\Null(G)$, this implies that $\Null(G)=\C\cdot\rho_{\tau}$.  Since $\Null(L)\subseteq\Null(G)$ (see Proposition \ref{lem:null-space-L}), we conclude that, under Condition $(U)$, 
			\begin{align}
				\Null(L)=\C\cdot\rho_{\tau},
			\end{align}
			which implies that the orthogonal projection $P$ in $\cS_{\tau}$ onto $\Null(L)$ is given by
			\begin{align}
				P\lambda=\Tr(\lambda)\cdot\rho_{\tau}.
			\end{align}
			This conclusion, together with Theorem \ref{thm:main}, implies Theorem \ref{thm:main1}.
		\end{proof}
%
%
%
%

\section{Remarks and extensions}\label{sec:remark-ext}
%


\begin{remark} 
	Unlike for the von Neumann dynamics, for the von Neumann-Lindblad one, we cannot pass to the Hilbert space framework by simply writing $\Tr(A\rho)=\Tr(\sqrt\rho^* A\sqrt\rho)\equiv \langle \sqrt\rho, A\sqrt\rho\rangle$. Indeed, since 
	\[\sqrt{\beta_t\rho}\neq \beta_t(\sqrt{\rho})\] (unlike for the unitary dynamics), $\sqrt{\beta_t\rho}$, entering  in 
	\[\Tr((\beta_t' A)\rho)=\Tr(A\beta_t\rho)=\Tr((\sqrt{\beta_t\rho})^* A\sqrt{\beta_t\rho})\equiv \langle \sqrt{\beta_t\rho}, A\sqrt{\beta_t\rho}\rangle,\] does not satisfy a simple evolution equation.
\end{remark}

\begin{remark}
Similarly to $\cS_{\tau}$, we can define another Hilbert space of states as 
\begin{align}\label{S*'}
	\cS_{\tau}^{\intercal}:=\{\lambda\in\cS_{1}\mid \rho_{\tau}^{-1/2}\lambda\in\cS_{2}\},
\end{align}
with the inner product
\begin{align}\label{ip-st'}
	\langle \lambda,\mu\rangle_{\rm st,\tau}^{\intercal}&:=\Tr(\lambda^{*}\rho_{\tau}^{-1}\mu).
\end{align}
\end{remark}


\begin{lemma}\label{lem:B*-S*-dual}
	The spaces $\cS_{\tau}$ and $\cS_{\tau}^{\intercal}$ are the dual of $\cB_{\tau}$ w.r.t. the couplings $\Tr(A\rho_{\tau}^{1/2}\lambda\rho_{\tau}^{-1/2})$ and $(A,\lambda):=\Tr(A\lambda)$, respectively. 
\end{lemma}
\begin{proof}
	We check only $\cS_{\tau}^{\intercal}$; the space $\cS_{\tau}$ is done similarly.  By the non-Abelian Cauchy-Schwarz inequality, for each $A\in\cB_{\tau}$ and $\lambda\in\cS_{\tau}^{\intercal}$, we have
	\begin{align}\label{coupling}
		|\Tr(A\lambda)|&=|\Tr(A\rho_{\tau}^{1/2}\rho_{\tau}^{-1/2}\lambda)|\leq \|A\rho_{\tau}^{1/2}\|_{\cS_{2}}\|\rho_{\tau}^{-1/2}\lambda\|_{\cS_{2}}\nonumber\\
		&=(\Tr(A^{*}A\rho_{\tau}))^{1/2}(\Tr(\lambda^{*}\rho_{\tau}^{-1}\lambda))^{1/2}\nonumber\\
		&=\|A\|_{\tau}\|\lambda\|_{\rm st,\tau}^{\intercal}
	\end{align}
	so that 
	\begin{align}
		\sup_{\|A\|_{\tau}=1}|\Tr(A\lambda)|&\leq \|\lambda\|_{\rm st,\tau}^{\intercal}.
	\end{align}
	We now show that this is in fact an equality.  By taking $A=\lambda^{*}\rho_{\tau}^{-1}$, we have
	\begin{align}
		\|A\|_{\tau}^{2}&=\Tr(A^{*}A\rho_{\tau})=\Tr(\rho_{\tau}^{-1}\lambda\lambda^{*})=(\|\lambda\|_{\rm st,\tau}^{\intercal})^{2}<\infty
	\end{align}
	so that $A\in\cB_{\tau}$ and, by taking $A'=A/\|\lambda\|_{\rm st,\tau}^{\intercal}$, we have $\|A'\|_{\tau}=1$ and 
	\begin{align}
		\Tr(A'\lambda)&=(\|\lambda\|_{\rm st,\tau}^{\intercal})^{-1}\Tr(\lambda^{*}\rho_{\tau}^{-1}\lambda)=\|\lambda\|_{\rm st,\tau}^{\intercal}.
	\end{align}
	Therefore, $\cS_{\tau}^{\intercal}$ is the dual space of $\cB_{\tau}$ w.r.t. the coupling $(\cdot,\cdot)$.
\end{proof}


Similarly to \eqref{rel-ips}, we define a map
\begin{align}\label{phi-t-def}
	\varphi^{\intercal}(A):=\rho_{\tau}^{1/2}A\rho_{\tau}^{1/2}.
\end{align}
By its definition, we have
\begin{align}\label{phi-t-unit}
	\langle\varphi^{\intercal}(A),\varphi^{\intercal}(B)\rangle_{\rm st,\tau}^{\intercal}&=\langle A,B\rangle_{\tau}.
\end{align}



\begin{remark}\label{rem:spaces}
	(a)	For each $r\in[0,1]$, the space $\cS_{\tau}^{(r)}$ and $\cS_{2}$ are in one-to-one correspondence.  Indeed, we define the map
	\begin{align}\label{weight}
		\pi^{(r)}:\cS_{2}\rightarrow\cS_{\tau}^{(r)}\quad\quad\pi^{(r)}(\kappa):= \rho_{\tau}^{r/2}\kappa\rho_{\tau}^{(1-r)/2}.
	\end{align}
	Obviously, the map $\pi^{(r)}$ is linear, bounded and invertible.  Moreover, for all $\lambda,\sigma\in\cS_{2}$, by \eqref{ip-st-r}, \eqref{weight} and the cyclicity of trace, we have
	\begin{align}\label{weight-uni}
		\langle\pi^{(r)}(\lambda),\pi^{(r)}(\sigma)\rangle_{{\rm st}, r}&=\langle\lambda,\sigma\rangle_{\cS_{2}}
	\end{align}
	so that $\pi^{(r)}$ is unitary.  Hence, $\cS_{\tau}^{(r)}$ is isomorphic to $\cS_{2}$.
	
	
	
	(b) Consider the family of maps $\varphi^{(r)}:\cB\rightarrow\cS_{\tau}^{(r)}$, $r\in [0,1]$, as $\varphi^{(r)}(A)=\rho_{\tau}^{r/2}A\rho_{\tau}^{1-r/2}$.  Since $\varphi^{(r)}$ maps $\cB$ into $\tilde{\cS}_{\tau}^{(r)}$ and, for each $A,B\in\cB$, 
	\begin{align}
		\langle\varphi^{(r)}(A),\varphi^{(r)}(B)\rangle_{{\rm st},r}&=\Tr((\rho_{\tau}^{1-r/2}A^{*}\rho_{\tau}^{r/2})\rho_{\tau}^{-r}(\rho_{\tau}^{r/2}B\rho_{\tau}^{1-r/2})\rho_{\tau}^{-1+r})\nonumber\\
		&=\Tr(A^{*}B\rho_{\tau})=\langle A,B\rangle_{\tau}.
	\end{align}
	By a density argument, we can extend $\varphi^{(r)}$ to a unitary map from $\cB_{\tau}$ to $\cS_{\tau}^{(r)}$, and we denote this extension also by $\varphi^{(r)}$.  Thus, $\cB_{\tau}$ and $\cS_{\tau}^{(r)}$ are isomorphic.
	
	
	Maps $\varphi$ and $\varphi^{\intercal}$ in Lemma \ref{lem:S*-dense} and in \eqref{phi-t-def} are the $r=0$ and $r=1$ cases of tNote that $\varphi^{(r)}$ is not $*$-map, unless $r=1$.
	
	
	
	(c) For any operator $K$ on $\cS_{1}$ and its dual $K'$ on $\cB$, we have
	\begin{align}\label{ad-K'}
		K'&=*\varphi^{-1} K^{*}\varphi *,
	\end{align}
	where $*:A\mapsto A^{*}$ and the adjoint $K^{*}$ is taken w.r.t. \eqref{ip-st}.  Indeed, for any $A\in\cB$ and $\lambda\in\cS_{1}$, we have
	\begin{align}
		(*\varphi^{-1}K^{*}\varphi *A,\lambda)&=\Tr((*\varphi^{-1}(K^{*}(A^{*}\rho_{\tau})))\lambda)=\Tr(((K^{*}(A^{*}\rho_{\tau}))\rho_{\tau}^{-1})^{*}\lambda)\nonumber\\
		&=\Tr(\rho_{\tau}^{-1}(K^{*}(A^{*}\rho_{\tau}))^{*}\lambda)=\langle K^{*}(A^{*}\rho_{\tau}),\lambda\rangle_{\rm st,\tau}\nonumber\\
		&=\langle A^{*}\rho_{\tau},K\lambda\rangle_{\rm st,\tau}=\Tr(\rho_{\tau}A(K\lambda)\rho_{\tau}^{-1})\nonumber\\
		&=\Tr(A(K\lambda))=(A,K\lambda)=(K'A,\lambda),
	\end{align}
	which implies \eqref{ad-K'}.
\end{remark}

\begin{remark}
	(i) The property $G'\leq 0$ is related to the contractivity of $\beta_{t}'$.  Indeed, for $A(t):=\beta_{t}'(A)$, using the Leibniz rule and HL equation \eqref{HLeq}, we find
	\begin{align}
		\partial_{t}\|A(t)\|_{\tau}^{2}/2&=(\langle L'A(t),A\rangle_{\tau}+\langle A(t),L'A(t)\rangle_{\tau})/2\nonumber\\
		&=\Re\langle A(t),L'A(t)\rangle_{\tau}.
	\end{align}
	Since $L_{0}'$ is anti-self-adjoint on $\cB_{\tau}$, we have
	\begin{align}
		\Re\langle A(t),L_{0}'A(t)\rangle_{\tau}&=0.
	\end{align}
	This, the relation $L'=L_{0}'+G'$ and the self-adjointness of $G'$ imply
	\begin{align*}
		\Re\langle A(t),L'A(t)\rangle_{\tau}&=2\langle A(t),G'A(t)\rangle_{\tau},
	\end{align*}
	and therefore,
	\begin{align}\label{contr-HLeqn}
		\partial_{t}\|A(t)\|_{\tau}^{2}&=2\langle A(t),G'A(t)\rangle_{\tau}\leq 0.
	\end{align}
	
	
	
	(ii) The property $D_{G'}(A,A)\geq 0$ for all $A\in\cB_{\tau}$ is related to the complete positivity of $\beta_{t}'$.  Indeed, we can generalize the operator function \eqref{dissip-G'} for $L'$ on $\sD(L')$ as
	\begin{align}
		D_{L'}(A,B)&=L'(A^{*}B)-A^{*}L'(B)-(L'(A))^{*}B.
	\end{align}
	Indeed, since $L_{0}'$ is a derivation, the domain $\sD(L')=\sD(L_{0}')$ is a $^{*}$-subalgebra in $\cB$ so that $AB$ and $A^{*}\in\sD(L')$, provided that $A,B\in\sD(L')$.   By the definition of derivations, we can show immediately that, for all $A,B\in\sD(L')$,
	\begin{align}
		D_{L'}(A,B)&=D_{G'}(A,B).
	\end{align}
	Since $G'$ is bounded on $\cB_{\tau}$, we can extend $D_{L'}$ to entire $\cB_{\tau}$.  It then follows from \eqref{G'A-ip} that $D_{L'}(A,A)\geq 0$ for all $A\in\cB_{\tau}$.
	
	
	Now, we show the positivity of $D_{L'}(A,A)$ follows from the property of complete positivity of $\beta_{t}'$.  To show this, since $\beta_{t}'$ is completely positive and $\beta_{t}'(\one)=\one$, by Kraus' theorem, we have, for each $t\geq 0$,
	\begin{align}
		\beta_{t}'(A)&=\sum_{k\geq 1}V_{k}^{*}(t)AV_{k}(t),\quad\quad\sum_{k\geq 1}V_{k}^{*}(t)V_{k}(t)=\one
	\end{align}
	where $V_{k}(t)\in\cB$ for all $k$.  Then, by the operator inequality \eqref{gen-Schwarz-2}, we obtain, for each $t\geq 0$, 
	\begin{align}
		0&\leq \beta_{t}'(A^{*}A)-\beta_{t}'(A)^{*}\beta_{t}'(A).
	\end{align}
	If we take $A\in\sD(L')$ and differentiate at $t=0$, we obtain
	\begin{align}
		0&\leq L'(A^{*}A)-A^{*}L'(A)-(L'(A))^{*}A\nonumber\\
		&=D_{L'}(A,A)=D_{G'}(A,A).
	\end{align}
	
	
	In fact, it was shown in \cite{Lind3} that, in the case that $L'$ is bounded, the positivity of $D_{L'}(A,A)$ for all $A\in\cB_{\tau}$ is also sufficient for $L'$ to generate a (norm continuous) completely positive semigroup.
\end{remark}



\appendix

		\section{Classes of jump operators satisfying (QDB)}\label{sec:jump-oprs}
Let $\{\psi_{s}\}$ be an orthonormal basis of eigenvectors of $\rho_{\tau}$ with corresponding eigenvalue $\rho_{s}$ and let $P_{ss'}:=\ket{\psi_{s}}\bra{\psi_{s'}}$.  Suppose
\begin{enumerate}
	\item[$(A)$] $\rho_{r}\rho_{r'}^{-1}\neq \rho_{s}\rho_{s'}^{-1}$ for all $(r,r')\neq (s,s')$ unless $r=r'$ and $s=s'$.
	\item[$(B)$]  For each $j$, the jump operator $W_{j}$ is of the form
	\begin{align}\label{W-coeff}
		W_{j}&:=\sum_{r,r'}W_{rr'}^{(j)}P_{r'r}.
	\end{align}
\end{enumerate}
Here $W_{rr'}^{(j)}$ are the infinite matrices given by
\begin{align}\label{W-KA}
	W_{rr'}^{(j)}&=\sqrt{K_{rr'}}A_{rr'}^{(j)},
\end{align}
where $K_{rr'}$ is the matrix, with non-negative entries, given by 
\begin{align}\label{cocycle20}
	K_{rr'}&=\begin{cases}
		\text{arbitrary},\quad	&	\quad\text{for }r'\geq r,\\
		K_{r'r}(\rho_{r}\rho_{r'}^{-1}),&\quad\text{for }r'<r,
	\end{cases}
\end{align}
with the infinite matrices $A_{rr'}^{(j)}$ satisfying
\begin{align}\label{A-symm}
	\overline{A_{rr'}^{(j)}}&=A_{r'r}^{(j)},\quad\quad\sum_{j\geq 1}|A_{rr'}^{(j)}|^{2}=1,
\end{align}
and
\begin{align}\label{A-vanish}
	\sum_{j\geq 1}\overline{A_{rr'}^{(j)}}A_{ss'}^{(j)}=0\quad\forall (r,r')\neq (s,s')\text{ unless }r=r'\text{ and }s=s'.
\end{align}


\begin{prop}\label{prop:G'W-qdbc}
	Assume Conditions $(A)$--$(B)$ and let $W_{j}$'s be given in \eqref{W-coeff}--\eqref{A-vanish}.  Then $G'$, defined by
	\begin{align}\label{G'(A)}
		G'(A)&:=\sum_{j\geq 1}\big(W_{j}^{*}AW_{j}-\frac{1}{2}\{W_{j}^{*}W_{j},A\}\big),
	\end{align}
	satisfies $(QDB)$.
\end{prop}


We give an example of $W_{j}$'s satisfying \eqref{W-coeff}--\eqref{A-vanish}:
\begin{align}\label{W-exp1}
	W_{j}&=\sum_{r,r'}\rho_{r'}\rho_{r}^{-1}A_{rr'}^{(j)}P_{r'r},
\end{align}
where  $A_{rr'}^{(j)}$, $j\geq 1$, are infinite matrices defined as
\begin{align}\label{sol-A}
	A_{rr'}^{(j)}&=f_{\omega_{rr'}}(j),\quad\quad f_{\omega_{rr'}}(j)\equiv f_{\varphi(\omega_{rr'})}(j),
\end{align}
where $\omega_{rr'}:=\log(\rho_{r}\rho_{r'}^{-1})$ for each $r,r'$, $\{f_{k}\}_{k\in \Z}$ is an orthonormal basis for $\ell^{2}(\mathbb{N})$ such that $\overline{f_{k}}=f_{-k}$ and $q:\{\omega_{rr'}\}_{r,r'}\rightarrow\Z$ is an injective function such that $q(-\omega)=-q(\omega)$. 


\begin{lemma}\label{lem:A-exp-verify}
	Assume Condition $(A)$.  Then the matrices $A_{rr'}^{(j)}$ defined in \eqref{sol-A} satisfy the relations \eqref{A-symm}--\eqref{A-vanish}.
\end{lemma}
\begin{proof}[Proof of Lemma \ref{lem:A-exp-verify}]
	First, by the definition \eqref{sol-A}, we have
	\begin{align}
		\overline{A_{rr'}^{(j)}}&=\overline{f_{\omega_{rr'}}(j)}=f_{-\omega_{rr'}}(j)=f_{\omega_{r'r}}(j)=A_{r'r}^{(j)},
	\end{align}
	and
	\begin{align}
		\sum_{j\geq 1}|A_{rr'}^{(j)}|^{2}&=\sum_{j\geq 1}|f_{\omega_{rr'}}(j)|^{2}=\|f_{\omega_{rr'}}\|_{\ell^{2}}^{2}=1.
	\end{align}
	Thus, the relations in \eqref{A-symm} are satisfied.  
	
	
	Next, for \eqref{A-vanish}, we compute
	\begin{align}\label{A-vanish-exp}
		\sum_{j\geq 1}\overline{A_{rr'}^{(j)}}A_{ss'}^{(j)}&=\sum_{j\geq 1}\overline{f_{\omega_{rr'}}(j)}f_{\omega_{ss'}}(j)=\langle f_{\omega_{rr'}},f_{\omega_{ss'}}\rangle_{\ell^{2}}=\delta_{\omega_{rr'},\omega_{ss'}}.
	\end{align}
	Under Condition $(A)$, if $r\neq r'$ or $s\neq s'$, then we must have $\omega_{rr'}\neq \omega_{ss'}$ for all $(r,r')\neq (s,s')$ so that, for such $r,r',s,s'$, the r.h.s. of \eqref{A-vanish-exp} vanishes.  Hence, \eqref{A-vanish} is satisfied.  Therefore, \eqref{sol-A} satisfies \eqref{A-symm}--\eqref{A-vanish}.
\end{proof}


\begin{proof}[Proof of Proposition \ref{prop:G'W-qdbc}]
	Recall $\Phi'(A):=\sum_{j\geq 1}W_{j}^{*}AW_{j}$ from \eqref{Phi'} and define
	\begin{align}
		K_{rr',ss'}&:=\Tr(P_{sr}\Phi'(P_{r's'}))=\sum_{j\geq 1}\langle\psi_{r},W_{j}^{*}\psi_{r'}\rangle\langle\psi_{s'},W_{j}\psi_{s}\rangle.
	\end{align}
	Let $W_{j}$'s be given by \eqref{W-coeff}--\eqref{A-vanish}.  Then 
	\begin{align}\label{K-coeff}
		K_{rr',ss'}&=\sum_{j\geq 1}\overline{W_{rr'}^{(j)}}W_{ss'}^{(j)}=\sqrt{K_{rr'}K_{ss'}}\sum_{j\geq 1}\overline{A_{rr'}^{(j)}}A_{ss'}^{(j)},
	\end{align}
	where $K_{rr'}:=K_{rr',rr'}$.  Eq. \eqref{K-coeff} implies 
	\begin{align}\label{K-coeff2}
		K_{ss'}&=\sum_{j\geq 1}|\langle\psi_{s},W_{j}\psi_{s'}\rangle|^{2}=\sum_{j\geq 1}|W_{rr'}^{(j)}|^{2}.
	\end{align}
	By construction \eqref{W-coeff}--\eqref{A-vanish}, $K_{rr'}$ satisfies
	\begin{align}\label{K-cocycle5}
		K_{rr'}\rho_{r}&=K_{r'r}\rho_{r'}.
	\end{align}
	

	\begin{lemma}[\cite{FGKV}, p.100]\label{lem:K-vanish}
		Assume Conditions $(A)$--$(B)$ and let $W_{j}$'s be defined by \eqref{W-coeff}--\eqref{sol-A}.  Then we have
		\begin{align}\label{K-vanish}
			K_{rr',ss'}=0=K_{s's,r'r},  \quad\forall(r,r')\neq (s,s') \text{ unless } r=r' \text{ and }s=s'.  
		\end{align}
	\end{lemma}
	\begin{proof}[Proof of Lemma \ref{lem:K-vanish}]
		Recall the definition of $K_{rr',ss'}$ from \eqref{K-coeff}.  By \eqref{A-vanish}, we have
		\begin{align}
			K_{rr',ss'}&=\sqrt{K_{rr'}K_{ss'}}\sum_{j\geq 1}\overline{A_{rr'}^{(j)}}A_{ss'}^{(j)}=0
		\end{align}
		for all $(r,r')\neq (s,s')$ unless $r=r'$ and $s=s'$.  Similarly, by \eqref{A-symm}--\eqref{A-vanish}, we have
		\begin{align}
			K_{s's,r'r}&=\sqrt{K_{s's}K_{r'r}}\sum_{j\geq 1}\overline{A_{s's}^{(j)}}A_{r'r}^{(j)}=\sqrt{K_{s's}K_{r'r}}\sum_{j\geq 1}\overline{A_{rr'}^{(j)}}A_{ss'}^{(j)}=0
		\end{align}
		for all $(r,r')\neq (s,s')$ unless $r=r'$ and $s=s'$.   This completes the proof.
	\end{proof}
	
	
	By Lemma \ref{lem:K-vanish}, we can write
	\begin{align}\label{Krr'-coeff}
		K_{rr',ss'}=K_{rr'}\delta_{(rr'),(ss')}(1-\delta_{rr'}\delta_{ss'})+K_{rr,ss}\delta_{rr'}\delta_{ss'}.
	\end{align}
	
	
	Let
	\begin{align}\label{C-coeff-def1}
		C_{rr',ss'}&:=\Tr(P_{sr}G'(P_{r's'})).
	\end{align}
	
	
	\begin{lemma}[\cite{FGKV}, p. 99]\label{lem:G'-qdbc}
		$G'$ satisfies $(QDB)$ if and only if $C_{rr',ss'}$
		\begin{align}\label{Crr'-coeff-qdb}
			C_{rr',ss'}\rho_{s}&=C_{s's,r'r}\rho_{s'}\quad(\text{equivalently, } C_{rr',ss'}\rho_{r}=C_{s's,r'r}\rho_{r'}.)
		\end{align}
	\end{lemma}
	\begin{proof}[Proof of Lemma \ref{lem:G'-qdbc}]
		Suppose $G'$ satisfies $(QDB)$.  Then, for all $r,r',s,s'$,
		\begin{align}\label{C-cocycle-comp1}
			\rho_{s}C_{rr',ss'}&=\rho_{s}\Tr(P_{sr}G'(P_{r's'}))=\Tr(\rho_{\tau}P_{sr}G'(P_{r's'}))\nonumber\\
			&=\Tr(P_{sr}G(P_{r's'}\rho_{\tau}))=\Tr(G'(P_{sr})P_{r's'}\rho_{\tau})\nonumber\\
			&=\rho_{s'}\Tr(P_{r's'}G'(P_{sr}))=\rho_{s'}C_{s's,r'r}.
		\end{align}
		
		
		Conversely, suppose \eqref{Crr'-coeff-qdb} holds for all $r,r',s,s'$.  We define the subspace
		\begin{align}
			\cF&:=\{A=\sum_{r,s=1}^{N}A_{rs}P_{rs}\mid N\in\mathbb{N},A_{rs}\in\C,\sup_{r,s}|A_{rs}|<\infty\}
		\end{align}
		of $\cB(\cH)$.  Since $\{\psi_{r}\}$ is an orthonormal basis for $\cH$, $\cF$ is weakly$^{*}$ dense, i.e., for every density operator $\rho$ on $\cH$ and $A\in\cB$,  there exists some sequence $\{A_{n}\}_{n\geq 1}$ in $\cF$ such that 
		\begin{align}
			\Tr((A-A_{n})\rho)\rightarrow 0.
		\end{align}
		The fact that $\cF$ is weakly$^{*}$ dense is due to the fact that $\cF$ is convex and $\cF\cap \cB_{1}$ is dense in $\cB_{1}$ the weaky operator topology and Theorem 2.4.7 of \cite{BrRo1}, where $\cB_{1}$ is the unit sphere of $\cB$.
		Then, for each $A,B\in\cF$, we can write 
		\begin{align}
			A&=\sum_{r,s=1}^{N}A_{rs}P_{rs},\quad B=\sum_{r',s'=1}^{M}B_{r's'}P_{r's'},
		\end{align}
		so that 
		\begin{align}\label{G'-F(H)}
			\langle A,G'(B)\rangle_{\rm obs,\tau}&=\Tr(A^{*}G'(B)\rho_{\tau})=\sum_{r,s=1}^{N}\sum_{r',s'=1}^{M}\overline{A_{rs}}B_{r's'}\Tr(P_{sr}G'(P_{r's'})\rho_{\tau})\nonumber\\
			&=\sum_{r,s=1}^{N}\sum_{r',s'=1}^{M}\overline{A_{rs}}B_{r's'}\rho_{s}\Tr(P_{sr}G'(P_{r's'}))\nonumber\\
			&=\sum_{r,s=1}^{N}\sum_{r',s'=1}^{M}\overline{A_{rs}}B_{r's'}\rho_{s}C_{rr',ss'}=\sum_{r,s=1}^{N}\sum_{r',s'=1}^{M}\overline{A_{rs}}B_{r's'}\rho_{s'}C_{s's,r'r}\nonumber\\
			&=\sum_{r,s=1}^{N}\sum_{r',s'=1}^{M}\overline{A_{rs}}B_{r's'}\rho_{s'}\Tr(G'(P_{sr})P_{r's'})\nonumber\\
			&=\sum_{r,s=1}^{N}\sum_{r',s'=1}^{M}\overline{A_{rs}}B_{r's'}\Tr(G'(P_{sr})P_{r's'}\rho_{\tau})\nonumber\\
			&=\Tr((G'(A))B\rho_{\tau})=\langle G'(A),B\rangle_{\rm obs,\tau}.
		\end{align}
		
		
		Now, since $G'$ is continuous in the weak$^{*}$ topology of $\cB$, we have $G'(A-A_{n})\rightarrow 0$ if $A_{n}\rightarrow A$ in the weak$^{*}$ topology.  Thus, for each $A,B\in\cB$, we choose sequences $\{A_{n}\}_{n\geq 1},\{B_{m}\}_{m\geq 1}$ in $\cF$ such that $A_{n}\rightarrow A$ and $B_{m}\rightarrow B$ weakly$^{*}$, by \eqref{G'-F(H)}, we have, as $n,m\rightarrow\infty$,
		\begin{align}
			|\langle A,&G'(B)\rangle_{\rm obs,\tau}-\langle G'(A),B\rangle_{\rm obs,\tau}|\nonumber\\
			&=|\langle A_{n},G'(B_{m})\rangle_{\rm obs,\tau}-\langle G'(A),B\rangle_{\rm obs,\tau}|+|\langle A-A_{n},G'(B)\rangle_{\rm obs,\tau}|\nonumber\\
			&\quad\quad\quad\quad\quad\quad\quad\quad\quad\quad\quad+\langle A_{n},G'(B-B_{m})\rangle_{\rm obs,\tau}\nonumber\\
			&=|\langle G'(A_{n}),B_{m}\rangle_{\rm obs,\tau}-\langle G'(A),B\rangle_{\rm obs,\tau}|+|\langle A-A_{n},G'(B)\rangle_{\rm obs,\tau}|\nonumber\\
			&\quad\quad\quad\quad\quad\quad\quad\quad\quad\quad\quad+|\langle A_{n},G'(B-B_{m})\rangle_{\rm obs,\tau}|\rightarrow 0.
		\end{align}
		Thus, for all $A,B\in\cB$, we have $\langle A,G'(B)\rangle_{\rm obs,\tau}=\langle G'(A),B\rangle_{\rm obs,\tau}$, which implies that $G'$ satisfies $(QDB)$.
	\end{proof}
	
	
	\begin{lemma}\label{lem:C-W1}
		The coefficients $C_{rr',ss'}$ and $K_{rr',ss'}$ in \eqref{C-coeff-def1} and \eqref{K-coeff} are related as
		\begin{align}\label{Crr-simp}
			C_{rr',ss'}&=K_{rr',ss'}-\frac{1}{2}\delta_{rr'}\delta_{ss'}\sum_{\ell}\big(K_{r\ell}+K_{s\ell}\big).
		\end{align}
	\end{lemma}
	\begin{proof}[Proof of Lemma \ref{lem:C-W1}]
		First, by \eqref{K-coeff}, we compute
		\begin{align}\label{Phi'(1)}
			\Phi'(\one)&=\sum_{j\geq 1}W_{j}^{*}W_{j}=\sum_{r,r',s,s'}\overline{W_{rr'}^{(j)}}W_{ss'}^{(j)}P_{rr'}P_{s's}=\sum_{r,r',s,s'}\overline{W_{rr'}^{(j)}}W_{ss'}^{(j)}\delta_{r's'}P_{rs}\nonumber\\
			&=\sum_{r,s,s'}\overline{W_{rs'}^{(j)}}W_{ss'}^{(j)}P_{rs}=\sum_{r,s,s'}K_{rs',ss'}P_{rs}.
		\end{align}
		Thus, we have
		\begin{align}\label{anti-comu-coeff1}
			\Tr(P_{sr}\{\Phi'(\one),P_{r's'}\})&=\sum_{k,\ell,\ell'}K_{k\ell',\ell\ell'}\Tr\big(P_{sr}(P_{k\ell}P_{r's'}+P_{r's'}P_{k\ell})\big)\nonumber\\
			&=\sum_{k,\ell,\ell'}K_{k\ell',\ell\ell'}\delta_{\ell r'}\Tr(P_{sr}P_{ks'})+\sum_{k,\ell,\ell'}K_{k\ell',\ell\ell'}\delta_{s' k}\Tr(P_{sr}P_{r'\ell})\nonumber\\
			&=\sum_{k,\ell,\ell'}K_{k\ell',\ell\ell'}\delta_{\ell r'}\delta_{rk}\delta_{ss'}+\sum_{k,\ell,\ell'}K_{k\ell',\ell\ell'}\delta_{s'k}\delta_{rr'}\delta_{s\ell}\nonumber\\
			&=\delta_{ss'}\sum_{\ell'}K_{r\ell',r'\ell'}+\delta_{rr'}\sum_{\ell'}K_{s'\ell',s\ell'}.
		\end{align}
		
		
		Recall the definition of $C_{rr',ss'}$ from \eqref{C-coeff-def1}.  Then, by \eqref{anti-comu-coeff1}, we have
		\begin{align}\label{C-coeff-exp1}
			C_{rr',ss'}&=\Tr(P_{sr}G'(P_{r's'}))=\Tr(P_{sr}\Phi'(P_{r's'}))-\frac{1}{2}\Tr(P_{sr}\{\Phi'(\one),P_{r's'}\})\nonumber\\
			&=K_{rr',ss'}-\frac{1}{2}\big(\delta_{ss'}\sum_{\ell}K_{r\ell,r'\ell}+\delta_{rr'}\sum_{\ell}K_{s'\ell,s\ell}\big).
		\end{align}
		By \eqref{Krr'-coeff}, we have, for each $r,r',\ell$, 
		\begin{align}\label{Krr-simp}
			K_{r\ell,r'\ell}&=K_{r\ell}\delta_{(r\ell),(r'\ell)}(1-\delta_{r\ell}\delta_{r'\ell})+K_{rr,r'r'}\delta_{rr'}\nonumber\\
			&=K_{r\ell}\delta_{rr'}(1-\delta_{r\ell})+K_{rr}\delta_{rr'}=K_{r\ell}\delta_{rr'}.
		\end{align}
		Substituting \eqref{Krr-simp} into \eqref{C-coeff-exp1} yields \eqref{Crr-simp}.
	\end{proof}
	
	
	\begin{lemma}\label{lem:C-W-re}
		The coefficients $C_{rr',ss'}$ in \eqref{Crr-simp} satisfy \eqref{Crr'-coeff-qdb} if and only if, for all $r,r',s$,
		\begin{align}\label{K-cocycle-symm}
			K_{rr'}\rho_{r}&=K_{r'r}\rho_{r'},\quad\quad K_{rr,ss}=K_{ss,rr}.
		\end{align}
	\end{lemma}
	\begin{proof}[Proof of Lemma \ref{lem:C-W-re}]
		By \eqref{Crr-simp}, \eqref{Crr'-coeff-qdb} holds for each $r,r',s,s'$ if and only if 
		\begin{align}
			K_{rr',ss'}\rho_{s}&-\frac{1}{2}\rho_{s}\delta_{rr'}\delta_{ss'}\sum_{\ell}(K_{r\ell}+K_{s\ell})\nonumber\\
			&=K_{s's,r'r}\rho_{s'}-\frac{1}{2}\rho_{s'}\delta_{rr'}\delta_{ss'}\sum_{\ell}(K_{r'\ell}+K_{s'\ell}).
		\end{align}
		Then, due to the presence of $\delta$-functions, we have
		\begin{align}
			\rho_{s}\delta_{rr'}\delta_{ss'}\sum_{\ell}(K_{r\ell}+K_{s\ell})&=\rho_{s'}\delta_{rr'}\delta_{ss'}\sum_{\ell}(K_{r'\ell}+K_{s'\ell}).
		\end{align}
		Thus, the above relations, together with \eqref{Krr'-coeff}, imply that \eqref{Crr'-coeff-qdb} holds if and only if 
		\begin{align}\label{CK-rel}
			\rho_{r}K_{rr'}&\delta_{(rr'),(ss')}(1-\delta_{rr'}\delta_{ss'})+\rho_{r}K_{rr,ss}\delta_{rr'}\delta_{ss'}\nonumber\\
			&=\rho_{r'}K_{r'r}\delta_{(rr'),(ss')}(1-\delta_{rr'}\delta_{ss'})+\rho_{r'}K_{ss,rr}\delta_{rr'}\delta_{ss'}.
		\end{align}
		If $s\neq s'$, then \eqref{CK-rel} becomes
		\begin{align}
			\rho_{r}K_{rr'}\delta_{(rr'),(ss')}&=\rho_{r'}K_{r'r}\delta_{(rr'),(ss')}\quad\Leftrightarrow\quad \rho_{r}K_{rr'}=\rho_{r'}K_{r'r}.
		\end{align}
		If $s=s'$ and $r=r'$, then \eqref{CK-rel} becomes, since $\rho_{r}>0$ for all $r$,
		\begin{align}
			\rho_{r}K_{rr,ss}=\rho_{r}K_{ss,rr}\quad\Leftrightarrow\quad K_{rr,ss}=K_{ss,rr}.
		\end{align}
		Hence, if $C_{rr',ss'}$ in \eqref{Crr-simp} satisfies \eqref{Crr'-coeff-qdb}, then \eqref{K-cocycle-symm} holds.
		

		Conversely, suppose \eqref{K-cocycle-symm} holds.  To show $C_{rr',ss'}$ in \eqref{Crr-simp} satisfies \eqref{Crr'-coeff-qdb}, it suffices to show \eqref{CK-rel} holds.  Since $K_{rr,ss}=K_{ss,rr}$, due to the presence of $\delta$-functions, we have
		\begin{align}
			\rho_{r}K_{rr,ss}\delta_{rr'}\delta_{ss'}&=\rho_{r'}K_{ss,rr}\delta_{rr'}\delta_{ss'}.
		\end{align}  
		Similarly, since $\rho_{r}K_{rr'}=\rho_{r'}K_{r'r}$, we have
		\begin{align}
			\rho_{r}K_{rr'}\delta_{rr'}\delta_{(rr'),(ss')}(1-\delta_{rr'}\delta_{ss'})&=\rho_{r'}K_{r'r}\delta_{(rr'),(ss')}(1-\delta_{rr'}\delta_{ss'}).
		\end{align}
		The last two relations imply that \eqref{CK-rel} holds, which further implies that $C_{rr',ss'}$ in \eqref{K-cocycle-symm} satisfies \eqref{Crr'-coeff-qdb}.
	\end{proof}
	

	
	
	\begin{cor}\label{cor:G'-qdbc-K}
		$G'$ satisfies $(QDB)$ if and only if the coefficients $K_{rr',ss'}$, defined in \eqref{K-coeff}, satisfy \eqref{K-cocycle-symm}.
	\end{cor}
	
	
	
	Now, we finish the proof of Proposition \ref{prop:G'W-qdbc}.  We distinguish the objects constructed in  \eqref{W-coeff}--\eqref{cocycle20} by $\sim$.  Then, by the construction \eqref{W-coeff}--\eqref{cocycle20} and definition \eqref{K-coeff2}, we have
	\begin{align}
		\sum_{j\geq 1}|\widetilde{W}_{rr'}^{(j)}|^{2}=\widetilde{K}_{rr'}\sum_{j\geq 1}|A_{rr'}^{(j)}|^{2}=\widetilde{K}_{rr'},
	\end{align}
	where the coefficients $\widetilde{K}_{rr'}$ are given by \eqref{cocycle20}.  By the construction, the coefficients $\widetilde{K}_{rr'}=\widetilde{K}_{rr',rr'}$ and $K_{rr,ss}$, given in \eqref{K-coeff}, satisfy \eqref{K-cocycle-symm}.  Hence, by Lemmas \ref{lem:G'-qdbc} and \ref{lem:C-W-re}, $G'$, given in \eqref{G'(A)}, satisfies $(QDB)$.
	%
	This completes the proof of Proposition \ref{prop:G'W-qdbc}.
\end{proof}



\section{Proof of Theorems \ref{thm:exist-uniq-S}} 
 \label{sec:pf-exist}

We will use the following abstract result (see \cite{BrRo1} Theorem 3.1.33):
\begin{thm}\label{thm:exist-evol}
	Let $X$ be a Banach space and let $V_{t}$ be a $C_{0}$- (resp. $C_{0}^{*}$-) semigroup on $X$ with generator $S$ and let $P$ be some bounded operator on $X$.  Then $S+P$ generates a $C_{0}$- (resp. $C_{0}^{*}$-) continuous semigroup $V_{t}^{P}$ on $X$.
\end{thm}


We begin with the next proposition:
\begin{prop}\label{prop:G-well-def}
	Under Condition $(W)$, the sum on the r.h.s. of \eqref{G} converges absolutely in $\cS_{1}$ and defines a bounded operator $G$ on the space $\cS_{1}$.
\end{prop}
\begin{proof}
	Clearly, $W_{j}\rho W_{j}^{*}\geq 0$ for $\rho\geq 0$.  Hence, by the triangle inequality, we have
	\begin{align}\label{sum-Wjrho}
		\big\|\sum_{j}W_{j}\rho W_{j}^{*}\big\|_{\cS_{1}}&\leq \sum_{j}\big\|W_{j}\rho W_{j}^{*}\big\|_{\cS_{1}}=\sum_{j}\Tr(W_{j}\rho W_{j}^{*})\nonumber\\
		&=\sum_{j}\Tr(W_{j}^{*}W_{j}\rho)
	\end{align}
	Let $Y:=\sum_{j\geq 1}W_{j}^{*}W_{j}$.  By standing assumption $(W)$, $Y$ is a bounded operator.  We show in Lemma \ref{lem:Wj-con} below that 
	\begin{align}
		\sum_{j}\Tr(W_{j}^{*}W_{j}\rho)&=\Tr(\sum_{j}W_{j}^{*}W_{j}\rho)=\Tr(Y\rho), 
	\end{align}
	which, together with \eqref{sum-Wjrho} and the fact that $Y$ is bounded, implies that
	\begin{align}
		\big\|\sum_{j}W_{j}\rho W_{j}^{*}\big\|\leq \left\|Y\right\|\|\rho\|_{\cS_{1}}<\infty.
	\end{align}
	Furthermore, since $Y$ is bounded, we have
	\begin{align}\label{lr-Y-multi}
		\left\|\{Y,\rho\}\right\|_{\cS_{1}}&\leq \left\|Y\rho\right\|_{\cS_{1}}+\left\|\rho Y\right\|_{\cS_{1}}\leq 2\|Y\|\|\rho\|_{\cS_{1}}<\infty.
	\end{align}
	Hence, the sum on the r.h.s. converges absolutely in $\cS_{1}^{+}$ and defines a bounded operator $G$ on $\cS_{1}^{+}$ with the bound
	\begin{align}
		\|G(\rho)\|_{\cS_{1}}\leq 2\|Y\|\|\rho\|_{\cS_{1}}.
	\end{align}
	This operator extends to $\cS_{1}$ by linearity.
\end{proof}


\begin{lemma}\label{lem:Wj-con}
	Under Condition $(W)$, we have, for all $\rho\in\cS_{1}$,
	\begin{align}\label{series-converg}
		\sum_{j}\Tr(W_{j}^{*}W_{j}\rho)=\Tr\big(\sum_{j}W_{j}^{*}W_{j}\rho\big).
	\end{align}
\end{lemma}
\begin{proof}
	Let $Y:=\sum_{j}W_{j}^{*}W_{j}$,  $\rho\in\cS_{1}^{+}$ and $\{\rho_{N}\}$ be a monotonically increasing sequence of rank-$N$ operators and that $\|\rho_{N}-\rho\|_{\cS_{1}}\rightarrow 0$ as $N\rightarrow\infty$.  Then we have
	\begin{align}
		\Tr(Y\rho)&=\lim_{N\rightarrow\infty}\Tr(Y\rho_{N}).
	\end{align}
	Let $\rho_{N}=\sum_{k=1}^{N}\lambda_{k}\ket{\psi_{k}}\bra{\psi_{k}}$, with $\{\psi_{k}\}$ orthonormal and $\lambda_{k}\geq 0$.  Since, for each $N$, $\Tr(Y\rho_{N})=\sum_{k=1}^{N}\lambda_{k}\langle\psi_{k},Y\psi_{k}\rangle$, we have, by Condition $(W)$, that
	\begin{align}\label{YrhoN}
		\Tr(Y\rho_{N})&=\lim_{M\rightarrow\infty}\sum_{j=1}^{M}\Tr(W_{j}^{*}W_{j}\rho_{N})=\lim_{M\rightarrow\infty}\sum_{j=1}^{M}\Tr(W_{j}\rho_{N}W_{j}^{*}).
	\end{align}
	Since $\Tr(W_{j}\rho_{N}W_{j}^{*})\geq 0$ and $\sum_{j=1}^{M}\Tr(W_{j}\rho_{N}W_{j}^{*})\leq \Tr(Y\rho_{N})$, the series on the r.h.s. of \eqref{YrhoN} converges and we have
	\begin{align}
		\Tr(Y\rho)=\sum_{j=1}^{\infty}\Tr(W_{j}\rho_{N}W_{j}^{*})&=\sum_{j=1}^{\infty}a_{j,N},
	\end{align}
	where $a_{j,N}:=\Tr(W_{j}\rho_{N}W_{j}^{*})$.  For all $j$ and $N$, $a_{j,N}$ are non-negative, monotonically increasing in $N$ and satisfy $\sum_{j=1}^{\infty}a_{j,N}=\Tr(Y\rho_{N})\leq \Tr(Y\rho)$.  Hence, by monotone convergence theorem for non-negative sums, we have
	\begin{align}
		\Tr(Y\rho)&=\lim_{N\rightarrow\infty}\sum_{j=1}^{\infty}a_{j,N}=\sum_{j=1}^{\infty}\lim_{N\rightarrow\infty}a_{j,N}\nonumber\\
		&=\sum_{j=1}^{\infty}\Tr(W_{j}\rho W_{j}^{*})=\sum_{j=1}^{\infty}\Tr(W_{j}^{*}W_{j}\rho).
	\end{align}
	This completes the proof.
\end{proof}


\begin{proof}[Proof of Theorem \ref{thm:exist-uniq-S}]
	We begin with the following lemma:
	\begin{lemma}\label{lem:S-str-con}
		Let $H$ be a self-adjoint operator on $\cH$.  Then the group $e^{L_{0}t}\rho:=e^{-iHt}\rho e^{iHt}$ is strongly continuous in $t$ on $\cS_{1}$.
	\end{lemma}
	\begin{proof}
		If $\rho=\sum_{j=1}^{\infty}\rho_{k}\ket{\psi_{k}}\bra{\psi_{k}}$ with $\|\psi_{k}\|=1$, $\rho_{k}\geq 0$ for all $k$ and $\sum_{k}\rho_{k}<\infty$, then, as $t\rightarrow 0$,
		\begin{align}\label{str-S1}
			\|e^{L_{0}t}\rho-\rho\|_{\cS_{1}}&\leq \|e^{-iHt}\rho-\rho\|_{\cS_{1}}+\|\rho-\rho e^{iHt}\|_{\cS_{1}}\nonumber\\
			&\leq 2\sum_{j=1}^{\infty}\|\psi_{k}\|\|e^{-iHt}\psi_{k}-\psi_{k}\|.
		\end{align}
		Since $H$ is self-adjoint, the group $e^{-iHt}$ is strongly continuous on $\cH$ by the Stone's theorem, so that the r.h.s. of \eqref{str-S1} converges to zero as $t\rightarrow 0$, by the dominated convergence theorem.  Thus, $e^{L_{0}t}$ is strongly continuous on $\cS_{1}$.
	\end{proof}
	
	
	Since $L_{0}$ generates a one-parameter, strongly continuous group of bounded operators on $\cS_{1}$, given explicitly by $e^{L_{0}t}\rho=e^{-iHt}\rho e^{iHt}$, and, by Proposition \ref{prop:G-well-def}, $G$ is bounded on $\cS_{1}$, it follows from Theorem \ref{thm:exist-evol}, with $X=\cS_{1}$, that $L=L_{0}+G$ generates a one-parameter, strongly continuous semigroup on $\cS_{1}$.  
%
%
%
%
\end{proof}

\begin{lemma}\label{lem:pos-vNL-evol}
	The semigroup $\beta_{t}$ is completely positive on $\cS_{1}$.
\end{lemma}
\begin{proof}
	The argument follows from Theorem 5.2 in \cite{Davies}.  For this,  we rewrite the vNL operator as
	\begin{align}
		L(\rho)&=-i[H,\rho]+\sum_{j\geq 1}(W_j\rho W_j^*-\frac{1}{2}\{W_j^*W_j,\rho\})\nonumber\\
		&=[-iH-Y,\rho]+\Phi(\rho),
	\end{align}
	where $Y=Y^*=\tfrac{1}{2}\sum\nolimits_{j\geq 1}W_j^*W_j$ and $\Phi(\rho)=\sum\nolimits_{j\geq 1}W_j\rho W_j^*$.  
	
	
	Let $B_t:=e^{-iHt-Y t}$, which is well-defined since $Y$ is bounded.  Then, the semigroup $S_t$ generated by $[-iH-Y,\cdot]$ is given by
	\begin{align}
		e^{L_{0}t}(\rho)&=B_t\rho B_t^{*},
	\end{align}
	which defines a completely positive semigroup by Kraus' theorem.  On the other hand, since, by Lemma \ref{prop:G-well-def}, the map $\Psi$ is bounded on $\cS_{1}$, $\Phi$ generates a semigroup on $\cS_{1}$
	\begin{align}
		e^{\Phi t}&:=\sum_{k=0}^{\infty}\frac{t^{k}}{k!}\Phi^{k}.
	\end{align}
	We note that, since $\Phi$ is completely positive by \cite{Kr2}, Theorem 3.3, so are $\Phi^{k}$ for any $k=0,1,2,...$ and any linear combination of them.  Therefore, $e^{\Phi t}$ is a completely positive semigroup.
	
	
	Then, it follows that, for any $n\in\mathbb{N}$, the operator $(T_{t/n}e^{\Psi t/n})^{n}$ is a completely positive semigroup on $\cS_{1}$.
	
	
	Finally, by Trotter-Lie formula, we have
	\begin{align}
		\beta_t(\rho)&=\exp(L_{0}+G)(\rho)=\lim_{n\rightarrow\infty}(T_{t/n}e^{\Psi t/n})^{n}(\rho),
	\end{align}
	where the limit is taken in the trace norm, so that the semigroup $\beta_t$ is completely positive.
\end{proof}
Note that the proof of Lemma \ref{lem:pos-vNL-evol} provides a different construction for the semigroup $\beta_{t}$.

		

		\section{Some additional  results}\label{sec:main-1}

		Let $Q'$ is the orthogonal projector in $\cB_{\tau}$ onto $\Null(G')$ and $Q'^{\perp}=\one-Q'$.  
		\begin{thm}\label{thm:RTE-dual1'}
			Assume Conditions $(H)$, $(W)$ and $(QDB)$ hold.  Then, the dual quantum evolution $\beta_{t}'(A)$ converges to $\Null(G')$ in the ergodic sense:
			\begin{align}\label{RTE-dual2-1}
				s\text{-}\lim_{T\rightarrow\infty}\frac{1}{T}\int_{0}^{T}\beta_{t}'dt&=Q',
			\end{align}
			strongly in $\cB_{\tau}$.
		\end{thm}
		Theorem \ref{thm:RTE-dual1'} is proven in Subsection \ref{sec:pf-RTE-dual1'}.


		Recall that, by \eqref{G0eigeneq}, $0$ is an eigenvalue of $G$ with the eigenvector $\rho_{\tau}$.  The relation between the resolvents of $G'$ and $G$ shows that:
		
		
		If $0$ is an isolated eigenvalue of $G'$, then the operator $G$ also has an isolated eigenvalue $0$ (see the proof of Lemma \ref{lem:property-G-qdbc} (c)).

		%
		
		
		\begin{remark}
			In the case where $\rho_{\tau}$ is the unique stationary solution of $\beta_{t}$, the asymptotic convergence of the dynamic $\beta_{t}(\rho)$ w.r.t. the trace-norm immediately follows from Theorem \ref{thm:RTE} and the inequality:
			\begin{align}\label{tr-norm-st}
				\|\rho-\rho_{\tau}\|_{\cS_{1}}&\leq \|\rho-\rho_{\tau}\|_{\rm st,\tau}
			\end{align}
			for $\rho\in\cS_{\tau}$, see \eqref{tr-st*}.
		\end{remark}


		\begin{thm}\label{thm:RTE-dual2}
			Assume Conditions $(H)$, $(W)$ and $(QDB)$.  Suppose further that
			\begin{enumerate}
				\item[$(Null)$] $\Null(G')\subseteq\Null(L_{0}')$.
			\end{enumerate}
			(a) If, in addition, the following condition is satisfied
			\begin{enumerate}
				\item[$(Spec)$] $G'$ has no singular continuous spectrum in a neighborhood of $0$,
			\end{enumerate}
			then the dual quantum evolution $\beta_{t}'(A)$ converges to $\Null(G')$ in the sense that, for all $A\in\sD(L')$,
			\begin{align}\label{RTE3-dual}
				\|\beta_{t}'(A)-Q'A\|_{\tau}&\rightarrow 0\quad\quad\text{as }t\rightarrow \infty.
			\end{align}
			
			(b) If, instead of Condition $(Spec)$, we have
			\begin{enumerate}
				\item[$(Gap)$] the eigenvalue $0$ of $G'$ is isolated,
			\end{enumerate}
			then the dual quantum evolution $\beta_{t}'(A)$ converges to the subspace $\Null(G')$ exponentially fast: for all $A\in\sD(L')$, we have
			\begin{align}\label{RTE1-dual}
				\|\beta_{t}'(A)-Q'A\|_{\tau}&\leq e^{-\theta t}\|A\|_{\tau},
			\end{align} 
			where $\theta:=\dist(0,\sigma(G')\setminus\{0\})$.

			(c) If, in addition to Condition $(Gap)$, the eigenvalue $0$ of $G'$ is simple, then 
			\begin{align}\label{RTE2-dual}
				\|\beta_{t}'(A)-c^{A}\one\|_{\tau}\leq e^{-\theta t}\|A\|_{\tau},
			\end{align}
			where $c^{A}=\langle\one,A\rangle_{\tau}=\Tr(A\rho_{\tau})$.
		\end{thm}
		

	
		
		Let $Q$ be the orthogonal projector onto $\Null(G)$ in $\cS_{\tau}$.  We have
		\begin{thm}\label{thm:RTE}
			Suppose the Conditions $(H)$, $(W)$, 
			$(QDB)$, $(Null)$ and $(Gap)$ hold.  Then,
			\begin{enumerate}
				\item[(a)] the vNL evolution $\beta_{t}(\rho)$ converges exponentially fast to the subspace $\Null(G)$: 
				\begin{align}\label{RTE1}
					\|\beta_{t}(\rho)-Q(\rho)\|_{\rm st,\tau}\leq e^{-\theta t}\|\rho\|_{\rm st,\tau},
				\end{align}
				for all $\rho\in\cS_{\tau}$, where, recall,  $\theta=\dist(0,\sigma(G')\setminus\{0\})$.
				\item[(b)] if, in addition, $0$ is a simple eigenvalue of $G'$, then, for all density operator $\rho\in\sD(L)$,
				\begin{align}\label{RTE2}
					\|\beta_{t}(\rho)-\rho_{\tau}\|_{\rm st,\tau}\leq e^{-\theta t}\|\rho\|_{\rm st,\tau},
				\end{align}
				where $\rho_{\tau}$ is the eigenvector of $G$ corresponding to the eigenvalue $0$.
			\end{enumerate}
		\end{thm}
		Theorem \ref{thm:RTE} is proven in Subsection \ref{sec:main-pf}.


The following result presents a variant of Theorem \ref{thm:RTE} for the space $\cS_{\tau}^{\intercal}$.
\begin{thm}\label{thm:iso-EV}
	Assume Conditions $(H)$, $(W)$, $(QDB)$ and that $0$ is a simple, isolated eigenvalue of $G'$.  Then, for all density operator $\rho\in\cS_{\tau}^{\intercal}$ and for  $\theta=\dist(0,\sigma(G')\setminus\{0\})$, we have
	\begin{align}
		\|\beta_{t}(\rho)-\rho_{\tau}\|_{\rm st,\tau}^{\intercal}\leq e^{-\theta t}\|\rho\|_{\rm st,\tau}^{\intercal}.
	\end{align}
\end{thm}
This theorem is proven below.


		
		\subsection{Proof of Theorem \ref{thm:RTE-dual1'}}\label{sec:pf-RTE-dual1'}

		Throughout this subsection, we omit the subindex ``${\tau}$" in the inner product and norm in $\cB_{\tau}$.  One should not confuse this norm with the operator norm on $\cB$.
		

						Let $Q'$ be the orthogonal projector onto the subspace $\Null(G')$ in $\cB_{\tau}$ w.r.t. its inner product and let $Q'^{\perp}=\one-Q'$.  
		
		
		Let $G'^{\perp}:=G'Q'^{\perp}$.  First, we show that $\Ran(G'^{\perp})=\sD((G'^{\perp})^{-1})$ is dense in $\Ran Q^{\perp}=:\cB_{\tau}^{\perp}$.  Let $E(\lambda)$ be the spectral resolution of the self-adjoint operator $G'^{\perp}$.  The set 
		\begin{align}
			\sD&:=\{\int f(\lambda)dE(\lambda)A\mid A\in\cB_{\tau}^{\perp},f\in C(\R)\cap L^{\infty}(\R),\nonumber\\
			&\quad\quad\quad\quad\quad\quad\quad\quad\quad\text{ and }f=0\text{ in a vicinity of }0.\}
		\end{align}
		is dense in $\Ran Q'^{\perp}$ and $(G'^{\perp})^{-1}$ is defined on $\sD$ and is given by 
		\begin{align}
			(G'^{\perp})^{-1}\int f(\lambda)dE(\lambda)A&=\int \lambda^{-1}f(\lambda)dE(\lambda)A.
		\end{align}
		(By the condition of $f$, the operator on the r.h.s. is bounded in $\cB_{\tau}$ by $\sup_{\lambda}|\lambda^{-1}f(\lambda)|\|A\|$.)
		
		
		Next, by Theorem \ref{thm:Gibbs}, we have $\beta_{t}'=e^{L_{0}'t}e^{G't}$ for all $t\geq 0$.  Then, the Cauchy-Schwarz and the H\"older inequalities yield, since $L_{0}'$ is anti-self-adjoint, for any $A\in\sD(L')$ such that $A^{\perp}:=Q'^{\perp}A\in \sD((G'^{\perp})^{-1})$ and any $B\in\cB_{\tau}$,
		\begin{align}\label{ergodic-CSH}
			|\langle B,&\frac{1}{T}\int_{0}^{T}\beta_{t}'(Q'^{\perp}A)dt\rangle|\leq\frac{1}{T}\int_{0}^{T}|\langle B,e^{L_{0}'t}e^{G't}A^{\perp}\rangle_{*}| dt\nonumber\\
			&=\frac{1}{T}\int_{0}^{T}|\langle e^{-L_{0}'t}B,e^{G'^{\perp}t}A^{\perp}\rangle| dt\leq\frac{1}{T}\int_{0}^{T}\|e^{-L_{0}'t}B\|\|e^{G'^{\perp}t}A^{\perp}\| dt\nonumber\\
			&=\frac{\|B\|}{T}\int_{0}^{T}\|e^{G'^{\perp}t}A^{\perp}\|dt\leq \frac{\|B\|}{\sqrt{T}}\left(\int_{0}^{T}\|e^{G'^{\perp}t}A^{\perp}\|^{2}dt\right)^{1/2}.
		\end{align}
		
		
		Now, since $G'$ is self-adjoint on $\cB_{\tau}$, we have
		\begin{align}\label{ergodic-dual}
			\int_{0}^{T}\|e^{G'^{\perp}t}A^{\perp}\|^{2}dt&=\int_{0}^{T}\langle A^{\perp},e^{2G'^{\perp}t}A^{\perp}\rangle dt\nonumber\\
			&=\langle A^{\perp},(e^{2G'^{\perp}T}-\one)(2G'^{\perp})^{-1}A^{\perp}\rangle.
		\end{align}
		Since $G'\leq 0$, we have $\|e^{2G'^{\perp}T}\|\leq 1$ and therefore, by \eqref{ergodic-dual}, 
		\begin{align}\label{ergodic-dual2}
			\int_{0}^{T}\|e^{G'^{\perp}t}A^{\perp}\|^{2}dt&\leq \|(G'^{\perp})^{-1/2}A^{\perp}\|^{2}.
		\end{align}
		By \eqref{ergodic-CSH} and \eqref{ergodic-dual2}, we have, for such $A,B\in\cB_{\tau}$,
		\begin{align}
			|\langle B,\frac{1}{T}\int_{0}^{T}\beta_{t}'(Q^{\perp}A)dt\rangle|&\leq \frac{1}{\sqrt{T}}\|B\|\|(G'^{\perp})^{-1/2}A^{\perp}\|,
		\end{align}
		which implies that, for all $A\in\cB_{\tau}$ such that $Q'^{\perp}A\in\sD((G'^{\perp})^{-1/2})$,
		\begin{align}\label{ergodic-con}
			\|\frac{1}{T}\int_{0}^{T}&\beta_{t}'(A)dt-Q'A\|=\|\frac{1}{T}\int_{0}^{T}\beta_{t}'(Q^{\perp}A)dt\|\nonumber\\
			&\leq \frac{1}{\sqrt{T}}\|(G'^{\perp})^{-1/2}Q'^{\perp}A\|\rightarrow 0
		\end{align}
		as $T\rightarrow\infty$.  Since $\sD((G'^{\perp})^{-1})$ is dense in $\cB_{\tau}^{\perp}$ and $\frac{1}{T}\int_{0}^{T}\beta_{t}'dt-Q'$ is bounded uniformly in $T$,
		\begin{align}
			\|\frac{1}{T}\int_{0}^{T}\beta_{t}'(A)dt-Q'A\|&\leq 2\|A\|,
		\end{align}
		for all $A\in\cB_{\tau}$, \eqref{ergodic-con} implies that, strongly in $\cB_{\tau}$,
		\begin{align}
			s\text{-}\lim_{T\rightarrow\infty}\frac{1}{T}\int_{0}^{T}\beta_{t}'dt=Q'.
		\end{align}
		This proves Theorem \ref{thm:RTE-dual1'}.\hfill$\qed$
	
		
		\begin{remark}\label{thm:RTE-dual1-Btau}			One can also prove a stronger result with $\Null(G')$ replaced by $\Null(L')$ (cf. Theorem \ref{thm:main}).
		\end{remark}

				\bigskip

		\subsection{Proof of Theorem \ref{thm:RTE}}\label{sec:main-pf}

		In 
		 this section, we drop the subscript ``${\rm st,\tau}$" in the inner product $\langle\cdot,\cdot\rangle_{\rm st,\tau}$ and norm $\|\cdot\|_{\rm st,\tau}$.
		
	

		Lemma \ref{lem:S*-dense} and Eq. \eqref{(QDB)-2} imply
		\begin{lemma}\label{lem:property-G-qdbc}
			Suppose $G$ satisfies $(QDB)$.  Then,
			\begin{enumerate}
				\item[(a)] $\sigma(G)=\sigma(G')$ and $\sigma_{d}(G)=\sigma_{d}(G')$, where $\sigma_{d}(A)$ denotes the discrete spectrum of operator $A$;
				\item[(b)] $\Null(G)\subseteq\Null(L_{0})$ if and only if $\Null(G')\subseteq\Null(L_{0}')$.
			\end{enumerate}
		\end{lemma}	
		

		
		\begin{proof}[Proof of Theorem \ref{thm:RTE}]
			(a) Recall that $Q$ denotes the orthogonal projector onto the subspace $\Null(G)\subseteq \cS_{\tau}$ w.r.t. the inner product $\langle\cdot,\cdot\rangle$ in the space of observables and define $Q^{\perp}=\one-Q$.  In the following, we denote $\rho_{t}=\beta_{t}(\rho)$ and $\bar{\rho}_{t}=Q^{\perp}\rho_{t}$.  Similar to the proof of Theorem \ref{thm:G'-property} in Subsection \ref{sec:pf-RTE-dual2}, we consider the Lyapunov functional $\|\bar{\rho}_{t}\|^{2}/2$.
			
			
			First, we note that, by Condition $(Null)$, $[L_{0},Q]=0$.  Since $L_{0}^{*}=-L_{0}$, $G^{*}=G$ on $\cS_{\tau}$ and $[L_{0},Q^{\perp}]=[L_{0},\one-Q]=0$, by Leibniz rule, we have
			\begin{align}\label{Lyapunov-diff}
				\partial_{t}\|\bar{\rho}_{t}\|^{2}/2&=\partial_{t}\langle\bar{\rho}_{t},\bar{\rho}_{t}\rangle/2=\langle\bar{\rho}_{t},(L^{*}+L)\bar{\rho}_{t}\rangle/2=\langle\bar{\rho}_{t},G\bar{\rho}_{t}\rangle.
			\end{align}
			By the Condition $(Gap)$ and Lemma \ref{lem:property-G-qdbc} (a), using spectral theory, there exists some $\theta>0$ such that $(-\theta,\theta)\cap(\sigma(G)\setminus\{0\})=\varnothing$ so that
			\begin{align}\label{spec-gap-2}
				G|_{\Ran Q^{\perp}}&\leq -\theta<0
			\end{align}
			so that, by substituting \eqref{spec-gap-2} into \eqref{Lyapunov-diff}, we obtain the following inequality
			\begin{align}
				\partial_{t}\|\bar{\rho}_{t}\|^{2}/2&=\langle\bar{\rho}_{t},G\bar{\rho}_{t}\rangle\leq -\theta\|\bar{\rho}_{t}\|^{2}
			\end{align}
			with the initial condition
			\begin{align}
				\|\bar{\rho}_{t}\|^{2}|_{t=0}&=\|\bar{\rho}_{0}\|^{2}.
			\end{align}
			We solve this inequality to obtain
			\begin{align}
				\|\bar{\rho}_{t}\|\leq e^{-\theta t}\|\bar{\rho}_{0}\|,
			\end{align}
			giving
			\begin{align}\label{rho-decay}
				\|\rho_{t}-Q\rho_{t}\|\leq e^{-\theta t}\|\rho_{0}\|.
			\end{align}
			
			
			To complete the proof, we compute $Q\rho_{t}$.  Since, by Condition $(Null)$ and Lemma \ref{lem:property-G-qdbc} (b), $\Ran Q=\Null(G)\subseteq\Null(L_{0})$ and $[L_{0},Q]=0$, we have $\Ran P\subseteq\Null(L)$ and $[L,Q]=[L_{0}+G,Q]=0$.  It follows that
			\begin{align}
				Q\rho_{t}&=Q\beta_{t}(\rho)=Q e^{Lt}\rho=e^{Lt}Q\rho=Q\rho,
			\end{align}
			which, together with \eqref{rho-decay}, implies \eqref{RTE1}.
			
			

			(b) Now, we suppose $0$ is a simple eigenvalue of $G$.  Then, we have $\Null(G)=\C\cdot\rho_{\tau}$ because $G(\rho_{\tau})=0$ by $(QDB)$.  Let $\rho\in\sD(L)$ with $\Tr\rho=1$.  Then, by cyclicity of trace, we have
			\begin{align}
				Q\rho&=\langle\rho_{\tau},\rho\rangle\rho_{\tau}=\Tr(\rho_{\tau}\rho\rho_{\tau}^{-1})\cdot\rho_{\tau}=\Tr(\rho)\cdot \rho_{\tau}=\rho_{\tau}.
			\end{align}
			By substituting this into \eqref{RTE1}, we obtain \eqref{RTE2}.
		\end{proof}

		\subsection{Proof of Theorem \ref{thm:iso-EV}}\label{sec:pf-iso-EV}		
		First of all, we note that, since $\Tr\rho=1$, 
		\begin{align}\label{coup-id}
			\Tr(A(\beta_{t}(\rho)-\rho_{\tau}))&=\Tr(\beta_{t}'(A)\rho)-\Tr(A\rho_{\tau})\nonumber\\
			&=\Tr((\beta_{t}'(A)-c^{A}\one)\rho),
		\end{align}
		where $c^{A}=\Tr(A\rho_{\tau})$.  
		
		
		%
		
		Recall that, by Theorem \ref{thm:G'-property} (b), $G'$ is self-adjoint and $G'\leq 0$ on the space $\cB_{\tau}$.  Since $0$ is an isolated eigenvalue of $G'$, $\sigma(G')\cap(-\theta,\theta)=\varnothing$ for $\theta:=\dist(0,\sigma(G')\setminus\{0\})$, and so, by \eqref{coup-id}  and \eqref{coupling}, we have
		\begin{align}
			\|\beta_{t}(\rho)-\rho_{\tau}\|_{\rm st,\tau}^{\intercal}&=\sup_{\|A\|_{\tau}=1}|\Tr(A(\beta_{t}(\rho)-\rho_{\tau}))|\nonumber\\
			&=\sup_{\|A\|_{\tau}=1}|\Tr((\beta_{t}'(A)-c^{A}\one)\rho)|\nonumber\\
			&\leq \sup_{\|A\|_{\tau}=1}\|\beta_{t}'(A)-c^{A}\one\|_{\tau}\|\rho\|_{\rm st,\tau}^{\intercal}\nonumber\\
			&\leq e^{-\theta t}\|\rho\|_{\rm st,\tau}^{\intercal}\sup_{\|A\|_{\rm st,\tau}=1}\|(\one-P')A\|_{\tau}\leq e^{-\theta t}\|\rho\|_{\rm st,\tau}^{\intercal}.
		\end{align}
		Now, using \eqref{RTE2-dual} and recalling that $P'$ is the rank-1 projector onto $\Null(G')=\C\cdot \rho_{\tau}$, we find $\|\beta_{t}(\rho)-\rho_{\tau}\|_{\rm st,\tau}^{\intercal}\leq e^{-\theta t}\|\rho\|_{\rm st,\tau}^{\intercal}$, as desired.\hfill\qed
%
%
		


		\subsection{Proof of Theorem \ref{thm:RTE-dual2}}\label{sec:pf-RTE-dual2}

		Throughout this subsection, we denote $A_{t}:=\beta_{t}'(A)$.

		(a) First we note that, by Leibniz rule, we have, for any $A\in\sD(L')$,
		\begin{align*}
			\partial_{t}\|A_{t}\|^{2}&=\partial_{t}\langle A_{t},A_{t}\rangle=\langle\partial_{t}A_{t},A_{t}\rangle+\langle A_{t},\partial_{t}A_{t}\rangle\nonumber\\
			&=\langle L'A_{t},A_{t}\rangle+\langle A_{t},L'A_{t}\rangle=\langle A_{t},(L'^{*}+L')A_{t}\rangle.
		\end{align*} 
		Since $L'^{*}+L^{*}=2G'$, this yields
		\begin{align}\label{Lyaponov-dual}
			\partial_{t}\|A_{t}\|^{2}&=2\langle A_{t},G'A_{t}\rangle.
		\end{align}
		

		Recall that $Q'^{\perp}:=\one-Q'$ and let $U_{0}$ be an open neighborhood of $0$ such that $G'$ has no singular continuous spectrum on $U_{0}$.  We denote $Q_{1}$ and $Q_{2}$ as the orthogonal projections onto the subspaces of point and absolutely continuous spectra of $G'$ in $U_{0}$, respectively.  Furthermore, we denote $Q_{3}$ the spectral projection of $G'$ onto $\sigma(G')\setminus U_{0}$.  It follows from Condition $(Spec)$ that $Q'^{\perp}=\sum_{i=1}^{3}Q_{i}$.  
		
		
		We note that, by Theorem \ref{thm:Gibbs}, $L_{0}'$ and $G'$ commute strongly, which then implies that $L_{0}'$ also commutes with any spectral projection of $G'$, which implies that $[L_{0}',Q_{i}]=0$ for each $i=1,2,3$.
		

		We denote $A_{t,i}:=Q_{i}A_{t}$ for $i=1,2,3$ and consider the Lyapunov functional $\|P'^{\perp}A_{t}\|^{2}$.  By triangle inequality, we have
		\begin{align}\label{Lyapunov-eqn}
			\|Q'^{\perp}A_{t}\|^{2}&\leq \sum_{i=1}^{3}\|A_{t,i}\|^{2}.
		\end{align}
		We will show that $\|A_{t,i}\|^{2}\rightarrow 0$ for $i=1,2,3$ as $t\rightarrow\infty$.
	
		
		
		We begin with $\|A_{t,1}\|^{2}$.  We observe that we can write $Q_{1}$ and $G'|_{\Ran Q_{1}}$ as
		\begin{align}
			Q_{1}&=\sum_{n}Q_{e_{n}},\quad\quad G'|_{\Ran Q_{1}}=\sum_{n}\lambda_{n}Q_{e_{n}},
		\end{align}
		where $\lambda_{n}$ are eigenvalues of $G'$ in $U_{0}$ and $e_{n}$ are the corresponding orthonormal eigenvectors of $G'$.  Since $G'< 0$ on $\Ran Q_{1}$, we have $\lambda_{n}<0$ for all $n$.  
		
		
		Let $A_{t}^{(n)}:=Q_{e_{n}}A_{t,1}$ for each $n$.  By \eqref{Lyaponov-dual}, the fact that $[e^{L_{0}'t},Q_{e_{n}}]=0$ and the relation $G'Q_{e_{n}}=\lambda_{n}Q_{e_{n}}$, we have
		\begin{align}\label{Lyapu-dual-pp}
			\partial_{t}\|A_{t}^{(n)}\|^{2}&=2\langle A_{t}^{(n)},G'A_{t}^{(n)}\rangle=\lambda_{n}|\langle e_{n},A_{t}\rangle|^{2}.
		\end{align}
		Introducing the notation $a_{n}(t):=|\langle e_{n},A_{t}\rangle|^{2}$ and using that $\|A_{t}^{(n)}\|^{2}=a_{n}(t)$, we find from \eqref{Lyapu-dual-pp} that
		\begin{align}
			\partial_{t}a_{n}(t)=2\lambda_{n}a_{n}(t)
		\end{align}
		with the initial condition
		\begin{align}
			\|A_{t}^{(n)}\|^{2}|_{t=0}&=\|A_{0}^{(n)}\|^{2}=a_{n}(0),
		\end{align}
		which is solved to be
		\begin{align}
			a_{n}(t)&=e^{2\lambda_{n}t}a_{n}(0).
		\end{align}
		Furthermore, by $Q_{1}=\sum_{n}Q_{e_{n}}$, we have
		\begin{align}
			\|A_{t,1}\|^{2}&=\sum_{n}\|A_{t}^{(n)}\|^{2}=\sum_{n}a_{n}(t)=\sum_{n}e^{2\lambda_{n}t}a_{n}(0).
		\end{align}
		This and $\lambda_{n}<0$ for each $n$ yield, by the dominated convergence theorem, that $\|A_{t,1}\|^{2}\rightarrow 0$ as $t\rightarrow \infty$.
		
		
		
		Now, we consider $\|A_{t,2}\|^{2}$.  Recall $Q_{2}$ is the orthogonal projection of $G'$ onto the subspace of absolutely continuous spectrum in $U_{0}$.  Let $dE_{G'}(\lambda)$ be the spectral measure of $G'$ on $\Ran P_{2}$.  Then, for each $A\in\Ran Q_{2}$, we have
		\begin{align}\label{Lyapu-dual-ac}
			\langle A,A\rangle&=\int_{U_{0}}d\mu_{A}(\lambda),\quad\quad \langle A,G'A\rangle=\int_{U_{0}}\lambda d\mu_{A}(\lambda),
		\end{align}
		where $d\mu_{A}(\lambda):=\langle A,dE_{G'}(\lambda)A\rangle$.  Since $d\mu_{A}(\lambda)$ is absolutely continuous w.r.t. the Lebesgue measure, there exists some positive, Lebesgue measurable function $f_{A}(\lambda)$ such that $d\mu_{A}(\lambda)=f_{A}(\lambda)d\lambda$.
		
		
		Let $f_{t}(\lambda):=f_{A_{t}}(\lambda)$.  Then, by \eqref{Lyaponov-dual} and \eqref{Lyapu-dual-ac}, we have
		\begin{align}
			\partial_{t}\int_{U_{0}}f_{t}(\lambda)d\lambda=\int_{U_{0}}2\lambda f_{t}(\lambda)d\lambda
		\end{align}
		with the initial condition
		\begin{align}
			\left(\int_{U_{0}}f_{t}(\lambda)d\lambda\right)|_{t=0}&=\int_{U_{0}}f_{0}(\lambda)d\lambda.
		\end{align}
		
	
		Next, we observe that, for any subset $V\subseteq U_{0}$ and $Q_{V}$ corresponding spectral projection onto the absolutely continuous spectrum in $V$, by Condition $(W)$, the fact that $[e^{L_{0}'t},Q_{V}]=0$ for all $V$ and the relation \eqref{Lyaponov-dual}, we have
		\begin{align}
			\partial_{t}\int_{V}f_{t}(\lambda)d\lambda&=\int_{V}2\lambda f_{t}(\lambda)d\lambda
		\end{align}
		with the initial condition
		\begin{align}
			\left(\int_{V}f_{t}(\lambda)d\lambda\right)|_{t=0}&=\int_{V}f_{0}(\lambda)d\lambda.
		\end{align}
		Now, for each $\lambda\in U_{0}$, we take a sequence of open neighborhoods $V_{n}\subseteq U_{0}$ of $\lambda$ such that $V_{n}\supseteq V_{n+1}$ and $V_{n}\rightarrow\{\lambda\}$ as $n\rightarrow\infty$.  By the Lebesgue differentiation theorem, we have, for almost all $\lambda\in U_{0}$,
		\begin{align}
			\partial_{t}f_{t}(\lambda)&=\lim_{n\rightarrow\infty}\frac{1}{m(V_{n})}\int_{V_{n}}f_{t}(\lambda)d\lambda\nonumber\\
			&=\lim_{n\rightarrow\infty}\frac{1}{m(V_{n})}\int_{V_{n}}2\lambda f_{t}(\lambda)d\lambda=2\lambda f_{t}(\lambda)d\lambda
		\end{align}
		and
		\begin{align}
			f_{0}(\lambda)&=\lim_{n\rightarrow\infty}\frac{1}{m(V_{n})}\int_{V_{n}}f_{0}(\lambda)d\lambda,
		\end{align}
		where $m(V)$ is the Lebesgue measure of $V$.  Then, for almost all $\lambda\in U_{0}$, we have
		\begin{align}
			f_{t}(\lambda)&=e^{2\lambda t}f_{0}(\lambda)
		\end{align}
		so that
		\begin{align}
			\|A_{t,2}\|^{2}&=\int_{U_{0}}f_{t}(\lambda)d\lambda=\int_{U_{0}}e^{2\lambda t}f_{0}(\lambda)d\lambda.
		\end{align}
		Since $U_{0}\subseteq (-\infty,0)$, we have, by the dominated convergence theorem, that $\|A_{t,2}\|^{2}\rightarrow 0$ as $t\rightarrow\infty$.
		
		
		
		Finally, we consider $\|A_{t,3}\|^{2}$.  Since $U_{0}$ is an open neighborhood of $0$, there exists some $\theta>0$ such that $[-\theta,\theta]\in U_{0}$ so that
		\begin{align}\label{spec-gap}
			G'|_{\Ran Q_{3}}&\leq -\theta.
		\end{align}
		Then, by \eqref{Lyaponov-dual} and \eqref{spec-gap}, we have
		\begin{align}\label{Lyapu-eqn}
			\partial_{t}\|A_{t,3}\|^{2}&=2\langle A_{t,3},G'A_{t,3}\rangle_{*}\leq -2\theta\|A_{t,3}\|^{2}
		\end{align}
		with the initial condition
		\begin{align}
			\|A_{t,3}\|^{2}|_{t=0}&=\|A_{0,3}\|^{2}.
		\end{align}
		Inequality \eqref{Lyapu-eqn} is solved as
		\begin{align}
			\|A_{t,3}\|&\leq e^{-\theta t}\|A_{0,3}\|,
		\end{align}
		which converges to $0$ as $t\rightarrow\infty$.  Therefore, \eqref{Lyapunov-eqn} gives, as $t\rightarrow\infty$,
		\begin{align}\label{perp-A-decay}
			\|Q'^{\perp}A_{t}\|&\leq \sum_{i=1}^{3}\|Q_{i}A_{t}\|\rightarrow 0.
		\end{align}
		
	
		
		To conclude this proof, we recall that $Q'$ is the orthogonal projection onto $\Null(G')$ and therefore $[Q',G']=0$.  By Theorem \ref{thm:Gibbs}, $[L_{0}',G']=0$ (in an appropriate sense) and therefore $[Q',L_{0}']=0$.  The last two relations imply
		\begin{align}
			[L',Q']=[L_{0}'+G',Q']=0.
		\end{align}
		By Condition $(Null)$, we have $\Ran Q'=\Null(G')\subseteq\Null(L_{0}')$, which implies that $\Ran Q'\subseteq\Null(L')$.  Thus, for all $A\in\cB_{\tau}$, we have
		\begin{align}\label{Null(G')-proj}
			Q'A_{t}&=Q'\beta_{t}'(A)=P'e^{L't}A=e^{L't}Q'A=Q'A.
		\end{align}
		Therefore, we have, by \eqref{perp-A-decay}, as $t\rightarrow \infty$,
		\begin{align}
			\|A_{t}-Q'A\|&=\|A_{t}-Q'A_{t}\|=\|Q^{\perp}A_{t}\|\rightarrow 0,
		\end{align}
		which proves \eqref{RTE3-dual}.  
		
				
	
		(b) By Condition $(Gap)$, we have $Q_{1}=Q_{2}=0$ so that $Q'^{\perp}=Q_{3}$.  It follows from proof for part (a) that there exists some $\theta>0$ such that
		\begin{align}\label{gap-conve}
			\|A_{t}-Q'A_{t}\|&=\|Q'^{\perp}A_{t}\|\leq e^{-\theta t}\|Q'^{\perp}A_{0}\|\leq e^{-\theta t}\|A\|.
		\end{align}
		It then follows from the same reason as in the last paragraph of proof for part (a) that $Q'A_{t}=Q'A$, which, together with \eqref{gap-conve}, implies \eqref{RTE1-dual}.

		
		(c) Since $0$ is a simple eigenvalue of $G'$ and $\one\in\Null(G')$, we have $\Null(G')=\C\cdot\one\subseteq\Null(L_{0}')$ so that $\Null(G')\subseteq\Null(L')$ and
		\begin{align}
			Q'A=\langle\one,A\rangle\cdot\one=\Tr(A\rho_{\tau})\one.
		\end{align}
		This and Theorem \ref{thm:RTE-dual2} (b) yield \eqref{RTE2-dual}.\hfill\qed

\section{GNS representation of $(\cB,\omega_{\tau})$}\label{sec:GNS}
In this section, we define the GNS representation associated with the pair $(\cB,\omega_{\tau})$ (by regarding $\cB$ as a von Neumann algebra and $\omega_{\tau}$ as a normal state on $\cB$).  This section follows mainly according to \cite{Haag}.



Recall the state $\omega_{\tau}$ is defined through a density operator $\rho_{\tau}>0$ and is given, for any $A\in\cB$, by
\begin{align}
	\omega_{\tau}(A)&=\Tr(A\rho_{\tau}).
\end{align}
The operator $\Omega_{\tau}:=\rho_{\tau}^{1/2}$ is Hilbert-Schmidt.  Note that, since $\rho_{\tau}>0$, the state $\omega_{\tau}$ is faithful.  Indeed, for each $A\in\cB$, we have
\begin{align}
	\omega_{\tau}(A^{*}A)&=\Tr(A^{*}A\rho_{\tau})=\|A\Omega_{\tau}\|_{\cS_{2}}^{2}
\end{align}
so that $\omega_{\tau}(A^{*}A)=0$ implies $A\Omega_{\tau}=0$.  Since $\Omega_{\tau}>0$, we have $A=0$.



Next, we define the representation $\pi_{\tau}$ of $\cB$ acting on $\cS_{2}$ by
\begin{align}
	\pi_{\tau}(A)\kappa&=A\kappa,\quad\quad \kappa\in\cS_{2},\quad A\in\cB,
\end{align}
i.e., the representation for left action by $\cB$.  Then, we have
\begin{align}
	\omega_{\tau}(A)&=\langle\Omega_{\tau},\pi_{\tau}(A)\Omega_{\tau}\rangle_{\cS_{2}}.
\end{align}



Since $\cS_{2}$ is a closed, two-sided ideal of $\cB$, there is another natural, anti-linear representation $\pi_{\tau}'$ for right action by $\cB$ on $\cS_{2}$, given by
\begin{align}
	\pi_{\tau}'(A)\kappa&=\kappa A^{*},\quad\quad \kappa\in\cS_{2},\quad A\in\cB,
\end{align}
i.e., the representation for right-action by $\cB$.  It is immediate from the definition that $\pi_{\tau}'$ satisfies
\begin{align}
	\pi_{\tau}'(AB)&=\pi_{\tau}'(A)\pi_{\tau}'(B),\quad\pi_{\tau}'(A^{*})=(\pi_{\tau}'(A))^{*}.
\end{align}


Recall the definition of the anti-unitary operator $J$: 
\begin{align}\label{mod-conj}
	J\kappa&=\kappa^{*},\quad\quad\text{for all }\kappa\in\cS_{2}.
\end{align}
Properties of $\Omega_{\tau},\pi_{\tau}$ and $\pi_{\tau}'$ are summarized in the following proposition:
\begin{prop}\label{prop:prop-GNS}(\cite{Haag}, Section V.1.4, Theorem 1.4.1)
	\begin{enumerate}
		\item[(a)] The operator norm of $\pi_{\tau}(A)$ and $\pi_{\tau}'(A)$ on $\cS_{2}$ are given by
		\begin{align}\label{rep-norm}
			\|\pi_{\tau}(A)\|&=\|\pi_{\tau}'(A)\|=\|A\|.
		\end{align}
		\item[(b)] $\Omega_{\tau}$ is cyclic and separating for both $\pi_{\tau}(\cB)$ and $\pi_{\tau}'(\cB)$.
		\item[(c)] $J^{*}=J$, $J^{2}=\one$ and $J\Omega_{\tau}=\Omega_{\tau}$.
		\item[(d)] $\pi_{\tau}$ and $\pi_{\tau}'$ are transformed to each other by the anti-unitary operator $J$ as
		\begin{align}
			J\pi_{\tau}(A)J&=\pi_{\tau}'(A),\quad\text{for each }A\in\cB.
		\end{align}
	\end{enumerate}
\end{prop}
\begin{proof}
	(a) By definition, we have
	\begin{align}
		\|\pi_{\tau}(A)\|_{\cS_{2}}^{2}&=\sup_{\|\kappa\|_{\cS_{2}}=1}\|\pi_{\tau}(A)\kappa\|_{\cS_{2}}^{2}=\sup_{\|\kappa\|_{\cS_{2}}=1}\Tr(\kappa^{*}A^{*}A\kappa)=\|A\|^{2},
	\end{align}
	where the last equality follows from by choosing $\kappa$ as some rank-1 projection operator.  The norm for $\pi_{\tau}'(A)$ can be found in a similar way using cyclicity of trace and the fact that $\|A^{*}\|=\|A\|$.
	
	
	
	%
	
	(b) Suppose $\kappa\in\cS_{2}$ is an element such that $\kappa\perp \pi_{\tau}(A)\Omega_{\tau}$ for all $A\in\cB$.  Then, by taking $A=\kappa$ (by viewing $\kappa$ as an element in $\cB$), we have
	\begin{align}
		0&=\langle\kappa,\pi_{\tau}(\kappa)\Omega_{\tau}\rangle_{\cS_{\tau}}=\Tr(\kappa^{*}\kappa\rho_{\tau}^{1/2}).
	\end{align}
	Since $\rho_{\tau}^{1/2}>0$, we must have $\kappa^{*}\kappa=0$, which implies that $\kappa=0$.  Together with the faithfulness of $\omega_{\tau}$, we see that $\Omega_{\tau}$ is a cyclic and separating vector for the representation $\pi_{\tau}$.  
	
	
	The cyclicity of $\Omega_{\tau}$ for $\pi_{\tau}'(\cB)$ follows from the separability of $\Omega_{\tau}$.  The separatibility of $\Omega_{\tau}$ for $\pi_{\tau}(\cB)$ and $\pi_{\tau}'(\cB)$ are proven similarly.
	

	
	(c) Let $J$ be as in \eqref{mod-conj}.  Since $J^{2}\kappa=(\kappa^{*})^{*}=\kappa$ for all $\kappa\in\cS_{2}$, we have 
	\begin{align}
		J^{2}&=\one.
	\end{align}
	
	Next, for each $\kappa,\sigma\in\cS_{2}$, by cyclicity of trace and by the fact that $J$ is anti-linear, we have
	\begin{align}
		\langle J^{*}\kappa,\sigma\rangle_{\cS_{2}}&=\overline{\langle \kappa,J\sigma\rangle_{\cS_{2}}}=\langle J\sigma,\kappa\rangle_{\cS_{2}}=\Tr((\sigma^{*})^{*}\kappa)\nonumber\\
		&=\Tr(\sigma\kappa)=\Tr((\kappa^{*})^{*}\sigma)=\langle\kappa^{*},\sigma\rangle_{\cS_{2}}=\langle J\kappa,\sigma\rangle_{\cS_{2}},
	\end{align}
	which implies that $J^{*}=J$.  
	
	Finally, by the definition \eqref{mod-conj}, we have $J\Omega_{\tau}=(\rho_{\tau}^{1/2})^{*}=\rho_{\tau}^{1/2}=\Omega_{\tau}$.
	
	
	
	(d) For each $A\in\cB$, since $J\Omega_{\tau}=\Omega_{\tau}$ from part (c), we have
	\begin{align}
		\pi_{\tau}'(A)\Omega_{\tau}&=\rho_{\tau}^{1/2}A^{*}=(A\rho_{\tau}^{1/2})^{*}=J(\pi_{\tau}(A)\Omega_{\tau})=(J\pi_{\tau}(A)J)\Omega_{\tau}.
	\end{align}
	Since $\Omega_{\tau}$ is cyclic for both $\pi_{\tau}(\cB)$ and $\pi_{\tau}'(\cB)$, we conclude that $J\pi_{\tau}(A)J=\pi_{\tau}'(A)$ for all $A\in\cB$.
\end{proof}

Thus, the triple $(\cS_{2},\pi_{\tau},\Omega_{\tau})$ gives the GNS representation associated with $(\cB,\omega_{\tau})$.



\begin{proof}[Proof of Theorem \ref{thm:GNS-TT2}]
	Recall the definitions of the unbounded, self-adjoint operator
	\begin{align}
		H_{\tau}:=-\ln\rho_{\tau},\quad\quad\text{so that }\rho_{\tau}&=e^{-H_{\tau}},
	\end{align}
	and the automorphism group $\alpha_{t}$ on $\cB$ by 
	\begin{align}
		\alpha_{t}(A)&=e^{iH_{\tau}t}Ae^{-iH_{\tau}t}.
	\end{align}
	We see that $\omega_{\tau}$ is an invariant state under $\alpha_{t}$.  Indeed, since $\alpha_{t}'(\rho_{\tau})\equiv e^{-iH_{\tau}t}\rho_{\tau}e^{iH_{\tau}t}=\rho_{\tau}$, we have, for each $A\in\cB$ and $t\in\R$,
	\begin{align}
		\omega_{\tau}(\alpha_{t}(A))&=\Tr(A\alpha_{t}'(\rho_{\tau}))=\Tr(A\rho_{\tau})=\omega_{\tau}(A).
	\end{align}
	
	
	
	We define the family of operators $U(t)$ on $\cS_{2}$, given by
	\begin{align}\label{U(t)}
		U(t)(\pi_{\tau}(A)\Omega_{\tau})&=\pi_{\tau}(\alpha_{t}(A))\Omega_{\tau},\quad\quad\text{for all }A\in\cB.
	\end{align}
	By \eqref{rep-norm} and $\|\alpha_{t}(A)\|=\|A\|$, for each $t$, the operator $U(t)$ is an isometry.  Since $\alpha_{t}$ is a $*$-automorphism group of $\cB$ and $\omega_{\tau}$ is invariant under $\alpha_{t}$, $U(t)$ is a one-parameter family of unitary operators and $U(t)\Omega_{\tau}=\Omega_{\tau}$ for each $t\in\R$ (see \cite{BrRo1}, Corollary 2.3.17).  Furthermore, since $\alpha_{t}$ is weakly$^{*}$ continuous on $\cB$, we have  
	\begin{lemma}\label{lem:sc-U(t)}
		The one-parameter group $U(t)$ of unitary operators defined in \eqref{U(t)} is strongly continuous.
	\end{lemma}
	\begin{proof}[Proof of Lemma \ref{lem:sc-U(t)}]
		For each $A\in\cB$, we have
		\begin{align}\label{sc-U}
			\|U(t)&(\pi_{\tau}(A)\Omega_{\tau})-\pi_{\tau}(A)\Omega_{\tau}\|_{\cS_{2}}^{2}\nonumber\\
			&=2\omega_{\tau}(A^{*}A)-\omega_{\tau}(A^{*}\alpha_{t}(A))-\omega_{\tau}((\alpha_{t}(A))^{*}A).
		\end{align}
		Since $\alpha_{t}$ is weakly$^{*}$ continuous, for each $\epsilon>0$, there exists some $\delta>0$ such that, for any $0<|t|<\delta$, the r.h.s. of \eqref{sc-U} is less than $\epsilon^{2}$.  Thus, for $0<|t|<\delta$, we have
		\begin{align}\label{sc-U2}
			\|U(t)(\pi_{\tau}(A)\Omega_{\tau})-\pi_{\tau}(A)\Omega_{\tau}\|_{\cS_{2}}&<\epsilon.
		\end{align}
		For general $\kappa\in\cS_{2}$, since $\sF$ is dense in $\cS_{2}$, for each $\epsilon>0$, there exists some $A\in\cB$ such that $\|\kappa-\pi_{\tau}(A)\Omega_{\tau}\|_{\cS_{2}}<\epsilon$.  By \eqref{sc-U2} and the fact that $U(t)$ is unitary for each $t\in\R$, for any $0<|t|<\delta$, we have
		\begin{align}
			\|U(t)\kappa-\kappa\|_{\cS_{2}}&\leq \|U(t)(\pi_{\tau}(A)\Omega_{\tau})-\pi_{\tau}(A)\Omega_{\tau}\|_{\cS_{2}}\nonumber\\
			&\quad\quad+2\|\kappa-\pi_{\tau}(A)\Omega_{\tau}\|_{\cS_{2}}<\epsilon+2\epsilon=3\epsilon.
		\end{align}
		Thus, $U(t)$ is strongly continuous on $\cS_{2}$.
	\end{proof}
	

	Now, by the cyclicity of $\Omega_{\tau}$ and the fact that $U(t)\Omega_{\tau}=\Omega_{\tau}$, we have, for all $A\in\cB$,
	\begin{align}
		U(t)\pi_{\tau}(A)U(t)^{*}&=\pi_{\tau}(\alpha_{t}(A)).
	\end{align}
	Let $L_{\tau}$ be the generator of $U(t)$.  Then, we have
	\begin{align}
		\pi_{\tau}(\alpha_{t}(A))&=e^{iL_{\tau}t}\pi_{\tau}(A)e^{-iL_{\tau}t},\quad\quad L_{\tau}\Omega_{\tau}=0,
	\end{align}
	which proves \eqref{mod-auto2}.
	

	To prove relation \eqref{mod2}, we introduce entire analytic elements for $\alpha_{t}$ (see \cite{BrRo1}, Section 2.5.3). 
	\begin{Def}\label{def:analytic}
		\normalfont
		We say an operator $A\in\cB$ is \textit{entire analytic} for $\alpha_{t}$ if there exists a function $f:\C\rightarrow\cB$ such that 
		\begin{enumerate}
			\item[(a)] $f(t)=\alpha_{t}(A)$ for $t\in\R$.
			\item[(b)] The function $z\mapsto\omega(f(z))$ is analytic on the entire plane $\C$ for all $\omega\in\cB'$, where $\cB'$ is the dual space of $\cB$.
		\end{enumerate}
		We denote the set of entire analytic operators for $\alpha_{t}$ by $\cB_{\rm ana}$.
	\end{Def}
	
	
	By Lemma \ref{lem:sc-U(t)}, $U(t)$ is strongly continuous, which leads to the following lemma:
	\begin{lemma}\label{lem:analytic-core3}
		For each $A\in\cB$ and any $\epsilon>0$, there exists some $B\in\cB_{\rm ana}$ such that
		\begin{align}
			\|(\pi_{\tau}(A)-\pi_{\tau}(B))\Omega_{\tau}\|_{\cS_{2}}<\epsilon.
		\end{align}
	\end{lemma}
	\begin{proof}
		We follow the argument in \cite{BrRo1}, Proposition 2.5.22.  Let $A\in\cB$ be fixed.  For each integer $n\geq 1$, we define
		\begin{align}
			A_{n}&=\sqrt{\frac{n}{\pi}}\int_{\R}\alpha_{t}(A)e^{-nt^{2}}dt.
		\end{align}
		Each $A_{n}$ is entire analytic for $\alpha_{t}$.  Indeed, for each $z\in\C$, the function 
		\begin{align}
			f_{n}(z)&=\sqrt{\frac{n}{\pi}}\int_{\R}\alpha_{t}(A)e^{-n(t-z)^{2}}dt,
		\end{align}
		is a well-defined function of $z\in\C$ since the function $t\mapsto e^{-(t-z)^{2}}$ is integrable for each $z$ and, when $z=s\in\R$, we have
		\begin{align}
			f_{n}(s)&=\sqrt{\frac{n}{\pi}}\int_{\R}\alpha_{t}(A)e^{-n(t-s)^{2}}dt=\sqrt{\frac{n}{\pi}}\int_{\R}\alpha_{t+s}(A)e^{-nt^{2}}dt\nonumber\\
			&=\alpha_{s}\left[\sqrt{\frac{n}{\pi}}\int_{\R}\alpha_{t}(A)e^{-nt^{2}}dt\right]=\alpha_{s}(A_{n}).
		\end{align}
		Also, for each $\omega\in\cB'$, we have 
		\begin{align}
			\omega(f_{n}(z))&=\sqrt{\frac{n}{\pi}}\int_{\R}\omega(\alpha_{t}(A))e^{-n(t-z)^{2}}dt
		\end{align}
		so that
		\begin{align}
			|\omega(f_{n}(z))|&\leq \|\omega\|\|A\|\sqrt{\frac{n}{\pi}}\int_{\R}|e^{-n(t-z)^{2}}|dt\nonumber\\
			&\leq \|\omega\|\|A\|\sqrt{\frac{n}{\pi}}\int_{\R}e^{-n(t-x)^{2}+ny^{2}}dt\leq e^{ny^{2}}\|\omega\|\|A\|,
		\end{align}
		where $z=x+iy$.  It then follows from the Lebesgue dominated convergence theorem that the function $z\mapsto \omega(f_{n}(z))$ is entire analytic.  Thus, $A_{n}$ is analytic.
		
		
		
		Next, we show that $\|(\pi_{\tau}(A_{n})-\pi_{\tau}(A))\Omega_{\tau}\|_{\cS_{2}}\rightarrow 0$ as $n\rightarrow\infty$.  For notational simplicity, we drop the subscript in the norm $\|\cdot\|_{\cS_{2}}$ in the rest of this proof.  Let $\xi_{A}=\pi_{\tau}(A)\Omega_{\tau}$.  Since $\sqrt{\frac{n}{\pi}}\int_{\R}e^{-nt^{2}}dt=1$, by \eqref{U(t)}, we have
		\begin{align}
			\|(\pi_{\tau}(A_{n})&-\pi_{\tau}(A))\Omega_{\tau}\|\nonumber\\
			&\leq \sqrt{\frac{n}{\pi}}\int_{\R}e^{-nt^{2}}\|\pi_{\tau}(\alpha_{t}(A))-\pi_{\tau}(A))\Omega_{\tau}\|dt\nonumber\\
			&=\sqrt{\frac{n}{\pi}}\int_{\R}e^{-nt^{2}}\|U(t)\xi_{A}-\xi_{A}\|dt
		\end{align}
		By Lemma \ref{lem:sc-U(t)}, $U(t)$ is strongly continuous so that, for any $\epsilon>0$, there exists some $\delta$ such that, for all $|t|<\delta$, $\|U(t)\xi_{A}-\xi_{A}\|<\epsilon$.  Note that the choice of $\delta$ is independent of $n$.  Thus, we choose $n$ large enough so that 
		\begin{align}\label{n-large}
			\sqrt{\frac{n}{\pi}}\int_{|t|\geq\delta}e^{-nt^{2}}dt<\frac{\epsilon}{2\|A\|}.
		\end{align}
		It follows that, since $\|U(t)\xi_{A}\|=\|\xi_{A}\|$,
		\begin{align}
			\sqrt{\frac{n}{\pi}}&\int_{\R}e^{-nt^{2}}\|U(t)\xi_{A}-\xi_{A}\|dt\nonumber\\
			&=\sqrt{\frac{n}{\pi}}\int_{|t|<\delta}e^{-nt^{2}}\|U(t)\xi_{A}-\xi_{A}\|dt\nonumber\\
			&\quad\quad\quad+\sqrt{\frac{n}{\pi}}\int_{|t|\geq \delta}e^{-nt^{2}}\|U(t)\xi_{A}-\xi_{A}\|dt\nonumber\\
			&\leq \epsilon\sqrt{\frac{n}{\pi}}\int_{|t|<\delta}e^{-nt^{2}}dt+\sqrt{\frac{n}{\pi}}\int_{|t|\geq \delta}e^{-nt^{2}}(\|U(t)\xi_{A}\|+\|\xi_{A}\|)dt\nonumber\\
			&\leq \epsilon+2\|\xi_{A}\|\sqrt{\frac{n}{\pi}}\int_{|t|\geq\delta}e^{-nt^{2}}dt\nonumber\\
			&<\epsilon+\epsilon=2\epsilon.
		\end{align}
		This completes the proof.
	\end{proof}
%
%
%

	
	\begin{lemma}\label{lem:analytic-core}
		Every element in $\sF_{\rm ana}:=\pi_{\tau}(\cB_{\rm ana})\Omega_{\tau}$ is entire analytic for $U(t)$.  Consequently, $\sF_{\rm ana}\subseteq\sD(e^{zL_{\tau}})$ for all $z\in\C$.
		
	\end{lemma}
	\begin{proof}[Proof of Lemma \ref{lem:analytic-core}]
		For all $A\in\cB_{\rm ana}$ and $\sigma\in\cS_{2}$, the function
		\begin{align}
			z\mapsto\langle\sigma,U(z)(\pi_{\tau}(A)\Omega_{\tau})\rangle_{\cS_{2}}&=\langle\sigma,\pi_{\tau}(\alpha_{z}(A))\Omega_{\tau}\rangle_{\cS_{2}}\nonumber\\
			&=\Tr(\sigma^{*}\alpha_{z}(A)\Omega_{\tau})
		\end{align}
		is analytic on $\C$.  Thus, every elements in $\sF_{\rm ana}$ is entire analytic for $U(t)$. 
	\end{proof}
	

	Now, by Lemma \ref{lem:analytic-core}, for each $A\in\cB_{\rm ana}$, $\pi_{\tau}(A)\Omega_{\tau}\in\sD(e^{-L_{\tau}/2})$ and
	\begin{align}
		e^{-L_{\tau}/2}(\pi_{\tau}(A)\Omega_{\tau})&=\pi_{\tau}(\alpha_{i/2}(A))\Omega_{\tau}=\alpha_{i/2}(A)\rho_{\tau}^{1/2}\nonumber\\
		&=(\rho_{\tau}^{1/2}A\rho_{\tau}^{-1/2})\rho_{\tau}^{1/2}=\rho_{\tau}^{1/2}A\nonumber\\
		&=(A^{*}\rho_{\tau}^{1/2})^{*}=J(\pi_{\tau}(A^{*})\Omega_{\tau}).
	\end{align}
	Since $J^{2}=\one$, this yields, for all $A\in\cB_{\rm ana}$,
	\begin{align}
		\pi_{\tau}(A^{*})\Omega_{\tau}&=Je^{-L_{\tau}/2}(\pi_{\tau}(A)\Omega_{\tau}).
	\end{align}
	
	
	Next, for any $A\in\cB$, we construct a sequence $\{A_{n}\}$ in $\cB_{\rm ana}$ as in the proof of Lemma \ref{lem:analytic-core3} so that, as $n\rightarrow\infty$,
	\begin{align}
		&\|\pi_{\tau}(A_{n})\Omega_{\tau}-\pi_{\tau}(A)\Omega_{\tau}\|_{\cS_{2}}\rightarrow 0,\\
		&\|Je^{-L_{\tau}/2}(\pi_{\tau}(A_{n})\Omega_{\tau})-\pi_{\tau}(A^{*})\Omega_{\tau}\|_{\cS_{2}}\nonumber\\
		&\quad\quad\quad\quad\quad\quad\quad=\|\pi_{\tau}(A_{n}^{*})\Omega_{\tau}-\pi_{\tau}(A^{*})\Omega_{\tau}\|_{\cS_{2}}\rightarrow 0.
	\end{align}
	By the closedness of the operator $Je^{-L_{\tau}/2}$, we have $\pi_{\tau}(A)\Omega_{\tau}\in\sD(e^{-L_{\tau}/2})$ and
	\begin{align}
		Je^{-L_{\tau}/2}(\pi_{\tau}(A)\Omega_{\tau})&=\pi_{\tau}(A^{*})\Omega_{\tau},
	\end{align}
	which proves \eqref{mod2}.
\end{proof}


	\begin{remark}
	In our case, we can also define entire analytic elements for $U(t)$ in $\cS_{2}$ in a similar way: An element $\kappa\in\cS_{2}$ is \textit{entire analytic} for $U(t)$ if, for all $\sigma\in\cS_{2}$, the function $z\mapsto\langle\sigma,U(z)\kappa\rangle_{\cS_{2}}$ is analytic on $\C$.
	
	
	Furthermore, we can define entire analytic elements for $U(t)$ equivalently using its generator $L_{\tau}$: An element $\kappa\in\cS_{2}$ is \textit{entire analytic} if $\kappa\in\sD(L_{\tau}^{n})$ for all $n\in\mathbb{N}$ and, for all $t>0$, the series
	\begin{align}
		\sum_{n=0}^{\infty}\frac{t^{n}}{n!}\|L_{\tau}^{n}\kappa\|_{\cS_{2}}<\infty.
	\end{align}
	Hence, $\kappa\in\sD(e^{zL_{\tau}})$ for all $z\in\C$ if $\kappa\in\cS_{2}$ is entire analytic for $U(t)$.  For a proof of the equivalence of the above two definitions for entire analytic elements, see \cite{BrRo1}, p.178--179.
\end{remark}


\begin{remark}
	In fact, the state $\omega_{\tau}$ is a KMS-state w.r.t. the automorphism group $\alpha_{t}$, i.e., $\omega_{\tau}$ is invariant under $\alpha_{t}$ and, for each $A,B\in\cB$, the function $F_{A,B}(z):=\omega_{\tau}(\alpha_{z}(A)B)$ is analytic on the strip $\cI:=\{z\in\C\mid 0<\Im(z)<1\}$ and continuous on $\overline{\cI}$ such that 
	\begin{align}
		F_{A,B}(t)&=\omega_{\tau}(\alpha_{t}(A)B),\quad\quad F_{A,B}(t+i)=\omega_{\tau}(B\alpha_{t}(A)).
	\end{align}
	In other words, $\omega_{\tau}$ is an equilibrium state of $\alpha_{t}$.
\end{remark}

\begin{remark}
	Formally, by regarding $e^{iH_{\tau}t}$ as an element in $\cB$, we can write 
	\begin{align}
		U(t)&=\pi_{\tau}(e^{iH_{\tau}t})\pi_{\tau}'(e^{iH_{\tau}t})
	\end{align}
	or, equivalently,
	\begin{align}\label{liouville}
		L_{\tau}&=\pi_{\tau}(H_{\tau})-\pi_{\tau}'(H_{\tau})
	\end{align}
	by extending the representations $\pi_{\tau}$ and $\pi_{\tau}'$ to those unbounded operators affiliated to $\cB$.  
\end{remark}

\end{document}